\documentclass[sn-aps,iicol]{sn-jnl}

\usepackage{graphicx}%
\usepackage{multirow}%
\usepackage{amsmath,amssymb,amsfonts}%
\usepackage{amsthm}%
\usepackage{mathrsfs}%
\usepackage[title]{appendix}%
\usepackage{xcolor}%
\usepackage{textcomp}%
\usepackage{manyfoot}%
\usepackage{booktabs}%
\usepackage{algorithm}%
\usepackage{algorithmicx}%
\usepackage{algpseudocode}%
\usepackage{listings}%
\usepackage{natbib}

\usepackage{lineno}

\newcommand{\offline}{$\overline{\mbox{Off}}\underline{\mbox{Line}}$}

\newcommand{\CC}{{C\nolinebreak[4]\hspace{-.05em}\raisebox{.4ex}{\tiny\bf ++}}}

\usepackage[symbol]{footmisc}

\jyear{2021}%

\theoremstyle{thmstyleone}%
\theoremstyle{thmstyletwo}%

\theoremstyle{thmstylethree}%

\usepackage{fancyhdr}

\begin{document}


\thispagestyle{plain}\cleardoublepage

\begin{titlepage}
\pagestyle{empty}\cleardoublepage
\thispagestyle{empty}


\title[ESAF: Developments and Results]{\vspace{-1.2cm}Developments and results in the context of the JEM-EUSO program obtained with the ESAF Simulation and Analysis Framework\vspace{0.8cm}}


\clearpage


\author[1]{\vspace{-0.5cm}S.~Abe} 
\author[2]{J.H.~Adams~Jr.} 
\author[3]{D.~Allard} 
\author[2]{P.~Alldredge}
\author[4]{L.~Anchordoqui}
\author[5,6]{A.~Anzalone}
\author[7,8]{\\E.~Arnone}
\author[3]{B.~Baret}
\author[7,8]{D.~Barghini}
\author[7,8]{M.~Battisti}
\author[10]{J.~Bayer}
\author[12,11]{R.~Bellotti}
\author[13]{\\A.A.~Belov}
\author*[7,8]{M.~Bertaina}\email{bertaina@to.infn.it}
\author[14]{P.F.~Bertone}
\author[7,8]{M.~Bianciotto}
\author[15]{P.L.~Biermann}
\author[7,16]{\\F.~Bisconti}
\author[17]{C.~Blaksley}
\author[3]{S.~Blin-Bondil}
\author[18]{P.~Bobik}
\author[19]{K.~Bolmgren}
\author[20]{\\S.~Briz}
\author[2]{J.~Burton}
\author[12,11]{F.~Cafagna}
\author[16,21]{G.~Cambi\'e}
\author[]{D.~Campana$^{\textmd{22}}$}
\author[23]{\\F.~Capel}
\author[6,24]{R.~Caruso}
\author[16,17,21]{M.~Casolino}
\author[7,8]{C.~Cassardo}
\author[7,25]{A.~Castellina}
\author[26]{\\K.~\v{C}ern\'{y}}
\author[14]{M.J.~Christl}
\author[27,22]{R.~Colalillo}
\author[28,16]{L.~Conti}
\author[7,8]{G.~Cotto}
\author[29]{H.J.~Crawford}
\author[8]{\\R.~Cremonini}
\author[3]{A.~Creusot}
\author[56]{A.~Cummings}
\author[20]{A.~de~Castro G\'onzalez}
\author[30]{\\C.~de~la~Taille}
\author[]{L.~del~Peral$^{\textmd{33}}$}
\author[20]{R.~Diesing}
\author[30]{P.~Dinaucourt}
\author[27]{A.~Di~Nola}
\author[15]{\\A.~Ebersoldt}
\author[17]{T.~Ebisuzaki}
\author[20]{J.~Eser}
\author*[7,8,53]{F.~Fenu}\email{francesco.fenu@gmail.com}
\author[7,8]{S.~Ferrarese}
\author[31]{\\G.~Filippatos}
\author[31]{W.W.~Finch}
\author[27]{F.~Flaminio}
\author[16,28]{C.~Fornaro}
\author[31]{D.~Fuehne}
\author[19]{\\C.~Fuglesang}
\author[32]{M.~Fukushima}
\author[7,25]{D.~Gardiol}
\author[13]{G.K.~Garipov}
\author[7,8]{A.~Golzio}
\author[3]{\\P.~Gorodetzky}
\author[27,22]{F.~Guarino}
\author[20]{C.~Gu\'epin}
\author[10]{A.~Guzm\'an}
\author[15]{A.~Haungs}
\author[31]{T.~Heibges}
\author[]{\\J.~Hern\'andez~Carretero$^{\textmd{33}}$}
\author[27,22]{F.~Isgr\`o}
\author[29]{E.G.~Judd}
\author[34]{F.~Kajino}
\author[17]{I.~Kaneko}
\author[17]{\\Y.~Kawasaki}\equalcont{Deceased. This paper is dedicated to Yoshiya Kawasaki who contributed significantly to the ESAF developments during the JEM-EUSO mission.}
\author[15]{M.~Kleifges}
\author[13]{P.A.~Klimov}
\author[35]{I.~Kreykenbohm}
\author[36]{J.F.~Krizmanic}
\author[31]{\\V.~Kungel}
\author[2]{E.~Kuznetsov}
\author[20]{F.~L\'opez~Mart\'inez}
\author[18]{S.~Mackovjak}
\author[37]{D.~Mand\'{a}t}
\author[7,8]{\\M.~Manfrin}
% ...




\affil[1]{Nihon University Chiyoda, Tokyo, Japan} 
\affil[2]{University of Alabama in Huntsville, Huntsville, AL, USA} 
\affil[3]{Universit\'e de Paris, CNRS, AstroParticule et Cosmologie, F-75013 Paris, France}
\affil[4]{Lehman College, City University of New York (CUNY), NY, USA}
\affil[5]{INAF - Istituto di Astrofisica Spaziale e Fisica Cosmica di Palermo, Italy} 
\affil[6]{Istituto Nazionale di Fisica Nucleare - Sezione di Catania, Italy} 
\affil[7]{Istituto Nazionale di Fisica Nucleare - Sezione di Torino, Italy}  
\affil[8]{\orgdiv{Dipartimento di Fisica}, 
\orgname{Universit\`{a} degli Studi di Torino}, \orgaddress{\street{Via Pietro Giuria 1}, \city{Torino}, \postcode{10125},  \country{Italy}}}
% ...
\affil[10]{Institute for Astronomy and Astrophysics, Kepler Center, University of T\"ubingen, Germany}
\affil[11]{Istituto Nazionale di Fisica Nucleare - Sezione di Bari, Italy}
\affil[12]{Universit\`{a} degli Studi di Bari Aldo Moro, Italy}
\affil[13]{Skobeltsyn Institute of Nuclear Physics, Lomonosov Moscow State University, Russia}
\affil[14]{NASA Marshall Space Flight Center, Huntsville, AL, USA}
\affil[15]{Karlsruhe Institute of Technology (KIT), Germany}
\affil[16]{Istituto Nazionale di Fisica Nucleare - Sezione di Roma Tor Vergata, Italy}
\affil[17]{RIKEN, Wako, Japan}
\affil[18]{Slovak Academy of Science, Institute of Experimental Physics, Kosice, Slovakia}
\affil[19]{KTH Royal Institute of Technology, Stockholm, Sweden}
\affil[20]{University of Chicago, IL, USA}
\affil[21]{Universit\`{a} di Roma Tor Vergata - Dipartimento di Fisica, Roma, Italy}
\affil[22]{Istituto Nazionale di Fisica Nucleare - Sezione di Napoli, Italy}
\affil[23]{Max Planck Institute for Physics, Munich, Germany}
\affil[24]{Dipartimento di Fisica e Astronomia ``Ettore Majorana", Universit\`{a} di Catania, Italy}
\affil[25]{Osservatorio Astrofisico di Torino, Istituto Nazionale di Astrofisica, Pino Torinese, Italy}
\thispagestyle{empty}
\clearpage
\affil[26]{Joint Laboratory of Optics, Faculty of Science, Palack\'{y} University, Olomouc, Czech Republic}
\affil[27]{Universit\`{a} di Napoli Federico II - Dipartimento di Fisica ``Ettore Pancini", Italy}
\affil[28]{Uninettuno University, Rome, Italy}
\affil[29]{Space Science Laboratory, University of California, Berkeley, CA, USA}
\affil[30]{Omega, Ecole Polytechnique, CNRS/IN2P3, Palaiseau, France}
\affil[31]{Colorado School of Mines, Golden, CO, USA}
\affil[32]{Institute for Cosmic Ray Research, University of Tokyo, Kashiwa, Japan}
\affil[33]{Universidad de Alcal\'a (UAH), Madrid, Spain}
\affil[34]{Konan University, Kobe, Japan}
\affil[35]{ECAP, University of Erlangen-Nuremberg, Germany}
\affil[36]{NASA Goddard Space Flight Center, Greenbelt, MD, USA}
\affil[37]{Institute of Physics of the Czech Academy of Sciences, Prague, Czech Republic}
\affil[38]{National Centre for Nuclear Research, Otwock, Poland}
\affil[39]{University of Utah, Salt Lake City, UT, USA}
\affil[40]{Lab. of Math. and Sub-Atomic Phys. (LPMPS), Univ. Constantine I, Constantine, Algeria}
\affil[41]{National Astronomical Observatory, Mitaka, Japan}
\affil[42]{University of Iowa, Iowa City, IA, USA}
\affil[43]{Joint Institute for Nuclear Research, Dubna, Russia}
\affil[44]{Sungkyunkwan University, Seoul, Republic of Korea}
\affil[45]{Faculty of Physics, University of Warsaw, Poland}
\affil[46]{Istituto Nazionale di Fisica Nucleare - Laboratori Nazionali di Frascati, Italy}
\affil[47]{Space Regatta Consortium, Korolev, Russia}
\affil[48]{Hokkaido University, Sapporo, Japan}
\affil[49]{Shinshu University, Nagano, Japan}
\affil[50]{ISDC Data Centre for Astrophysics, Versoix, Switzerland}
\affil[51]{Centre for Development of Advanced Technologies (CDTA), Algiers, Algeria}
\affil[52]{Fairfield University, Fairfield, CT, USA}
\affil[53]{Agenzia Spaziale Italiana, Via del Politecnico, 00133, Roma, Italy}
\affil[54]{IRAP, Universit\'e de Toulouse, CNRS, Toulouse, France}
\affil[55]{Center for Pulsed Lasers (CLPU), Spain}
\affil[56]{Pennsylvania State University, PA, USA}
\affil[57]{Department of Physics and Astronomy, University of California, Irvine, USA}


\abstract{JEM--EUSO is an international program for the development of space-based
Ultra-High Energy Cosmic Ray observatories. The program consists of a series
of missions which are either under development or in the data analysis phase.
All instruments are based on a wide-field-of-view
telescope, which operates in the near-UV range, designed to detect the
fluorescence light emitted by
extensive air showers in the atmosphere. We describe the simulation software ESAF
in the framework of the JEM--EUSO program and explain the physical assumptions used. 
We present here the implementation of the JEM--EUSO, POEMMA, K--EUSO,
TUS, Mini--EUSO, EUSO--SPB1 and EUSO--TA configurations in ESAF. 
For the first time ESAF simulation outputs are compared with experimental data. 
}


\keywords{JEM--EUSO, Monte Carlo Simulations, ESAF}

\clearpage\maketitle
\pagestyle{empty}
\end{titlepage}

\twocolumn
\pagestyle{headings}
\section{Introduction}\label{sec:intro}
The study of Ultra-High Energy Cosmic Rays (UHECRs) and the understanding of particle
acceleration in the cosmos is of utmost importance for astro-particle 
physics as well as for fundamental physics. The current main 
goals are to identify sources of UHECRs and their composition. 
UHECRs above $5\times10^{19}$~eV have a flux lower than 1 event per century per square
kilometer~\cite{auger}, therefore, huge sensitive volumes are necessary to collect enough statistics.

A space-based detector for UHECR research has the advantage of a very large exposure and a uniform coverage of the celestial sphere. The idea of space-based observation of UHECRs was first proposed by
John Linsley in the late 70s with the SOCRAS concept~\cite{linsley}.
The principle of observation is based on the detection of UV light emitted by isotropic fluorescence of atmospheric nitrogen
excited by Extensive Air Showers (EAS) in the Earth’s atmosphere and forward-beamed Cherenkov radiation diffusely reflected from 
the Earth’s surface or dense cloud tops. 
In 1995 Linsley’s original idea was re-adapted by Yoshiyuki Takahashi into
the MASS concept~\cite{takahashi} which evolved in 1996, in the US, into the OWL mission~\cite{owl}.
In parallel the MASS, or Airwatch concept as it was later on renamed, evolved in Europe into EUSO, the Extreme Universe Space
Observatory. Livio Scarsi first proposed EUSO as a free-flyer to the European Space Agency's (ESA)  F2/F3 \cite{f2f3Esa} call
in 2000. ESA selected the mission but recast it as a payload for the Columbus
module of the International Space Station (ISS)~\cite{euso}. The phase-A study for the feasibility of EUSO,
started in 2001 
and was successfully completed in March 2004. Although
EUSO was found technically ready to proceed into phase B, ESA did not continue
the program. 

\begin{figure*}[ht]
\centering
\includegraphics[width=1.\textwidth]{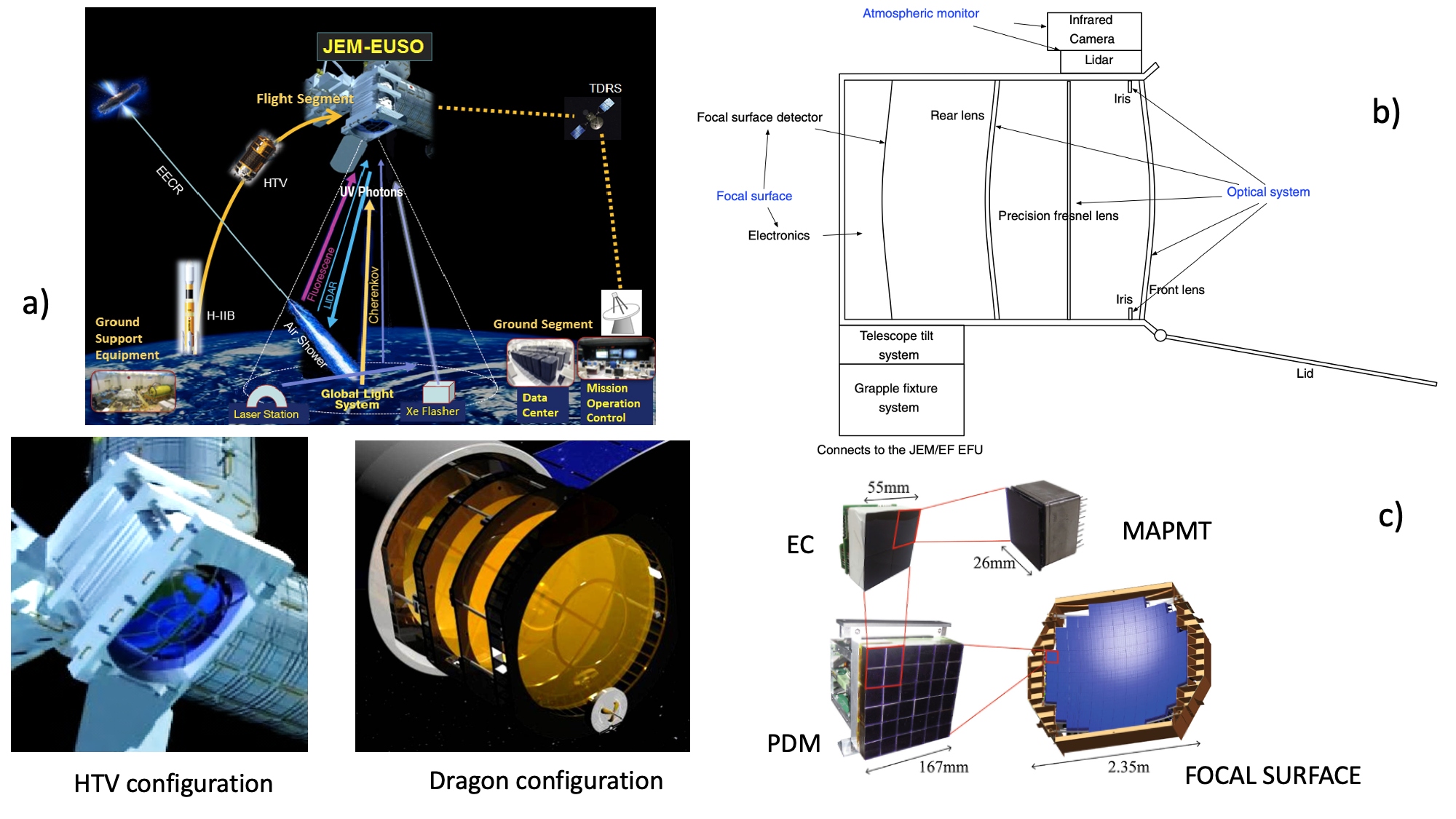}
\caption{Panel a): Conceptual view of the whole JEM--EUSO system. Panel b): Conceptual design of the JEM--EUSO system: 
three Fresnel lenses focus the light on the focal surface. Panel c): Conceptual design of the JEM--EUSO Focal Surface with
its elements and sub-elements parts. 
The focal surface of the detector is formed by 137 Photo Detector Modules (PDMs) comprising of $\sim5000$ Multi-Anode
Photo-Multiplier Tubes (MAPMTs) in total (36 MAPMTs per PDM, 64 pixels each). 
Each PDM is composed by 9 Elementary Cells (ECs). 
See the text for details. Figure adapted from~\cite{ea-instrument}.}
\label{fig:jemeuso}
\end{figure*}

\begin{figure}[h]
\centering
\includegraphics[width=\columnwidth]{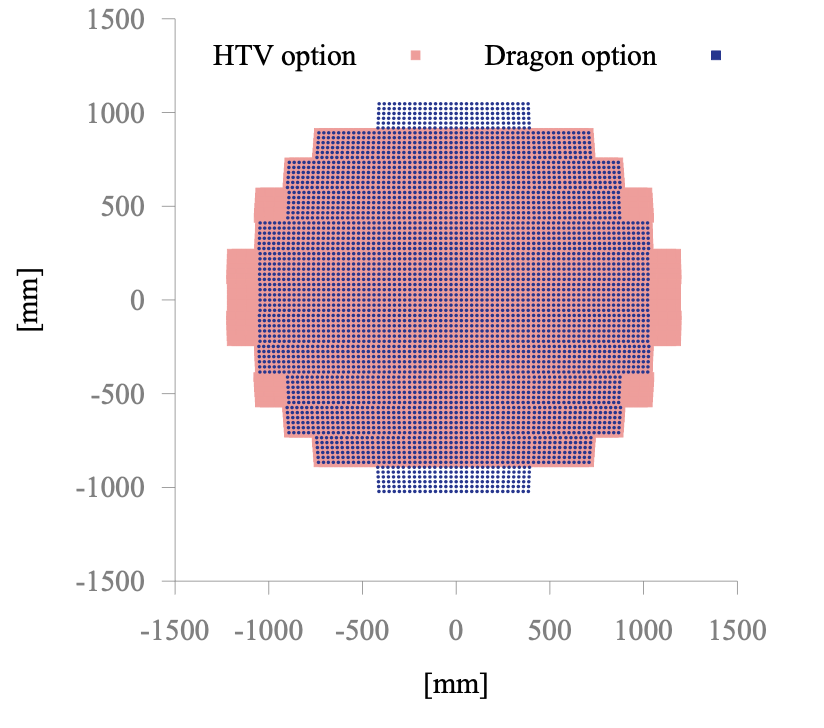}
\caption{A diagram of the position of the MAPMTs of the Dragon configuration (blue) of JEM--EUSO superimposed on the same diagram for the HTV configuration (red). Image taken from~\cite{guzman-icrc2015}.}
\label{fig:dragon-htv-fs}
\end{figure}

In 2006, the Japanese and US teams, under the leadership of Yoshiyuki Takahashi,
redefined the mission as an observatory attached to KIBO, the Japanese Experiment
Module (JEM) of the ISS. They renamed the mission JEM--EUSO and started a new
phase-A study targeting launch in 2013 in the framework of the second utilization
phase of the JEM/EF (Exposure Facility)~\cite{yoshi-jemeuso}.
The Phase A/B1 study of JEM--EUSO led by Japan Aerospace Exploration Agency (JAXA) continued with extensive
simulations, design, and prototype developments that significantly improved the
JEM--EUSO mission profile (see Fig.~\ref{fig:jemeuso}), targeting a launch in 2016~\cite{jemeuso}.
As a result of this study, the main telescope was designed with a wide Field-of-View 
(FoV;  $\sim0.85$~sr)
optics composed of three Fresnel lenses. 
Different configurations have been studied: the
``side-cut'' version of the instrument with a 2.65~m major axis and 1.9~m
minor axis (4.5~m$^2$ optical aperture) to fit in the cargo of the JAXA HTV
rocket as described in~\cite{exposure} and the SpaceX Dragon option with 2.5~m diameter 
circular optics~\cite{ea-instrument}. 
The difference between the two configurations is shown in Fig.~\ref{fig:dragon-htv-fs}. The use of Fresnel lenses realized in Poly(Methil MethAcrylate) - PMMA - allows building a refractor telescope capable of meeting the requirements of a large optical system with wide FoV with the constraints of a spaceborne experiment. Furthermore, the reduced thickness of the lenses allows to reduce the mass of the optics resulting in a light system capable of withstanding launch vibration and thermal expansion.

The telescope records the EAS-induced tracks with a
time resolution of $2.5~\mu$s, known as a Gate Time Unit (GTU). 
The Focal Surface (FS) detector is formed by 137 Photo Detector Modules (PDMs) comprising of $\sim5000$ Multi-Anode
Photo-Multiplier Tubes (MAPMTs) in total (36 MAPMTs per PDM, 64 pixels each). 
Each PDM is composed by 9 Elementary Cells (ECs). 
One EC is composed by 4 MAPMTs, 64 pixels each.
The FS detector is highly pixelated in $\sim3\times10^5$ channels providing a pixel FoV
of $\sim0.074^\circ$, equivalent to $\sim0.5$~km at ground seen from an altitude of
$\sim400$~km. An optical filter is placed in front of the MAPMTs to select photons mainly in the
fluorescence bandwidth (290--430~nm).

In 2013 it became clear that JEM--EUSO could not proceed further within the JAXA leasdership.
The mission was put on hold status and JEM--EUSO was reoriented as an extensive pathfinder program, with the acronym redefined as the 
Joint Experiment Missions for Extreme Universe Space Observatory~\cite{EUSO-program}.
The program includes 
several missions from ground (EUSO--TA~\cite{eusota}),
on board of stratospheric balloons (EUSO--Balloon~\cite{eusobal}, EUSO--SPB1~\cite{spb1}, and EUSO--SPB2~\cite{spb2}),
and from space (TUS~\cite{tus} and Mini--EUSO~\cite{minieuso}). Each employs fluorescence
detectors to test the observational technique, and to validate the technology. The final goal of the program is the realization of a
large space-based mission following the concepts 
developed in the past decades, namely the medium-size
K--EUSO~\cite{keuso} and the large-size POEMMA~\cite{poemma} missions.

All these detectors demand extensive simulation work to estimate their
performance and to support the analysis of the 
collected data. It was clear at the time of the JEM--EUSO mission that the most
efficient way was to re-adapt 
the existing software instead of developing totally new code. For this reason,
the two official software packages adopted by JEM--EUSO are the 
Euso Simulation and Analysis Framework (ESAF)~\cite{berat}, originally developed
within the EUSO project, and the 
\offline\ package~\cite{offline} designed for the Pierre Auger 
Observatory~\cite{auger-nim}. A comparison between the two frameworks and their
designs is reported in~\cite{spb2} while examples of cross-checks carried out in the past on their relative performance is summarized in Appendix~\ref{sec:esaf-offline}.
The main motivations to adopt both packages are:
a)~it is straightforward to re-adapt the EUSO code to the JEM--EUSO configuration; b)~\offline\ output is extensively tested  within
the Auger project and thus with experimental data;
c)~the possibility to adopt both packages gives opportunities for 
cross-checks. 
A synthetic description of the developments performed with the \offline\ software 
to accommodate the different configurations of the telescopes of the JEM--EUSO program
can be found in~\cite{tom-offline}.
In this paper we summarize developments and performance results obtained with ESAF, including 
some of those already discussed in earlier publications. 

The main objective of this paper is to demonstrate the potential and flexibility of the ESAF software and its
applications in the context of the JEM--EUSO program.
A comparison with experimental data is provided to show the utility of the ESAF software in the interpretation of the data.
The paper is structured in
the following way. 
Section~\ref{sec:missionsoverview} outlines the detector characteristics of the different projects of the JEM--EUSO program
which have been implemented in ESAF.
Section~\ref{sec:esaf} summarizes the main features of the ESAF framework while 
Section~\ref{sec:esaf-jemeuso} reports the new ESAF developments performed within the JEM--EUSO program.
Section~\ref{sec:results} provides the key results obtained with simulations for the different detector configurations implemented
in the software, namely JEM--EUSO, POEMMA, K--EUSO, Mini--EUSO, TUS, EUSO--SPB1 and EUSO--TA as well as comparisons with data.
The conclusions and perspectives are the subject of Section~\ref{sec:conclusions}.

\section{The missions of the JEM-EUSO program}
\label{sec:missionsoverview}

In this section we summarize the different configurations implemented in ESAF to simulate the performance of the various projects of the JEM-EUSO program which have been defined since the original JEM-EUSO mission was put on hold. A comparison of the main parameters of the different configurations is presented in table \ref{tab:parameters}.

\begin{figure*}[!ht]
\centering
\includegraphics[width=1.\textwidth]
{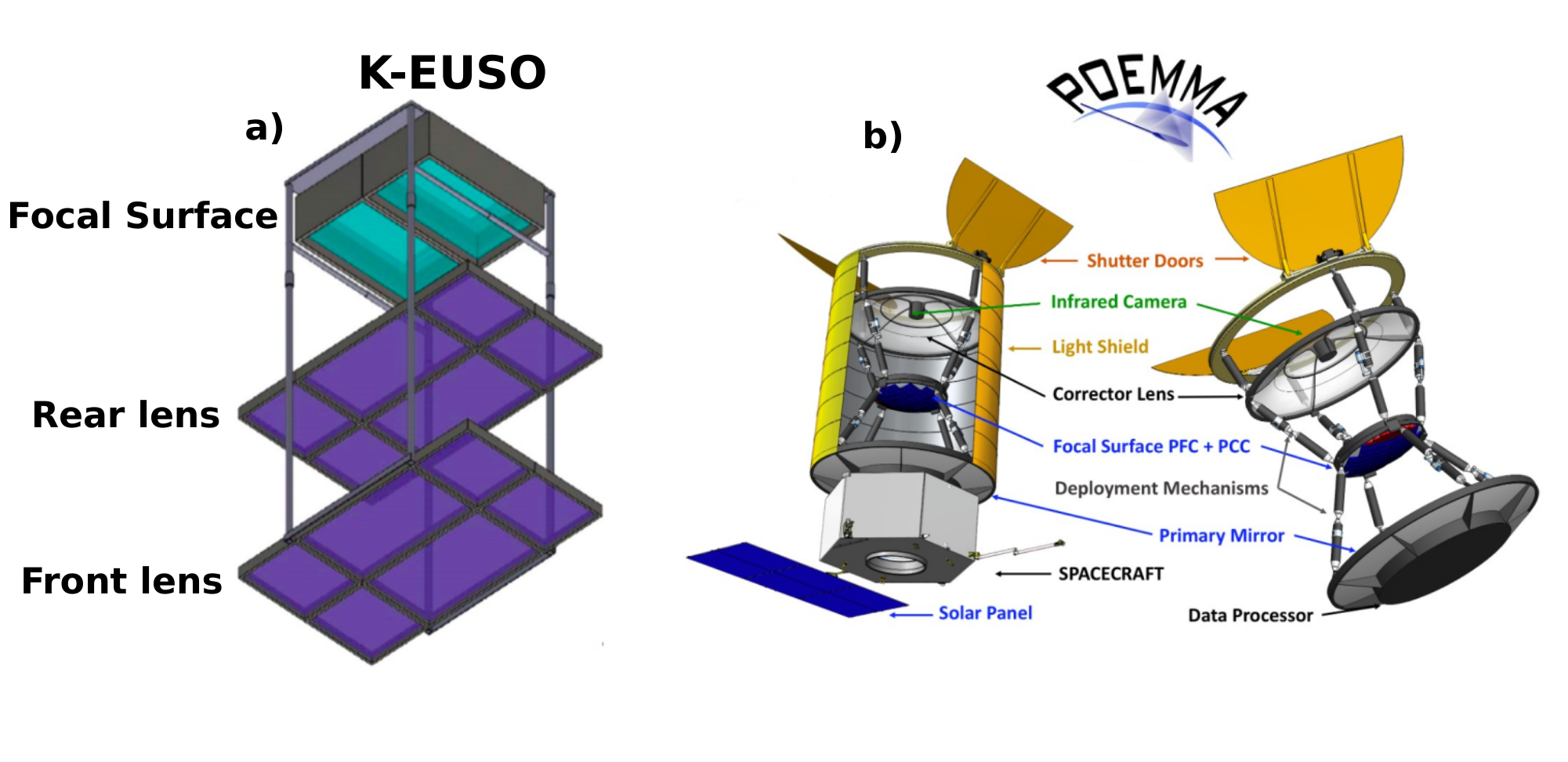}
\caption{Scheme of the simulated K--EUSO detector (panel a) and of the POEMMA telescope (panel b). Figure adapted from~\cite{keuso} and \cite{poemma}.}
\label{fig:keuso-poemma}
\end{figure*}

K--EUSO is the result of the joint efforts to improve the performance of the Russian KLYPVE
mission~\cite{klypve} by employing the technologies developed for the JEM--EUSO mission,
such as the focal
surface detectors and the readout electronics. The KLYPVE mission, named after Russian words
``ultra-high energy cosmic rays''
has undergone pre-phase A study, including launch and accommodation on the ISS. Since its first
conception as KLYPVE, the K--EUSO project has passed various modifications aimed to increase
the FoV
and UHECR statistics~\cite{keuso1,keuso2}, compatibly with shipping possibilities using
the Soyuz spacecraft and to decrease the number of
external vehicle activities by the astronauts. All the different improved solutions have been implemented in ESAF. We report in Section~\ref{sec:results} on the expected performance
of the latest version of the
instrument under study~\cite{fenu-keuso2021,keuso} (see Fig.~\ref{fig:keuso-poemma}a)).


In this configuration, the detector consists of a refractive optical system of
$1400~\text{mm}\times2400~\text{mm}$ size. 
The optics is based on two Fresnel lenses that focus the light onto a focal surface of 
$1300~\text{mm}\times1000~\text{mm}$ size. The FS consists of 44 PDMs. Each pixel has a field of view of 0.1$^\circ$ which corresponds 
to $\sim$700 m on the ground. The time resolution is in the process of being optimized and ranges from 1 $\mu$s to 2.5 $\mu$s. The exact value will be based on a trade-off between the limited hardware and telemetry budgets and the need of a good time resolution. In the following, the 2.5 $\mu$s GTU has been adopted for simulations.

The Probe Of Extreme Multi-Messenger Astrophysics (POEMMA) design~\cite{poemma} combines the concept developed for the OWL
mission and the experience of designing the JEM--EUSO fluorescence detection camera. POEMMA is composed of two identical satellites flying in formation at an altitude of 525 km with the ability to observe overlapping regions during moonless nights at angles ranging from nadir to just above the limb of the Earth, but also with independent pointing strategies to exploit at maximum the scientific program of the mission. 
Each telescope (see Fig.~\ref{fig:keuso-poemma}b)) is composed of a wide (45$^\circ$) FoV Schmidt optical 
system with an optical collecting area of over 6 m$^2$. The focal surface of POEMMA is composed of a hybrid of two types
of cameras: about 90\% of the FS is dedicated to the POEMMA fluorescence camera (PFC), while
the POEMMA Cherenkov camera (PCC) occupies the crescent moon shaped edge of the FS which
images the limb of the Earth. The PFC is composed of 55 JEM--EUSO PDMs based on MAPMTs for a total of
$\sim$130,000 channels. The GTU for the PFC is 1 $\mu$s. 
The much faster POEMMA Cherenkov camera (PCC) is composed of Silicon Photo-Multipliers (SiPMs) which is tested with EUSO--SPB2.

\begin{figure}[!ht]
    \centering
    \includegraphics[width=0.47\textwidth]{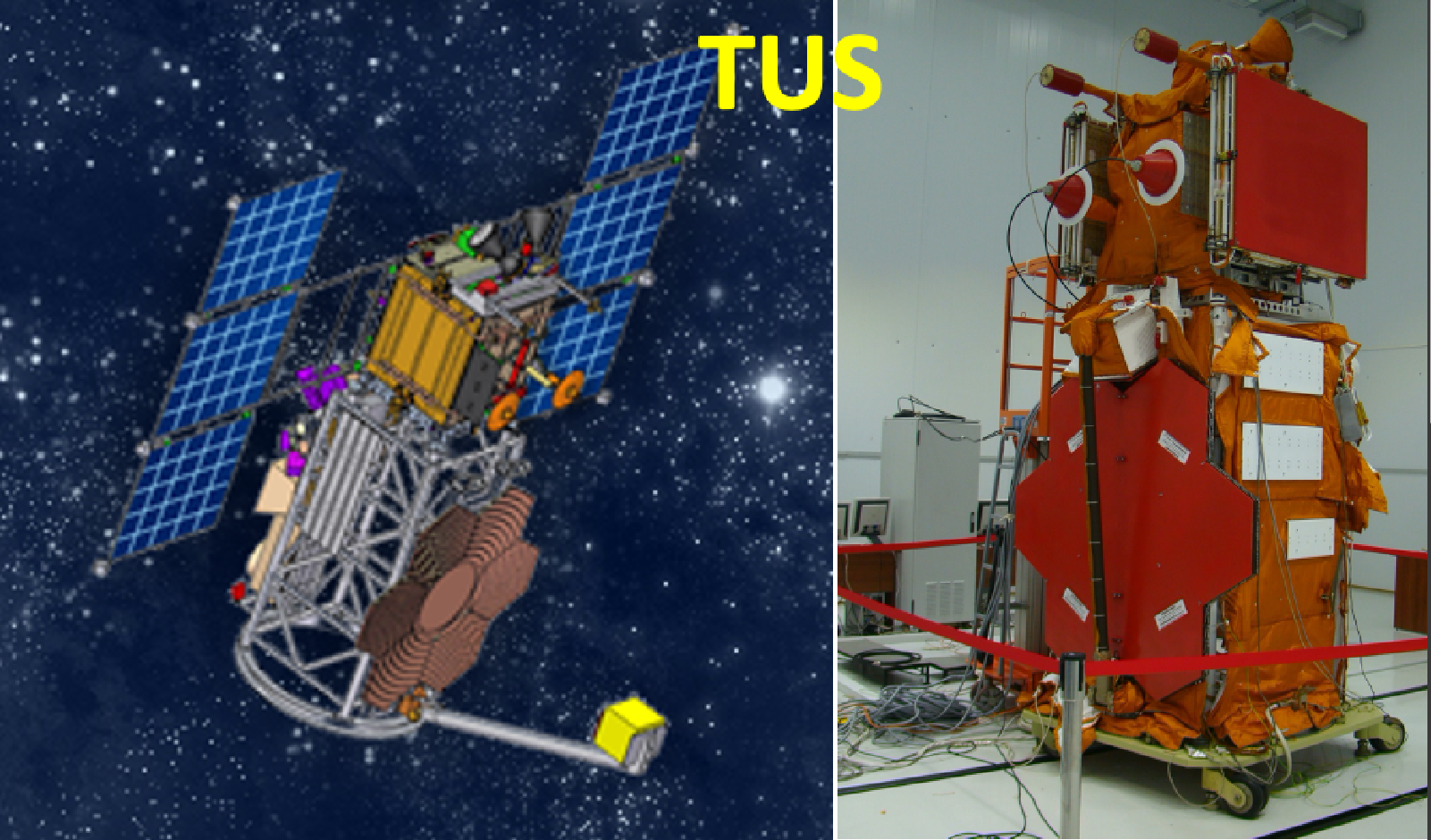}
    \caption{Artist's view of the TUS detector on board the Lomonosov
    satellite (left) and preflight
    preparations at the Vostochny cosmodrome (right).}
    \label{fig:tus}
\end{figure}

The world's first orbital detector aiming at detecting UHECRs was
the Tracking Ultraviolet Setup (TUS) UV telescope,
launched on April~28, 2016 as a
part of the scientific payload of the Lomonosov satellite~\cite{tus},
see Fig.~\ref{fig:tus}.

TUS provides the first opportunity to compare ESAF simulations to real data taken from space.
   Some examples can be found in~\cite{tus-jcap} and are reported here.
The instrument recorded data until the end of November 2017. Different
scientific modes were tested: cosmic ray, lightning and meteor modes.
The satellite had a sun-synchronous orbit (i.e. passing over any given point of the earth surface at the same local mean solar time) with an inclination of 97.3$^{\circ}$,
a period of $\sim94$~min, and a height of 470--500~km.
The TUS telescope consisted of two main parts: a modular Fresnel mirror-concentrator
with an area of $\sim2~\text{m}^2$ and 256 PMTs
arranged in a $16\times16$ photo-receiver matrix located in the focal plane of
the mirror. The FoV of one pixel was 10~mrad, which corresponds
to a spatial spot of $\sim5~\text{km}\times5~\text{km}$ at sea level.
Thus, the full area observed by TUS at any moment was
$\sim80~\text{km}\times80~\text{km}$. TUS was sensitive to the near-UV band
and had a time resolution of 0.8~$\mu$s in the cosmic ray mode, with a full
temporal window of 256 time steps. TUS data offer the opportunity to
develop strategies in the analysis and reconstruction of events which will be
essential for future space-based missions.

\begin{figure*}[!ht]
    \centering
    \includegraphics[width=0.52\textwidth]{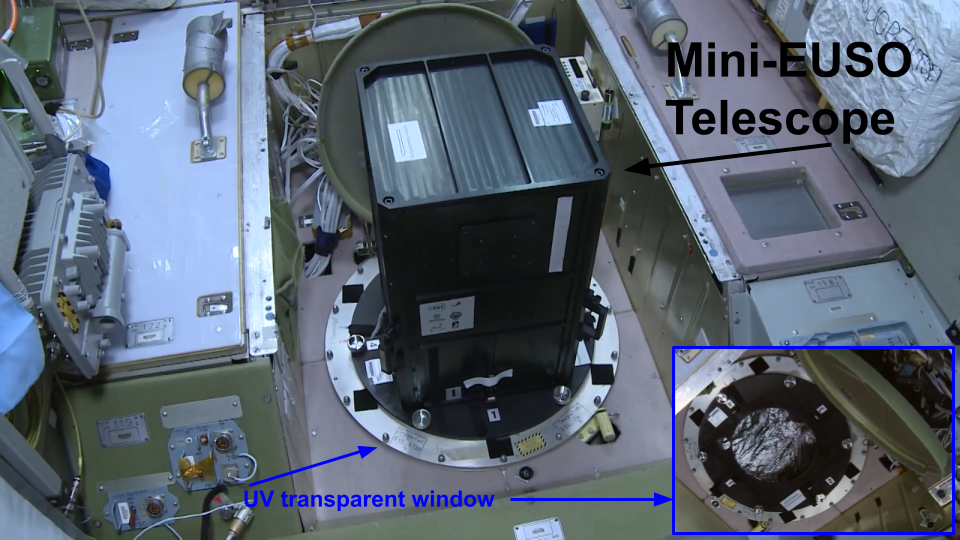}~
    \includegraphics[width=0.46\textwidth]{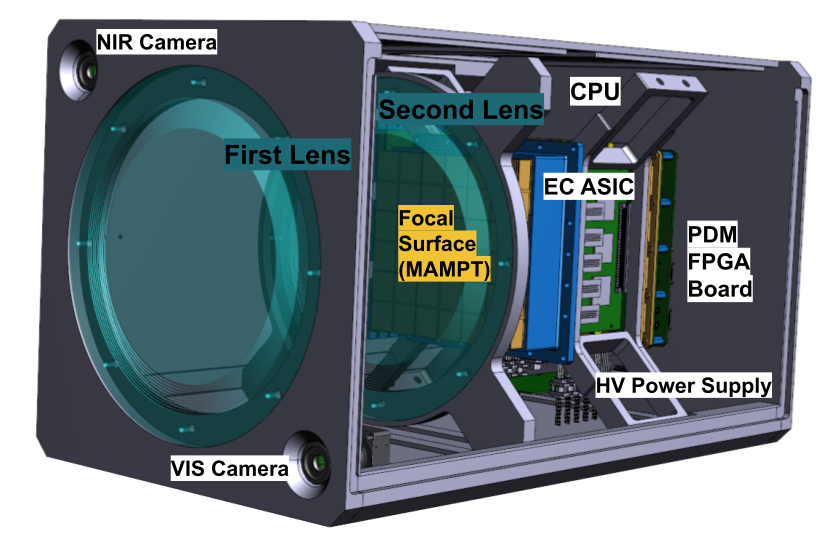}
    \caption{Mini--EUSO
attached to the Zvezda module on the ISS (left) and schematic view of the different parts of the Mini--EUSO detector (right). 
}
    \label{fig:minieuso}
\end{figure*}

Mini--EUSO~\cite{minieuso} is an UV telescope launched in August 2019
and installed periodically inside the ISS, with installations occurring every couple of weeks since October 2019. Mini-EUSO is installed looking down on the Earth from a nadir-facing
window in the Russian Zvezda module (see Fig.~\ref{fig:minieuso}). So far $\sim$80 sessions of about 12 hours of data 
taking have been performed. 
Mini--EUSO maps the Earth in the UV range (290 - 430 nm) with a
spatial resolutions of $\sim$6 km (similar to TUS) and three different temporal resolutions of
2.5 $\mu$s, 320 $\mu$s, and 41 ms operating simultaneously. While the 41 ms time range allows continuous video-taking, the other
two modes allow acquisitions of 4 packets of 128 GTUs each every 5.24 s when the trigger condition is satisfied, to catch fast luminous transients (flashes,
lightning, etc.). The optical system consists of 2 Fresnel lenses of 25 cm diameter each
with a large FoV of $\sim$44$^{\circ}$.
Data carried down to Earth from the ISS allowed for
the first analyses showing that Mini--EUSO observes different Earth emissions depending on the surface visible, e.g.,
ground, sea, or clouds as well as slow transients such as meteors. Tens of thousands of meteor events have been identified in the data
with absolute magnitude lower than +5 (the events last typically in the order of 1 second).
At shorter times scales (typically hundreds of $\mu$s), several hundreds of lightning associated signals (among them 26 elves) 
have been detected. In addition, many anthropogenic flashes presumably related to airport lights or other flashing tower lights have been acquired.
A summary of the most recent results of Mini--EUSO can be found in
~\cite{minieuso-uv}.


While TUS was conceived mainly to prove the observation of UHECRs from space
with a FS-instrumentation similar to ground-based detectors,
Mini--EUSO has been developed in order to test the same FS-instrumentation
foreseen for K--EUSO and POEMMA. 
Moreover, it is important to recall that Mini--EUSO was designed to detect
similar photoelectron counts per pixel as JEM--EUSO in the case of diffused light sources.
This is done by compensating the $\sim$$10^{-2}$ times optics aperture with $\sim$$10^2$
times wider pixel FoV.
Therefore, these results on diffuse light sources are representative of observations of
the future large missions K--EUSO and POEMMA, which have similar apertures and instantaneous FoV.
As a consequence, Mini--EUSO energy threshold for UHECRs is well above $10^{21}$~eV as explained in Section~\ref{sec:results}.

\begin{figure*}[!ht]
\begin{center}
\includegraphics[width=1.\textwidth]{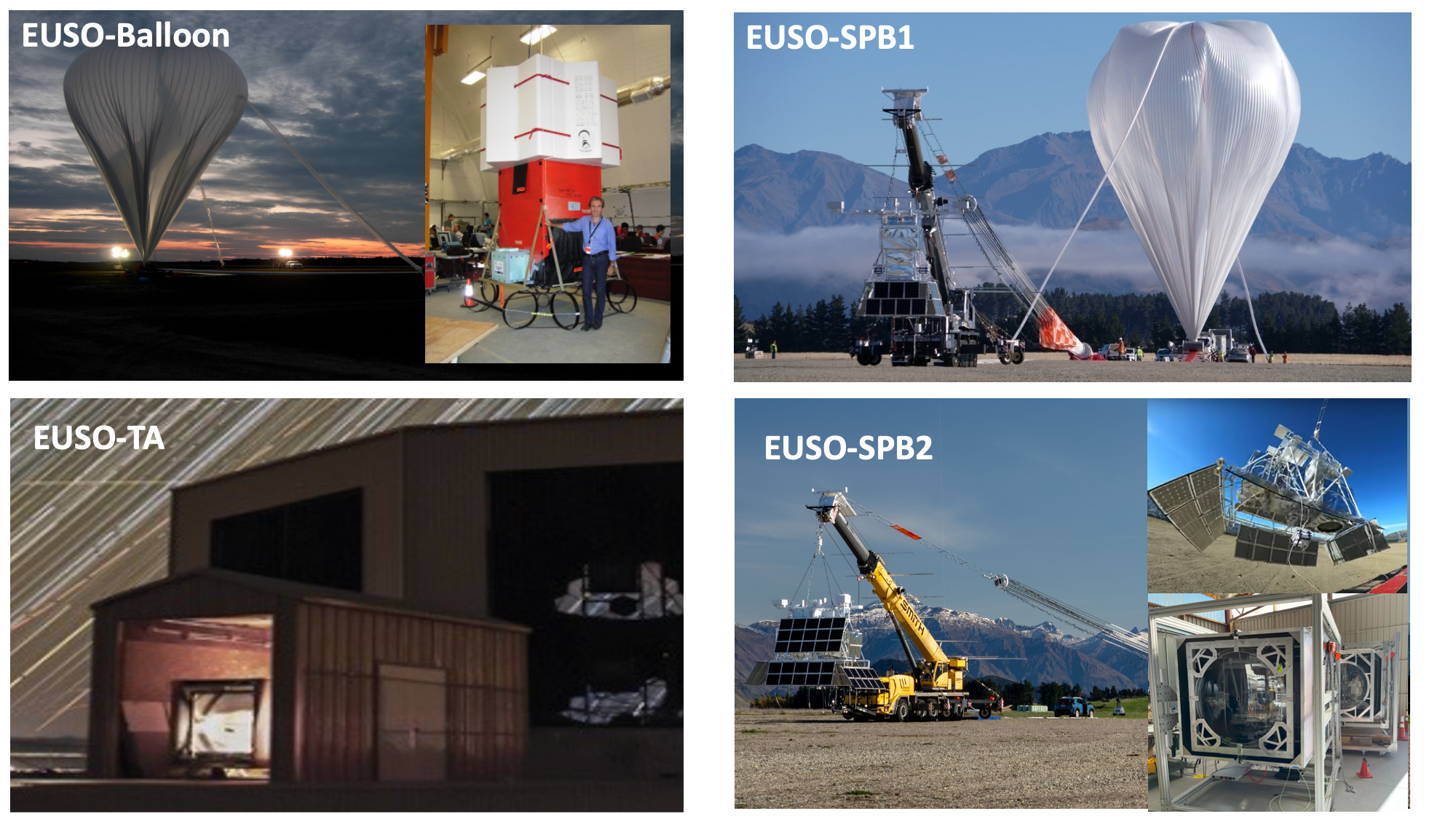}
\end{center}
\caption{Stratospheric balloon missions EUSO--Balloon, EUSO--SPB1, and EUSO--SPB2, and ground mission EUSO--TA of the 
JEM--EUSO program. Image adapted from~\cite{EUSO-program}.}
\label{fig:detector}
\end{figure*}

The JEM--EUSO program includes stratospheric balloon missions
with increasing level of performance and upgraded designs (see Fig.~\ref{fig:detector}). In addition to
demonstrating the capabilities of the JEM--EUSO instruments to detect and
reconstruct EASs from the edge of space, they also give access to direct
measurement of the UV nightglow emission and artificial UV contributions above
ground and oceans, which are important information to optimize the design of
the space-based missions.
Three balloon flights have been performed so far: EUSO--Balloon (Canada, 1 night),
EUSO--SPB1 (Pacific Ocean, 12 nights), and EUSO--SPB2 (Pacific Ocean, 37 hours).
The telescope configuration of EUSO--Balloon and EUSO--SPB1 is similar: two Fresnel lenses of $\sim1$~m$^2$ each and
one PDM as FS with 2.5~$\mu$s time resolution.

EUSO--Balloon~\cite{eusobal} was launched by National Centre for Space Studies in France (CNES) from the Timmins base
in Ontario (Canada) on the moonless night of August 25, 2014.
After reaching the floating altitude of $\sim$38 km, EUSO--Balloon imaged the UV
intensity with a spatial and temporal resolutions of 130 m and
2.5 $\mu$s, respectively, in the wavelength range 290 - 430 nm for more than 5 hours before
descending to ground. The full FoV in nadir mode was $\sim11^{\circ}$.
During 2.5 hours of EUSO--Balloon flight, a
helicopter circled under the balloon operating UV flashers and
a UV laser to simulate the optical signals from UHECRs, to calibrate the
apparatus, and to characterise the optical atmospheric conditions. Data collected
by EUSO--Balloon have been analyzed
to infer different information among which the response of the detector to the
UV flasher and laser events, and the UV radiance from the Earth atmosphere and
ground in different conditions: clear and cloudy atmosphere, forests,
lakes, as well as city lights~\cite{kenji-uv}.
The helicopter events proved to be extremely useful to understand the system's
performance and to test the capability of EUSO--Balloon to detect and reconstruct signals similar
to EASs~\cite{eser-jinst}. A summary of the results of the EUSO--Balloon mission is reported in~\cite{balloon-results}.

EUSO--SPB1~\cite{spb1}
was launched on April 25, 2017 from Wanaka (New Zealand) as a mission of opportunity
on a NASA SPB test flight planned to circle the southern
hemisphere. The telescope was an upgraded version of that used in the EUSO--Balloon
mission. The JEM--EUSO first level trigger was implemented with adaptations for a balloon observation~\cite{nim-trigger}.
Prior to flight, in October 2016, the fully assembled EUSO--SPB1 detector was
tested for a week at the EUSO--TA site to measure its response and for calibrations by means of a
portable Ground Laser System (GLS). Observations of the Central Laser Facility
(CLF), stars and meteors were performed. Unfortunately, although the instrument was showing
nominal behaviour and performance, the flight was terminated prematurely in the
Pacific Ocean about 300 km SE of Easter Island after only 12 days
aloft of the $\sim$100 scheduled, due to a leak in the balloon, and the payload was lost. During flight, $\sim$30 hours of data were collected~\cite{SCOTTI201994}. 

EUSO--SPB2~\cite{spb2}
was launched on May 13, 2023 from Wanaka on a NASA SPB test flight. 
It was equipped with 2
telescopes. One telescope was devoted to UHECR measurements using the fluorescence technique.
EUSO--SPB2 employed a Schmidt camera with 3 PDMs, a 1 $\mu$s GTU, and a more efficient balloon-oriented 
trigger logic to improve the sensitivity of the instrument~\cite{george-asr}. The FS of the second telescope
was based on SiPM sensors and a dedicated electronics to detect the Cherenkov emission in air by UHECR EASs. 
%
Unfortunately, the balloon developed a leak and was terminated over the Pacific Ocean after only about 37 hours of flight. Despite the very short flight, all instruments performed very well. Analysis of the flight data is ongoing. Preliminary results were presented in ~\cite{johannes_icrc2023}.

\onecolumn
\begin{sidewaystable}[h!]
\sidewaystablefn
\small
\begin{center}
\resizebox{22cm}{!} 
{ 
\begin{minipage}{\textheight}
\caption{The main parameters of different configurations of JEM-EUSO experiments}\label{tab:parameters}
\begin{tabular*}{\textheight}
{@{\extracolsep{\fill}}l|c|c|c|c|c|c|c|c|c@{\extracolsep{\fill}}}
Experiment & JEM-EUSO  & K-EUSO  & POEMMA & Mini-EUSO & TUS & EUSO-Balloon & EUSO-SPB1 & EUSO-SPB2 & EUSO-TA\\
\midrule
Optics type & lenses & lenses & mirror & lenses & mirror & lenses & lenses & mirror & lenses\\
\midrule
Optics \\aperture (m$^2$) & $\sim$4.5 & $\sim$3 & $\sim$6 & $\sim$0.05 & $\sim$2 & $\sim$1 & $\sim$1 & $\sim$1 & $\sim$1\\
\midrule
Height (km) & 400 & 400 & 525 & 400 & $\sim$500 & $\sim$38 & $\sim$33 & $\sim$33 & $\sim$0\\
\midrule
FoV ($^\circ$) &$\sim$64$\times$45 & $\sim$20$\times$15 & $\sim$45 & $\sim$44 & $\sim$9 & $\sim$11 & $\sim$11 & 12$\times$36 & $\sim$11\\
\midrule
Area at ground \\(km$^2$) & 1.4$\times$10$^5$ & 4.8$\times$10$^4$ & 1.5$\times$10$^5$ & $\sim$8$\times$10$^4$ & $\sim$6.4$\times$10$^3$ & $\sim$54 & $\sim$40 & $\sim$150 & ---\\
\midrule
PDMs & 137 & 44 & 55 & 1 & 1 & 1 & 1 & 3 & 1\\
\midrule
Pixels & 3.2$\times$10$^5$ & 1$\times$10$^5$ & 1.3$\times$10$^5$ & 2304 & 256 & 2304 & 2304 & 6912 & 2304\\
\midrule
Spatial ang. \\resolution ($^\circ$) & $\sim$0.074 & $\sim$0.1 & $\sim$0.084 & $\sim$0.9 & $\sim$0.7 & $\sim$0.2 & $\sim$0.2 & $\sim$0.25 & $\sim$0.2\\
\midrule
Pixel size \\at ground (km) &$\sim$0.5 - 0.6 & $\sim$0.6 - 0.7 & $\sim$0.8 & $\sim$6.3 & $\sim$5.0 & $\sim$0.13 & $\sim$0.12 & $\sim$0.14 & ---\\
\midrule
GTU ($\mu$s) & 2.5 & 2.5 & 1.0 & 2.5 & 0.8 & 2.5 & 2.5 & 1.0 & 2.5\\
\midrule
Bckg level \\(cts/pix/GTU) & 1.1 & 0.6 & 1.5 & $\sim$1.0 & $\sim$18.0 & $\sim$0.5 - 1.0 & 1 - 2 & $\sim$1 & 1 - 2\\
\midrule
Reference & \cite{exposure,ea-instrument} & \cite{keuso} & \cite{poemma,poemma-prd} & \cite{minieuso,mini-trigger} & \cite{tus,tus-jcap} & \cite{eusobal,balloon-results} & \cite{spb1} & \cite{spb2,johannes_icrc2023} & \cite{eusota}\\
\botrule
\end{tabular*}
\end{minipage}
}
\end{center}
\end{sidewaystable}
\twocolumn

EUSO--TA~\cite{eusota} is a ground-based telescope, installed at the Telescope Array~\cite{ta} (TA) site in Black Rock
Mesa, Utah, USA (see Fig.~\ref{fig:detector}). This is the first detector to successfully use a Fresnel lens
based optical system and one PDM foreseen for JEM--EUSO with a 2.5 $\mu$s GTU. 
The FoV is $10.6^{\circ} \times 10.6^{\circ}$.
The telescope is located in front of one of the fluorescence detector stations of the TA experiment. Between 2015 and 2016, a few campaigns of
joint observations with TA allowed EUSO--TA to detect 9 UHECR events in $\sim$140 hours of data taking, all lasting 1 or 2 GTUs at maximum, as well as
a few meteors. The limiting magnitude of +5.5 on summed frames ($\sim$3 ms) was established.
These observations provided important data to optimize the detector technology
in view of subsequent balloon and space missions.
The current upgrade of the detector includes a new acquisition system based on the Zynq board, like in Mini--EUSO, and the implementation of a self-triggering system which has become
operational in June 2022.

\section{An overview of the Euso Simulation and Analysis Framework}
\label{sec:esaf}
The Euso Simulation and Analysis Framework (ESAF) is a simulation and analysis software 
specifically designed for the performance assessment of space-based cosmic ray
observatories. It has been developed in the framework of the 
EUSO project~\cite{berat}. 
The software consists of $\sim 2 \cdot 10^{5}$ lines of code written in \CC~following
an object oriented approach and makes use of 
the ROOT data analysis framework developed at European Organization for Nuclear Research (CERN)~\cite{root-official}.

In this section, we briefly summarize the key aspects of the ESAF software while in the following one
we report on the new developments to support the JEM--EUSO program. A short technical description of the ESAF design is reported in Appendix~\ref{sec:esaf-design}. See reference~\cite{berat} for a detailed and comprehensive description of the ESAF software.

The ESAF software performs the simulation of an UHECR event, its detection and
the shower parameter reconstruction.
In more detail, the ESAF code includes: a)~EAS simulation both by means of internally
developed algorithms (e.g., the SLAST shower generator, which includes the Greisen--Ilina--Linsley (GIL) parameterization~\cite{gil}) and interfaces to existing widely-used codes (e.g., CORSIKA~\cite{corsika} or CONEX~\cite{conex}); b)~a complete description of the atmosphere, including aerosols, ozone and Rayleigh scattering and clouds;
c)~fluorescence and Cherenkov light production; d)~a complete simulation of photon propagation, from the production
point up to the telescope, including diffuse reflection interactions with ground and atmosphere, and a Monte-Carlo code dealing with
multiple scattering; e)~simulation of optics, geometry and a photodetector of a telescope; f)~simulation of the
electronics, trigger, and background; g)~pattern recognition and shower signal identification above background; h)~reconstruction of direction, energy, and slant depth of the shower maximum ($X_{\max}$), 
i.e., the atmospheric depth of the shower maximum from the top of the atmosphere.

The ESAF software produces two distinct executable files called 
\textit{Simu} and \textit{Reco} respectively.
The first one performs the simulation of the entire physical process,
the second one activates
the entire reconstruction chain. The two codes work independently.

\section{Development of the ESAF simulation and reconstruction frameworks for JEM-EUSO}
\label{sec:esaf-jemeuso}

Since 2007, when ESAF was adopted also for the JEM--EUSO mission, new functionality was developed to include specific characteristics of the JEM--EUSO telescope. More recently, the configurations of 
almost all the projects conceived within the JEM--EUSO program have been developed as well. 
These new implementations include: a)
detector configurations and the new trigger algorithms; b) two new track recognition algorithms named 
{\it Linear Track Trigger} (LTT) and {\it Peak and WIndow SEarching} (PWISE); c) new reconstruction algorithms for the
energy and $X_{\max}$ EAS parameters; d) light transients emitted by other classes of events such as ground flashers,
space debris, Transient Luminous Events (TLEs)~\cite{tles}, meteors and nuclearites~\cite{derujula}.

In what follows, the ESAF software is described in more details and a few
examples of its functionality are provided with
specific emphasis on the new implementations.

\subsection{The EAS simulation framework}

The \textit{Simu} framework of ESAF performs the simulation of the entire physical process from shower to telemetry. Both an extensive air shower and the photon
propagation are simulated.
CONEX~\cite{conex} and CORSIKA \cite{corsika} interfaces have been implemented in the framework in addition to the ESAF EAS generators such as SLAST~\cite{slast}. All these generators are currently adopted in ESAF, depending on the objective of the simulation. CORSIKA and CONEX guarantee more carefully tuned to experimental results, however, the computation time is usually considerable. On the other hand, SLAST is a very fast simulator which can provide a sufficiently good approximation of the EAS development when the user is interested in an overall performance result. This is the simulator which has been adopted in the results presented in this paper, unless differently mentioned.

\begin{figure*}[!ht]
\centering
\includegraphics[width=1.\textwidth]{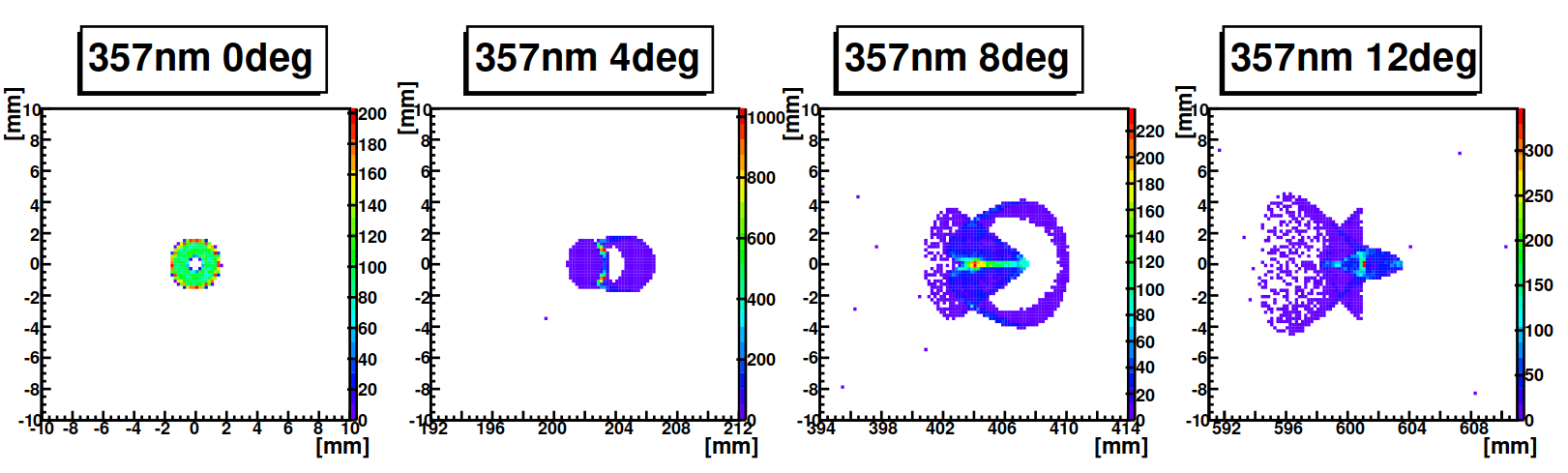}
\caption{PSF diagrams of one version of the K--EUSO optics.
Panels correspond to light at 357~nm with incident angles from the 
optical axis 0$^\circ$, 4$^\circ$, 8$^\circ$ and 12$^\circ$ respectively.
The size of each frame is $20~\text{mm}\times20~\text{mm}$.
The number of arrived photons increases as the color changes from blue to red.
An arbitrary point-like source has been used to test the proper implementation of the optics response in ESAF code.
}
\label{fig:optics}
\end{figure*}

An atmosphere model according to the 1976 Standard US 
Atmosphere~\cite{US_Atmosphere} is implemented,
as well as the fluorescence yield parameterization~\cite{nagano} and the standard Cherenkov production theory.
The LOWTRAN~7 package~\cite{lowtran7} is embedded into ESAF to simulate the atmospheric transmission.
Both Rayleigh and Mie scattering as well as the ozone absorption are taken into account.
The presence of clouds is simulated in a parametric way as a uniform layer with
predefined optical depth, altitude and thickness. 
Photon spectral distribution, timing and direction are produced for each step of the shower development (10g/cm$^{2}$). According to such distributions we generate the single photons that are propagated to the instrument depending on the solid angle covered by the telescope pupil. Each of such photons is followed individually and the transmittance is calculated accordingly. No weighting is applied since photons are few and therefore we can afford to perform the calculation of each of them.
The optics simulation is performed through a ray trace code developed at RIKEN~\cite{naoto}.
Fig.~\ref{fig:optics} shows an example of typical sizes of the optical Point Spread Function (PSF). The images refer to
one of the different K--EUSO configurations that have been designed along the years.

For the RIKEN Monte-Carlo simulator and the GEANT4 optics interface~\cite{svetlana}, the real Fresnel structure is implemented, so photons can be reflected or refracted on the surface of grooves. A cross section drawing of the Mini-EUSO optics with grooves implemented in the ``RIKEN simulator'' is reported in \cite{minieuso}.
In the latter case, the position of the spot on the focal surface is parameterized
in an analytical way as a function of the entrance direction.
An additional random component of Gaussian shape is added to parameterize the point spread function. An efficiency factor
takes into account the transmittance of lenses.
In particular, this is the procedure that has been adopted to estimate the performance of 
POEMMA~\cite{poemma-prd} (see Fig.~\ref{fig:keuso-poemma}) but it could be applied to any detector.

%

\begin{figure*}[!ht]
    
\centering
\includegraphics[width=0.50\textwidth, height=0.30\textwidth]{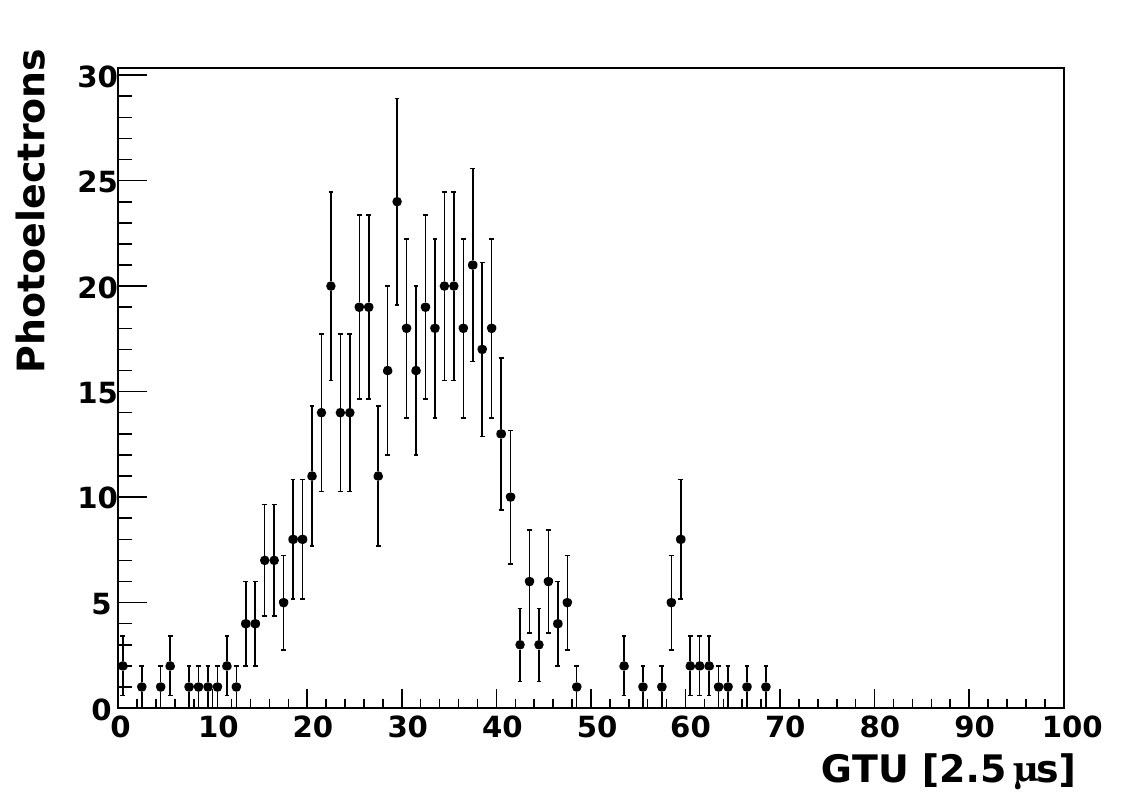}
\includegraphics[width=0.46\textwidth]{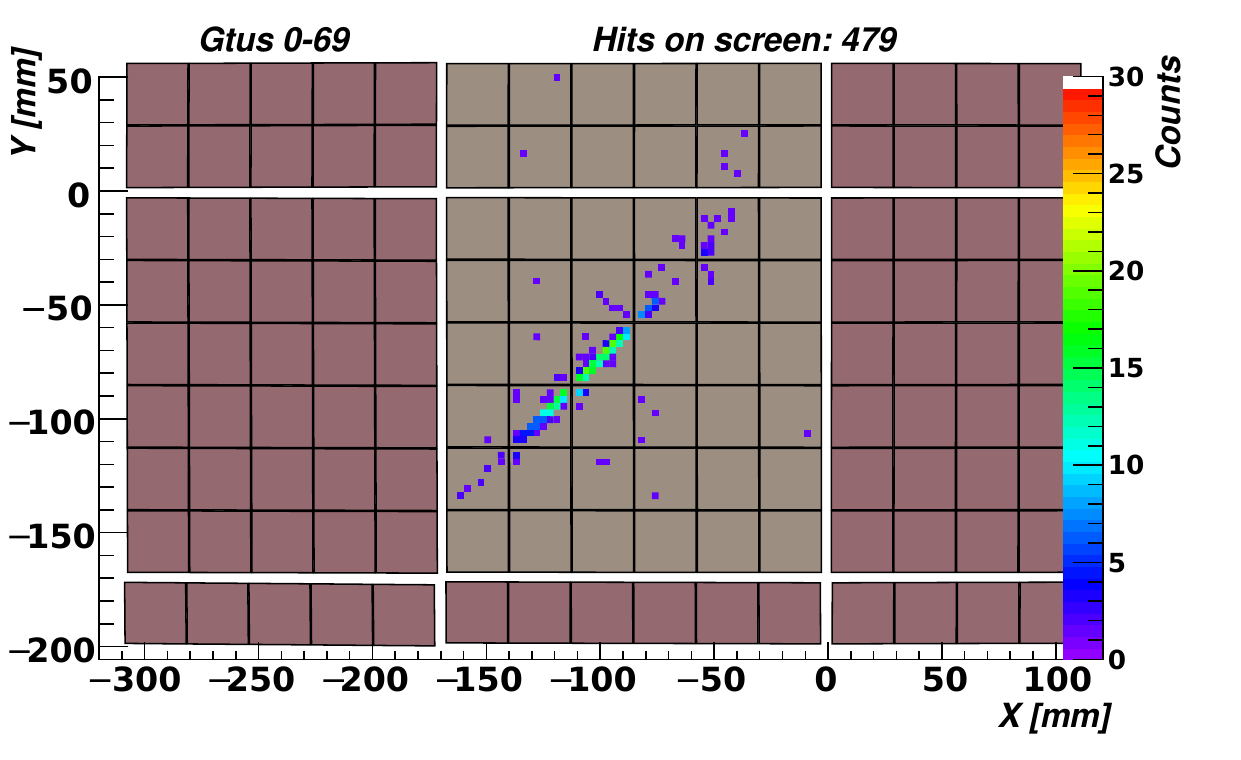}
\caption{Simulation of a $10^{20}$~eV, $60^\circ$ zenith angle proton event
performed for one of the proposed and studied configurations of the K-EUSO detector.
On the left, a photoelectron profile at the focal surface is shown.
The photoelectron image of the same event on the FS is shown in the right
panel. 
}
\label{fig:signal-FS}
\end{figure*}

The Focal Surface (FS) structure (see Fig.~\ref{fig:signal-FS}) is read as a parameter file,
where the position and orientation of each single MAPMT is 
defined.
The overall detector efficiency is parameterized at the single pixel level
as product of quantum and collection efficiency.
The MAPMTs have an average gain set by a parameter and the front-end
electronics is treated in a simplified way with a threshold on the current pulse delivered by the MAPMT. The current pulse associated to each photoelectron is varied according to a gaussian distribution.

The background can be added at either the signal from the front-end electronics or to the pixel counts.
The electronics simulation is then concluded by the trigger.
Several algorithms, specific for each detector, have been implemented in ESAF
and can be used in combination.
JEM--EUSO, K--EUSO and POEMMA adopt the same trigger scheme on two levels operated on a PDM basis. 
The first level trigger looks for concentrations of the signal localized in space 
and time. The second level trigger is activated each time the first level trigger conditions are satisfied and integrates 
the signal intensity in a sequence of test directions. 

Directions close to the one of the simulated EAS are expected to have the larger signal over noise ratio (SNR), overcoming, therefore, the preset trigger threshold.


Whenever such condition is met, the second level trigger is issued. The activation 
of the second level trigger starts the transmission and data storage procedure. 

Thresholds are set to have a rate of spurious triggers from background fluctuations at the order of a trigger every few seconds of background simulations at most, to make the rate consistent with the telemetry requirements.

Details of the logic can be found 
in~\cite{nim-trigger} and in~\cite{joerg-trigger} for the first and second levels trigger, respectively. The trigger logic for balloon missions~\cite{matteo-spb1} is a modified version of the first level trigger of JEM--EUSO while Mini--EUSO~\cite{mini-trigger} and TUS~\cite{tus} missions adopt totally different approaches described in Sec.~\ref{sec:tus} and Sec.~\ref{sec:minieuso}.


\subsection{The EAS reconstruction framework}

\begin{figure}[!ht]
\centering
\includegraphics[width=\columnwidth]{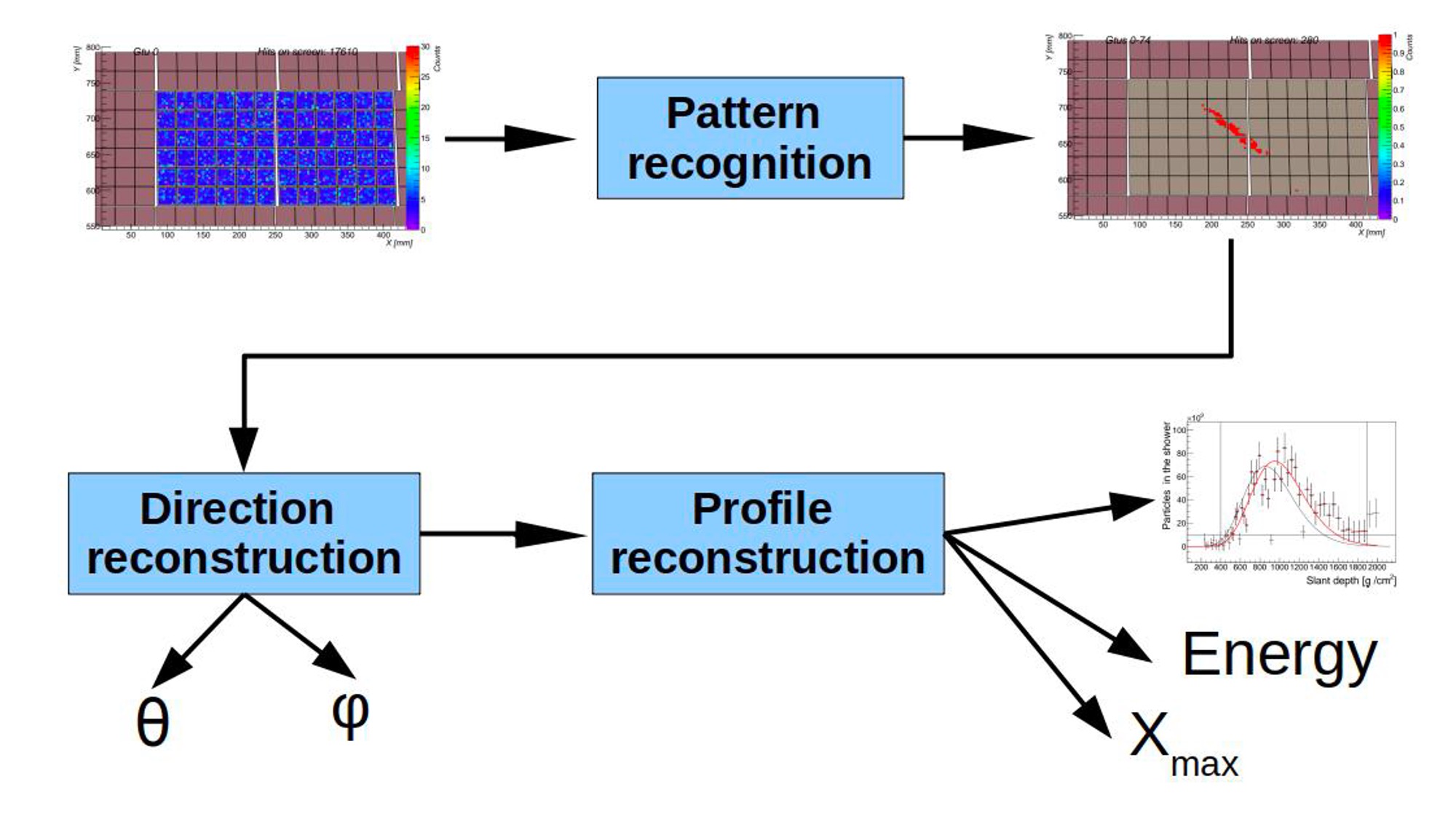}
\caption{The general scheme of the reconstruction framework.}
\label{fig:RecoScheme}
\end{figure}

The first task to accomplish in the reconstruction phase (see Fig.~\ref{fig:RecoScheme}) is to identify the signal from the shower in the recorded data.

A number of track recognition algorithms has been implemented in the code. The \textit{PWISE} algorithm~\cite{guzman} searches for high concentrations of the signal on single pixels.
The algorithm selects pixels and GTUs in which the signal is above a certain threshold and checks whether this signal excess is persistent over time.
The \textit{LTTPatternRecognition}~\cite{mario-asr} is modeled following  the second level trigger philosophy of the JEM--EUSO 
project. The integration of the signal is performed
in a set of predefined test directions and the one which maximizes the integral is chosen to reconstruct the event.
Both algorithms exploit the morphology of the signal, which can be seen as a spot of light moving on the focal surface at the projected speed of light.

The next step is the reconstruction of the shower
 geometry which is an essential step, as it allows the search for anisotropy and provides input
 required for good energy and X$_{max}$ reconstruction.


The \textit{TrackDirection2} is the angular reconstruction module~\cite{mernik-ea}. Several algorithms are implemented in it but two main families of algorithms are in use; analytical and numerical. In the first group, a fit to the speed of the shower is performed to determine its inclination after the Track Detector Plane (TDP) has been identified.  




The inclination of the shower in this plane with respect to the horizontal is inferred from the projected speed of the signal on the focal surface.
The algorithms can be used in an iterative way to improve the knowledge of the arrival direction of the shower.

In the numerical approach, a series of test geometries are chosen and the deviations with respect to the timing and arrival direction of the measured event are calculated.
The test direction that best describes the measured event properties is taken to be the arrival direction. 
The so-called Numerical Exact Method 1 has proven to have  the best performance and is currently used as default. Such method minimizes the deviation of the arrival times of the signal from the test shower w.r.t. the data.

The energy reconstruction is performed in the \textit{PmtToShowerReco} module~\cite{fenu}. In this algorithm, the shower profile is reconstructed starting from the signal after correcting for detector effects, atmospheric absorption and fluorescence yield. The shower profile is then fit with a parameterization to obtain the primary particle parameters.

The count profile of the shower is reconstructed by selecting a collection area that follows the cluster of pixels selected by the PWISE or the LTTPatternRecognition.
The size of this selection area is a trade-off between
the need to collect the highest possible fraction of the signal and the need to limit the background. A stricter selection is indeed more appropriate for the direction reconstruction. For this reason, the PWISE algorithm is more suited for the angular reconstruction, given the typically narrower track selected.
On the other hand, the LTTPatternRecognition, with its very wide selection of pixels, is more appropriated for the shower profile reconstruction.
The detector modeling allows to take into account the detector efficiency and to retrieve a photon curve at the entrance pupil. %
The modeling is performed through a series of lookup tables, produced with an extensive Monte-Carlo simulation of the detector response. 
The arrival direction associated to each pixel is extracted from the very large number of photons simulated. 
The efficiency of the detector as a function of the arrival direction and on the wavelength can be retrieved then depending on the arrival direction of photons. 


%
%

The reconstruction currently implemented is designed for the JEM--EUSO mission, which was meant to operate in monocular mode. As such, the method is particularly challenging and has to rely on some iterative procedures. Two methods are described
in~\cite{fenu-ea}, one with and the other without a Cherenkov reflection peak.
The presence of a peak gives a good constraint on the position of $X_{\rm max}$. The absence of the reflection peak requires using an iterative procedure starting from the reconstructed arrival direction and the parameterized maximum slant depth of a standard shower. Such assumptions are used as starting conditions of an iterative process which minimizes the biases caused by these first choices. Parameters like arrival direction, altitude of the shower maximum and slant depth of the maximum can be varied and the region of the parameter space which best describes the data can be identified. The atmospheric absorption is then modeled according to the LOWTRAN7
package~\cite{lowtran7} and, after estimating the shower position, the luminosity of the shower can be calculated.


As a final step, a parameterization of the energy distribution of the secondary
particles is used to calculate the fluorescence \cite{nagano} and Cherenkov yields.
At the end of the procedure, it is possible to reconstruct the secondary particles profile of the shower. 



The reconstruction of energy and $\mathrm X_{\max}$ is then performed through
a fit with a shower profile function to 
the reconstructed shower profile~\cite{fenu-ea}.
A fit with the so called GIL function~\cite{gil} is adopted. The GIL is a simplified parameterization based on older hadronic interaction models. The difference between GIL and other common descriptions e.g. Gaisser-Hillas is negligible for our application. 

\subsection{Implementation of other classes of events}

The science versatility of experiments like Mini--EUSO and TUS
requires the simulation of 
many different light transients originating from different physical phenomena.
For this reason, different
types of signals have been included in the capabilities of the ESAF software  along the years.
In these cases the time profile, the intensity of light emission and the track extension of the event have been included without simulating the physical process responsible for such light emission.


The first class of events implemented in ESAF are Transient Luminous Events (TLEs) such as blue jets, sprites and
elves which have been discovered relatively
recently and are still not well understood~\cite{pasko}. They have typical durations of
ms or tens of ms. Mini--EUSO has a dedicated trigger algorithm to capture TLEs and
other millisecond scale phenomena at high resolution~\cite{mini-trigger}. These data
could help
improve the understanding of the formation mechanisms of filamentary plasma
structures, complementing atmospheric science experiments.
The TLE simulator in ESAF is composed of the {\it TLEGenerator, TLELightSource, TLEBunchRadiativeTransfer} classes which
are subroutines of the {\it EventGenerator, 
LightSource} and {\it RadiativeTransfer} interfaces respectively. The {\it TLESpectrumHisto}
class is responsible for the spectral distributions of the phenomena.

\begin{figure*}[ht]
\centering
\includegraphics[width=0.43\textwidth]{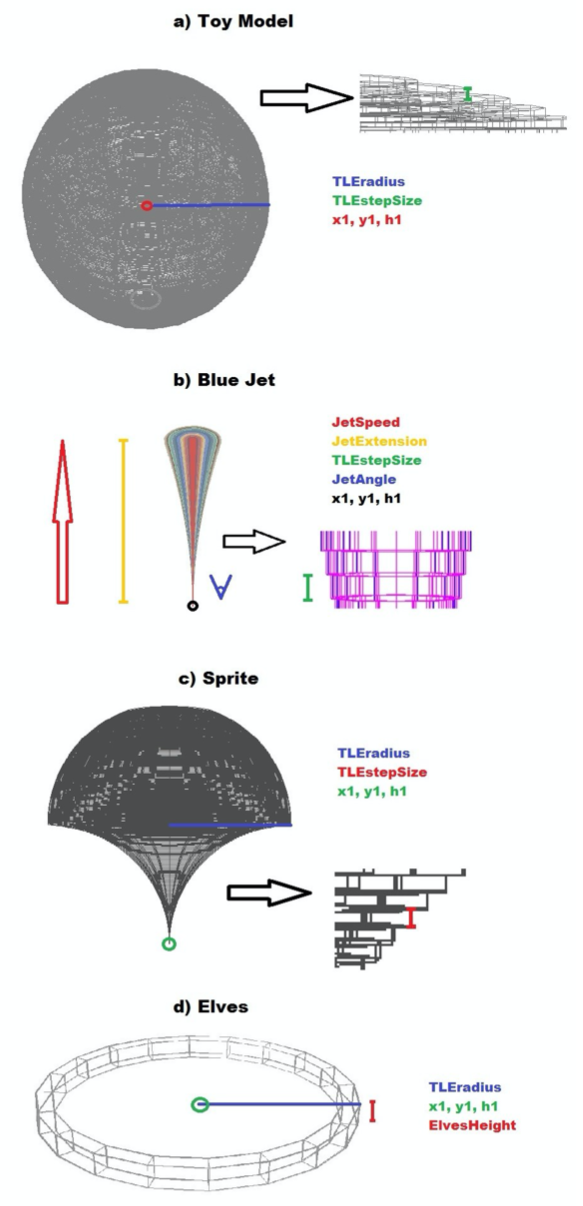}
\includegraphics[width=0.55\textwidth]{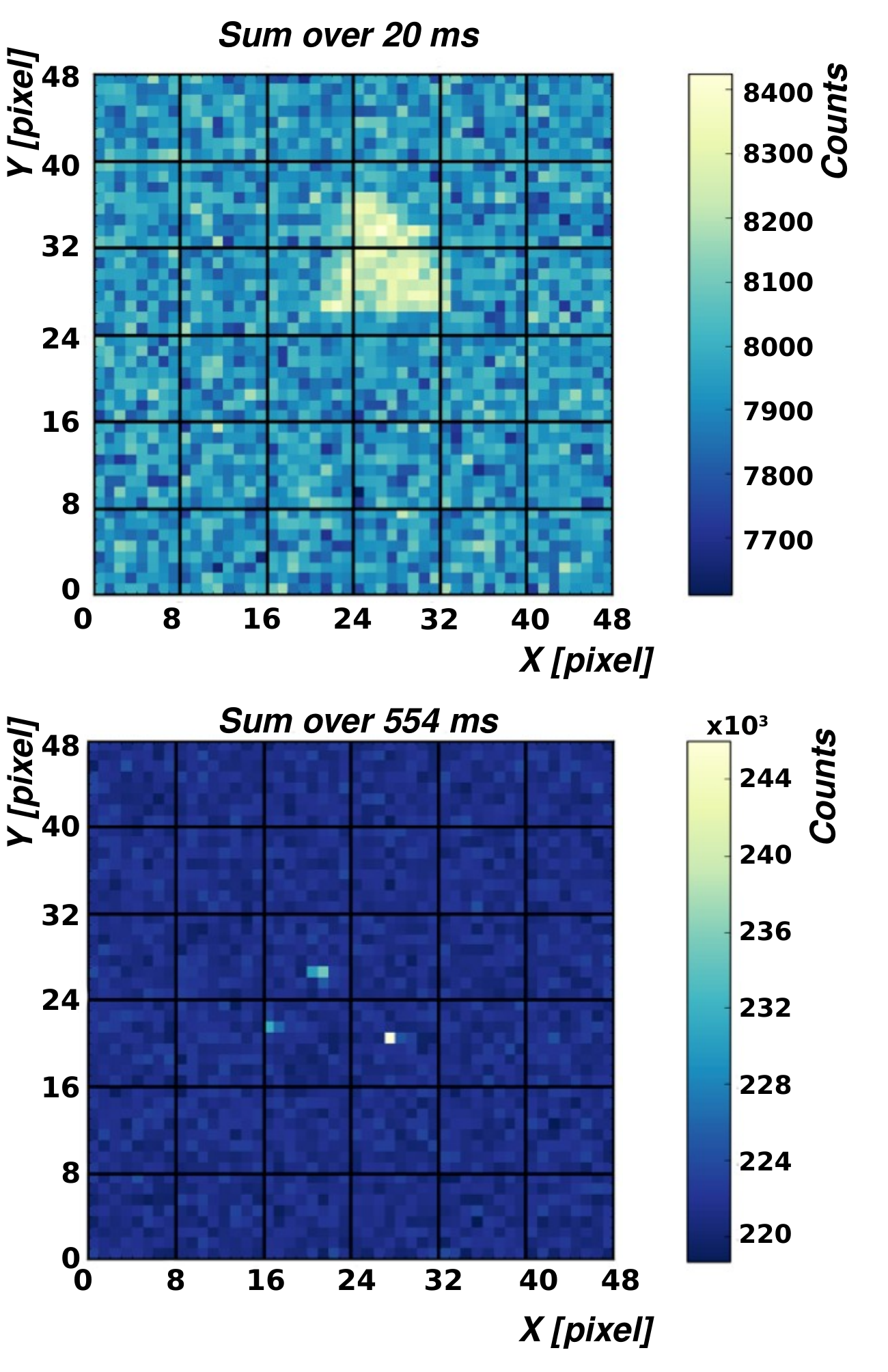}
\caption{Left: Graphical representation of the configurable parameters of the four different
classes of events that can be generated in ESAF: a) a Toy Model; b) Blue Jet; c) Sprite; d) Elves.
Right: Expected light track of a typical diffused elve (top) and 3 localized blue
jet events (bottom) generated by ESAF and detectable by Mini--EUSO. Background emission of 1~count/pixel/GTU is also included. Right plot adapted from~\cite{CAPEL20182954}. 
}
\label{fig:tle-representation}       
\end{figure*}

The left side of Fig.~\ref{fig:tle-representation} shows a graphical representation of the configurable parameters of the four different
classes of events that can be generated in ESAF: a) a Toy Model; b) Blue Jet; c) Sprite; d) Elves.

The right side of 
Fig.~\ref{fig:tle-representation} 
shows examples
of typical TLEs as they are simulated for Mini--EUSO.
The simulations employ toy models which in a simplified way
reproduce the size, shape and wavelength spectra of the different physical phenomena.
Details of the implementation of these phenomena in ESAF can be found
in~\cite{perdichizzi}.

Among the scientific objectives of the JEM--EUSO program is the study of slower events such as meteors and fireballs.
The simulation of meteor-induced light tracks is described in~\cite{nardelli}, 
which inherits the approach described in~\cite{ea-meteors}.
Similarly to the TLE case, a new class has been developed for meteors called {\it MeteorLightSource}. 
Meteors produce tracks which are very slow moving
compared to UHECR events, 
resulting in more than 1000 times more data produced than for UHECR events.
The solution that has been adopted is to 
fully track only a fraction of the produced photons and to re-weight them at the detector level.
Regarding the simulation of the light signal of the meteor, the starting position, direction, speed, duration and magnitude
can be chosen randomly or set as input parameters.
The variation of the brightness of the meteor as a function of time, or the meteor
lightcurve, is also chosen randomly, or provided as an input. Since the shape of the
lightcurve can be highly variable in the real world, the model adopts a very flexible
approach, representing it with a 9$^{th}$ degree polynomial. In most practical applications
performed so far, simulated lightcurves look reasonably realistic, taking into
account the large intrinsic variability of the phenomenon, as shown, for instance, in
the analysis by~\cite{pecina}. Moreover, since it is known that real meteors can exhibit one or
more secondary bursts, the model includes the possibility to simulate one secondary
burst, occurring at some instant before the end of the event, and having a morphology
which is also represented by a 9$^{th}$ degree polynomial (again, fixed or randomly
chosen).
Moreover, ESAF now also includes the implementation
of a formula by Jacchia~\cite{jacchia} which links the magnitude, mass
and the velocity of the meteoroid. Given the meteor's velocity and
magnitude, that are free parameters in the simulations, one can derive the corresponding mass of the meteoroid. By assuming
a value of the density $\rho$ (so far a fixed value $\rho$ = 3.55 g/cm$^3$ has been
assumed in preliminary tests) it is possible to compute the
corresponding size of the meteoroid. More details about the simulation of meteors for JEM--EUSO, Mini--EUSO and EUSO--TA can be found in~\cite{meteoroids2016}, while an example of a simulated meteor with ESAF is reported in Sec~\ref{sec:minieuso}.

\section{Simulation of various missions of the JEM--EUSO program and derived performance}
\label{sec:results}

The complexity of the JEM--EUSO program requires an extensive effort to study
the performance of all different detectors of the program. The detectors are either space-based, like JEM--EUSO,
Mini--EUSO, TUS, K--EUSO, and POEMMA or balloon based, like EUSO--Balloon, EUSO--SPB1 and EUSO--SPB2.
EUSO--TA is the only one located on ground. EUSO--SPB2 is not implemented yet in ESAF in the final configuration and can be simulated
only using the \offline\ package. For this reason it will not be discussed in detail in this paper.

All the detectors point normally in nadir mode or with slightly
tilted configurations, except
for EUSO--TA which points northwest with an elevation typically of 15--25$^\circ$. All
the detectors have different focal surfaces: Mini--EUSO, EUSO--TA, EUSO--Balloon and EUSO--SPB1 are single-PDMs detectors while JEM--EUSO, POEMMA and K--EUSO have a multi-PDM layout. TUS on the other hand,
consists of an array of 256 PMTs on a square of $16\times16$ PMTs. The time frame adopted in the simulations
is $2.5~\mu$s with the exception of TUS, which has a frame of 0.8~$\mu$s and a totally different electronics
configuration. In the following subsections, the main simulation results for each configuration are summarized starting from the JEM--EUSO case, which has been studied more extensively than other configurations.
The description of the standard parameters adopted in the simulations is reported in Appendix~\ref{sec:esaf-param}.
This section includes information which has already been the subject of previous publications.

\subsection{The JEM--EUSO configuration}
\label{sec:jemeuso}

\begin{figure*}[!ht]
\centering
\includegraphics[width=0.46\textwidth]{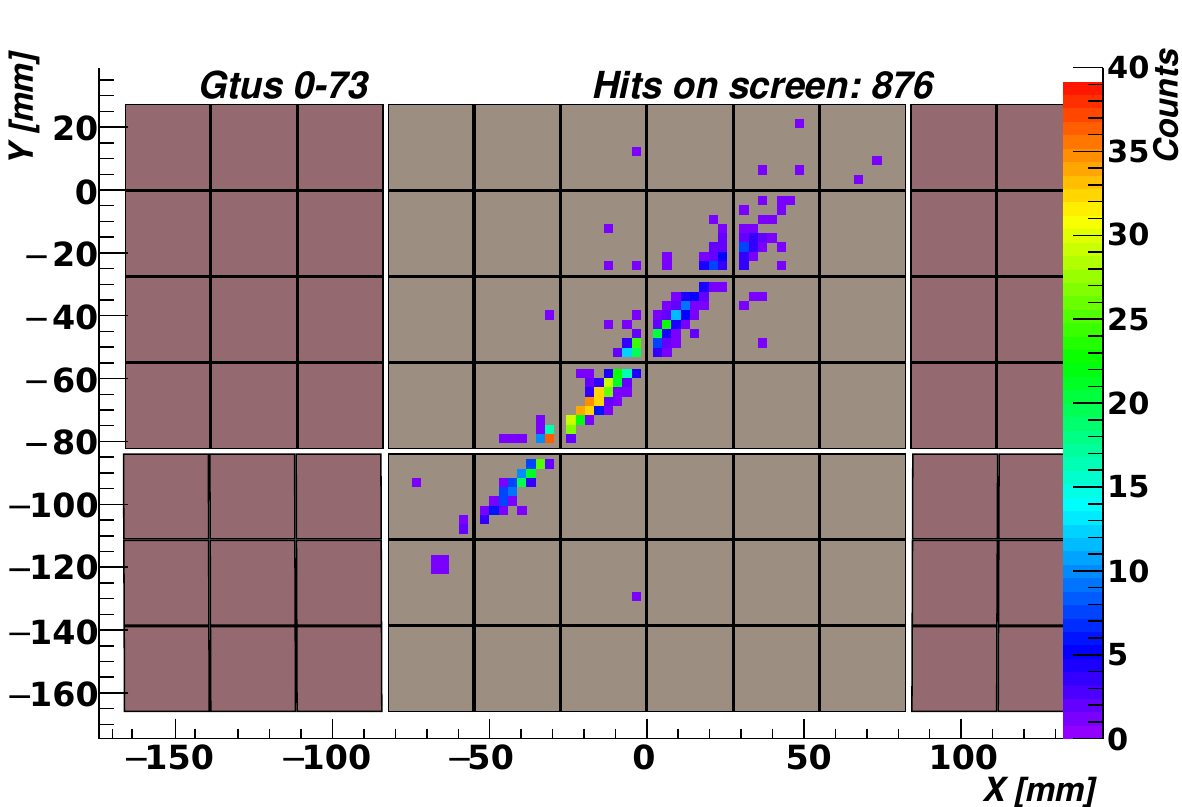}
\includegraphics[width=0.52\textwidth]{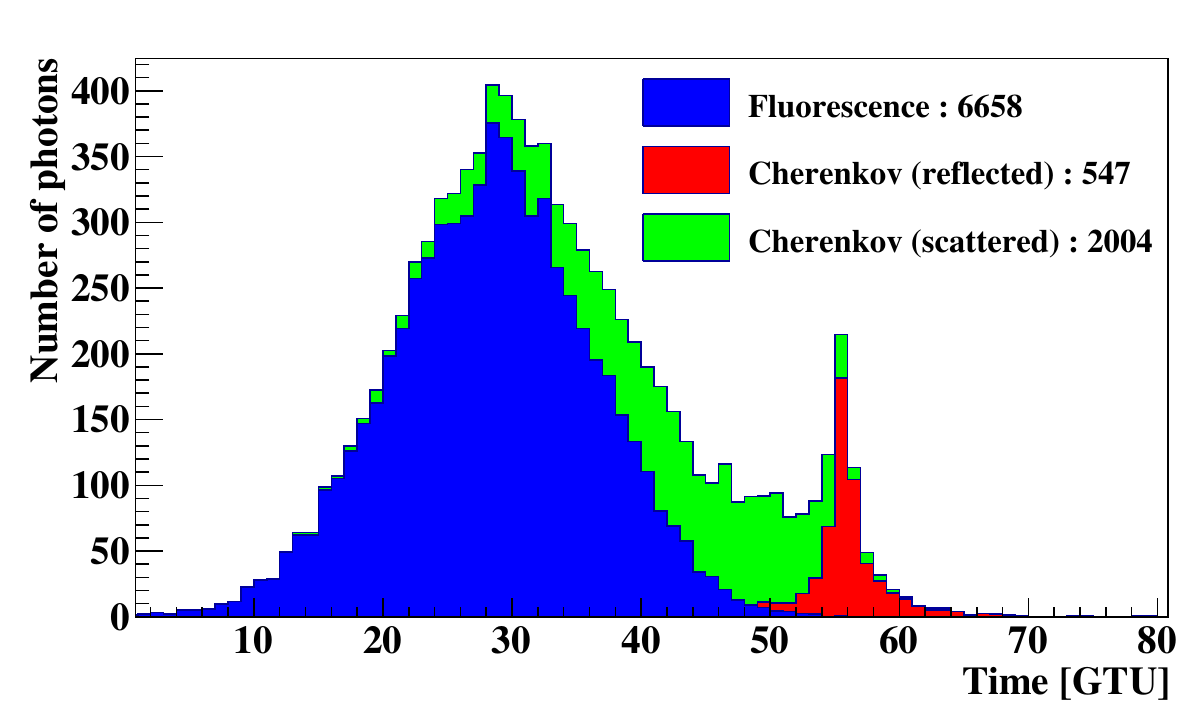}
\caption{Left: Spatial profile of photon counts from an EAS caused by a 10$^{20}$~eV
proton with zenith angle 60$^\circ$.
Each square (dark and light brown) represents a MAPMT. 
A parts of 6 PDMs are shown.
The small coloured
squares show the number of photon counts detected by each channel of the MAPMT. This event is crossing two PDMs, with the shower starting to develop in the bottom central PDM and continuing in
the top central PDM. Right: Time profile of photons, obtained adding all photons of the previous picture
detected on each GTU of 2.5~$\mu$s ($\sim$80 GTUs for a total signal duration of $\sim$200 $\mu$s). It is possible to see the contribution to the signal from
the three components: the UV fluorescence light (blue), scattered Cherenkov (green) and Cherenkov diffusely reflected peak (red).  
}
\label{fig:jemeuso-signal}
\end{figure*}

JEM--EUSO has been conceived as a mission to be installed on the ISS orbiting at $\sim$400~km height. 
The main characteristics of the telescope are summarized in Section~\ref{sec:intro}.
Fig.~\ref{fig:jemeuso-signal} shows the spatial and temporal profile of an EAS generated by a 10$^{20}$~eV proton with zenith 
angle 60$^{\circ}$ simulated with ESAF. The contribution of the fluorescence and Cherenkov components at the pupil level are indicated on the right in different colors.

\begin{figure}[!htb]
\centering
\includegraphics[width=1.\columnwidth]{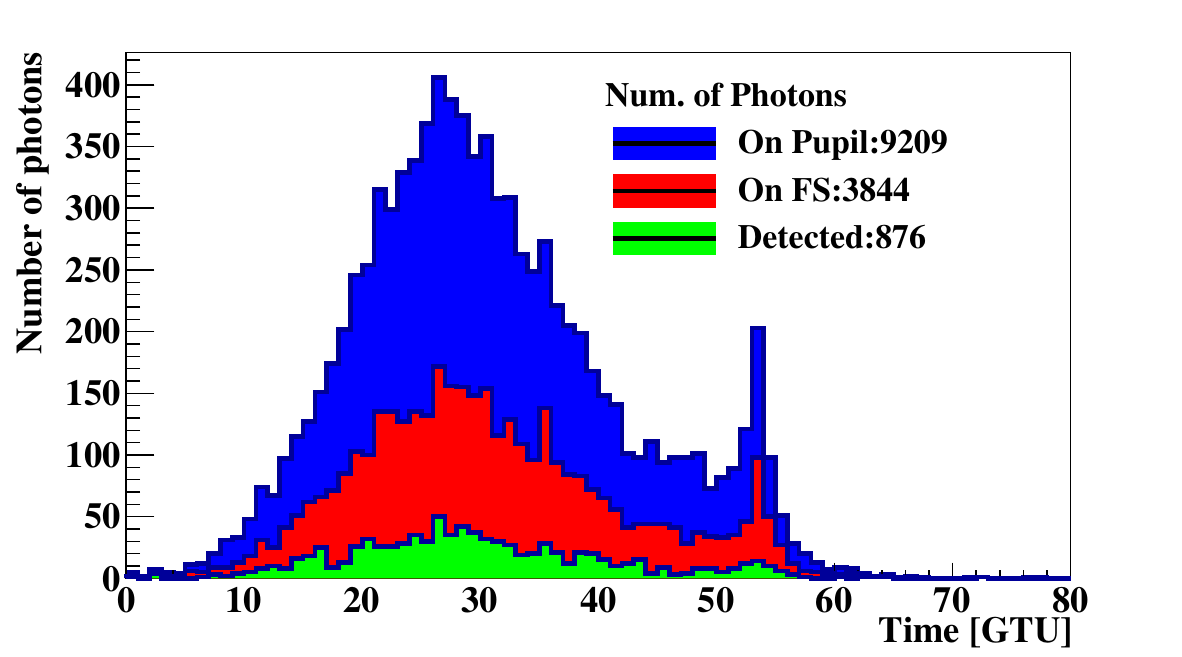}
\caption{Photons coming to the detector, photons intersecting the focal surface (FS) and photons detected
as a function of time (in GTU). Due to the covering factor and quantum efficiency taken into account, the
fraction of photons creating a signal (`Detected’) is about 0.3. In this example: 10$^{20}$eV proton event with
60$^\circ$ zenith angle. 
}
\label{fig:jemeuso-photons}
\end{figure}

The optics and detector response are then simulated and Fig.~\ref{fig:jemeuso-photons} shows the photons arriving at the detector (blue), the photons arriving at the focal surface (FS, red) and the photons detected at pixel level (Detected, green).

By applying the trigger algorithm after introducing the nightglow background, the geometrical aperture is determined
in nadir mode.
A tilted configuration of the telescope is available in ESAF as well.
In the tilt mode, the 
observation area is scaled by $\propto\sec^{3}(\xi) $ as a function of tilting angle $\xi$ of the optical axis from 
the nadir.
This increases the sample of events at the highest energies, however, 
showers will be seen at larger distances, therefore they will appear dimmer compared to nadir mode.

\begin{figure*}[!htb]
\centering
\includegraphics[width=1.\textwidth]{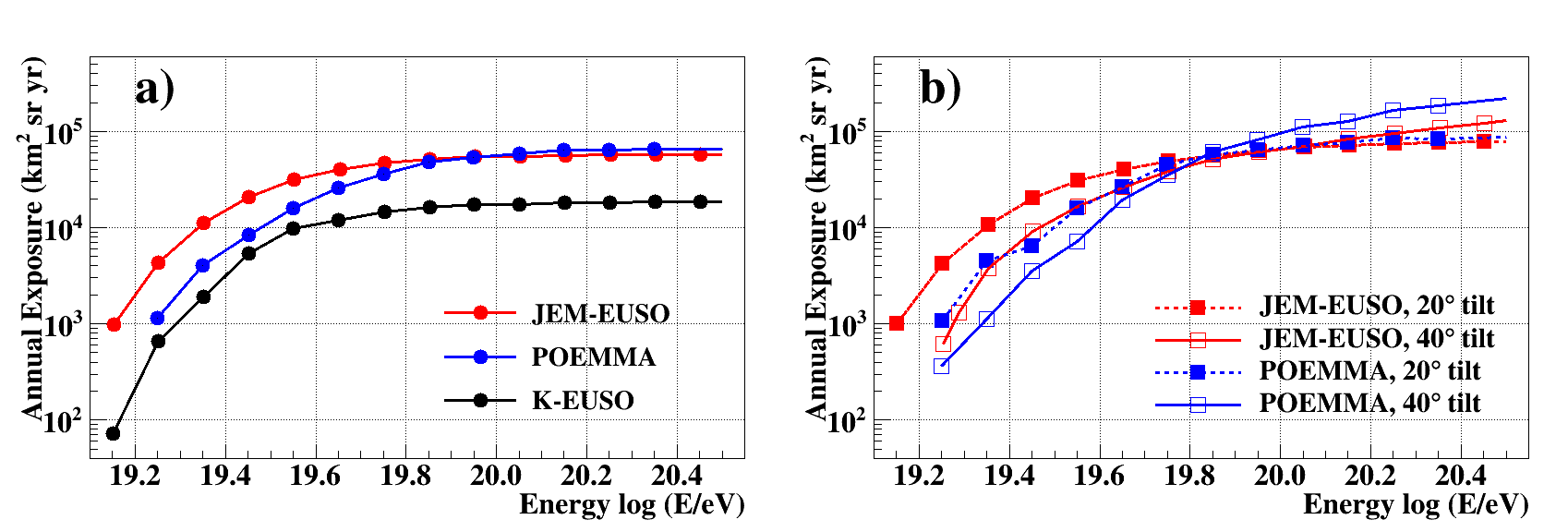}
\caption{JEM--EUSO, K--EUSO and POEMMA exposures as a function of energy for nadir (left) and different tilting angles (right).
In case of K--EUSO only the nadir configuration has been simulated as the tilt option is not considered in the present mission design.} 
\label{fig:comparison-exposure}
\end{figure*}

In order to estimate the aperture, a specific nightglow emission has to be assumed. A
background level ($I_{\mathrm{BG}}$) of 500 photons/(m$^{2}\cdot$sr$\cdot$ns) is considered.
This is a typically
 measured value according to literature \cite{garipov,barbier,catalano}, and it represents also the value assumed
 in earlier JEM-EUSO simulations. 
This corresponds to an average count level of $\sim$1.1 counts/GTU/pixel.
The assumption of a constant background over the entire FoV, most likely, 
is an overly simplistic assumption since the background radiance depends on the tilting angle
under which the atmosphere is observed. However, at a first-order approximation
and especially for low tilting angles we can consider the shower-to-detector distance
to be the leading factor affecting the threshold in energy. 
Fig.~\ref{fig:comparison-exposure} shows the exposure as a function of energy for nadir and different tilting angles.
As expected by tilting the telescope, both the threshold energy and exposure increase at the highest 
energies. 

The quasi-nadir configuration of $\xi$ = 20$^\circ$ allows keeping at the lowest energies an almost similar exposure to the nadir configuration, while increasing it moderately at $\sim 10\% - 20\%$ level in the $E\gtrsim10^{20}$~eV.


Compared 
to nadir mode, the tilted mode is suitable to increase the exposure at energies $E\gtrsim 2 \times 10^{20}$~eV, where 
the flux is particularly low.
The exposure calculation is based on a Monte Carlo simulation of proton EASs of variable energy and direction. To avoid border 
effects, cosmic rays are injected in an area $A_{simu}$ larger than the FoV. The ratio of the triggered $N_{trigg}$ over 
simulated $N_{simu}$ events is then calculated for each energy bin. 
The effects of day-night cycle and Moon phases are taken into account in 
$\eta$, the astronomical duty cycle. Effects of clouds and artificial lights are also taken into account by $\eta_{clouds}$ (see later on in this section) 
and $\eta_{city}$, respectively. In this formula, we assumed $\eta$ = 0.2, $\eta_{clouds}$ = 0.72 and $\eta_{city}$ = 0.9 (which includes also lightning and aurorae contributions)
as estimated in~\cite{exposure}. The exposure $\mathcal{E}(E)$ is then calculated over time 
$t$, which is assumed to be 1 year in the following:

\begin{equation}
\mathcal{E}(E) = \frac{N_{trigg}}{N_{simu}}(E) \times A_{simu} \times \Omega \times \eta \times \eta_{clouds} \times
\eta_{city} \times t
\label{eqn:exposure}
\end{equation}

An overall conversion factor between geometrical aperture and exposure of
$\sim$13\% is obtained as a product of $\eta \times \eta_{clouds} \times
\eta_{city}$.

\begin{figure}[h]
\centering
\includegraphics[width=1.\columnwidth]{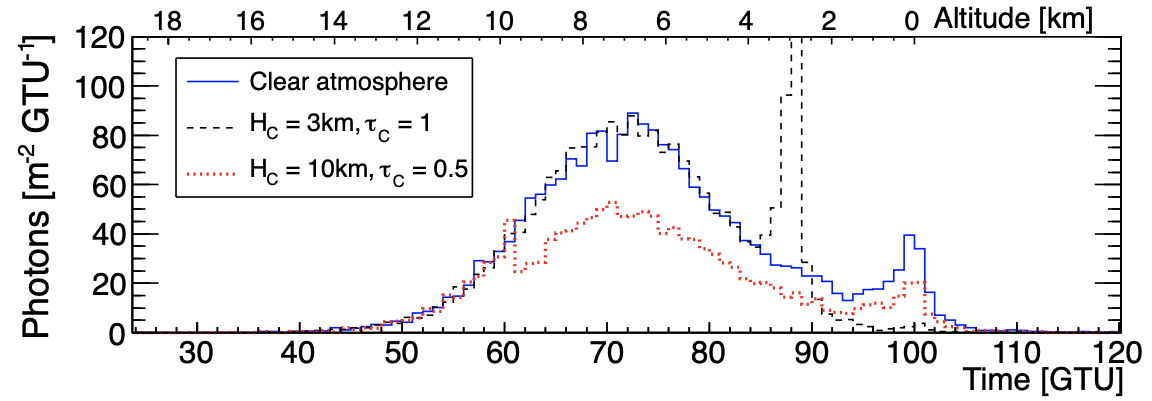}
\caption{Arrival time distribution of photons from a proton induced EAS of $E_0$ = 10$^{20}$ eV and
$\theta$ = 60$^\circ$ for different atmospheric conditions. The solid line represents the case for the clear atmosphere. 
Dashed and dotted lines denote the cloudy cases for optical depth $\tau_C$ = 1 at cloud height H$_C$ = 3 km and $\tau_C$ = 0.5 at H$_C$ = 10 km, respectively.
The axis on the top indicates the altitude where photons originate for the given arrival time.
The photon number is indicated per m$^2$.
In order to compare to the Figure~\ref{fig:jemeuso-photons} the reader has to multiply by the optics size ($\sim$4.5 m$^2$). 
Also the starting time of the event is different because it depends on the first photon reaching the optics, however, the duration of the event in both cases is ~55-60 GTU in clear sky atmosphere.
Image adapted from~\cite{ea-clouds}.
 }
\label{fig:jemeuso-clouds}
\end{figure}

As ESAF provides the opportunity to simulate the presence of clouds, extensive simulations have been conducted
to understand their impact on the aperture reduction and reconstruction capabilities. 
Fig.~\ref{fig:jemeuso-clouds} shows the light curves of EAS with different
atmospheric conditions for the case of proton EASs with $E = 10^{20}$ eV and $\theta$ = 60$^{\circ}$. 
The solid line represents the case for clear atmosphere. Dashed and
dotted lines show the cases for clouds optical depth $\tau_C$ = 1 at height $H_C$ = 3 km and $\tau_C$ = 0.5
at $H_C = 10$~km, respectively. The horizontal axis is the absolute time in GTUs. 
The axis on the top indicates the
altitude where photons originate for the given arrival time. 
In the clear atmosphere condition, the light curve indicates the EAS development
followed by the Cherenkov footprint reflectively diffused by the Earth's surface at GTU number 100. 
For $\theta$ = 60$^\circ$ as in this
example, the apparent shower development 
lasts $\sim$60 GTUs (=150 $\mu$s).
Using these data, the EAS parameters are reconstructed.  
In presence of clouds, EAS light curves are affected. If the
optical depth of the cloud is large enough, the shower track is truncated.
Upward photons scattered or emitted below the cloud are extinguished and do not
contribute to the signals at the telescope. For this example, with a cloud at 3 km,
the apparent signals 
last 40 GTUs. It is still feasible to apply the
reconstruction techniques used in the case of the clear atmosphere by using only the
measurements taken above the cloud.
As seen in the figure for the case of a small optical depth, photon signals originating
below the cloud are attenuated. This likely lowers the estimated energy of the
EAS if the same techniques for the clear atmosphere are applied. Alternatively, the
Cherenkov footprint is still observable and the orientation and apparent velocity are
not affected, thus the consequence on arrival direction determination is limited.
As described in~\cite{ea-clouds} using of ESAF simulations of EASs in variable cloudy conditions together with the analyses of 
satellite measurements of the cloud distribution indicate that more than 60\% of the cases allow for conventional EAS 
observation, while an additional $\sim$20\% can be observed with reduced quality. The combination of the relevant factors results in an 
effective trigger aperture of $\sim$72\% of the aperture in clear atmosphere condition.

The scientific outputs of such a mission rely on the quality of the reconstructed EAS parameters. Therefore,
the performance on angular, energy and $X_{\max}$ reconstruction have been extensively studied for the JEM--EUSO telescope using different
configurations, namely the HTV and Dragon layouts, nadir and tilt options. 
The study presented in the following adopts either the SLAST-GIL shower generator or CONEX with EPOS-LHC~\cite{epos-lhc} hadronic interaction model. 
While SLAST-GIL has the advantage of considerably reducing the simulation time, it adopts a simplified
approach based on a parameterized shower simulation in which shower-to-shower fluctuations are not completely taken
into account. This is not an issue in specific studies such as the aperture curve determination, however, it might have
a more significant impact in case of reconstruction of EAS parameters. Another advantage of using the CONEX simulation 
program is that it is possible to study the behavior of the detector response to different primary particles. 
In the following we report on proton, iron and photon studies. 

\begin{figure*}[h]
\centering
\includegraphics[width=2.\columnwidth]{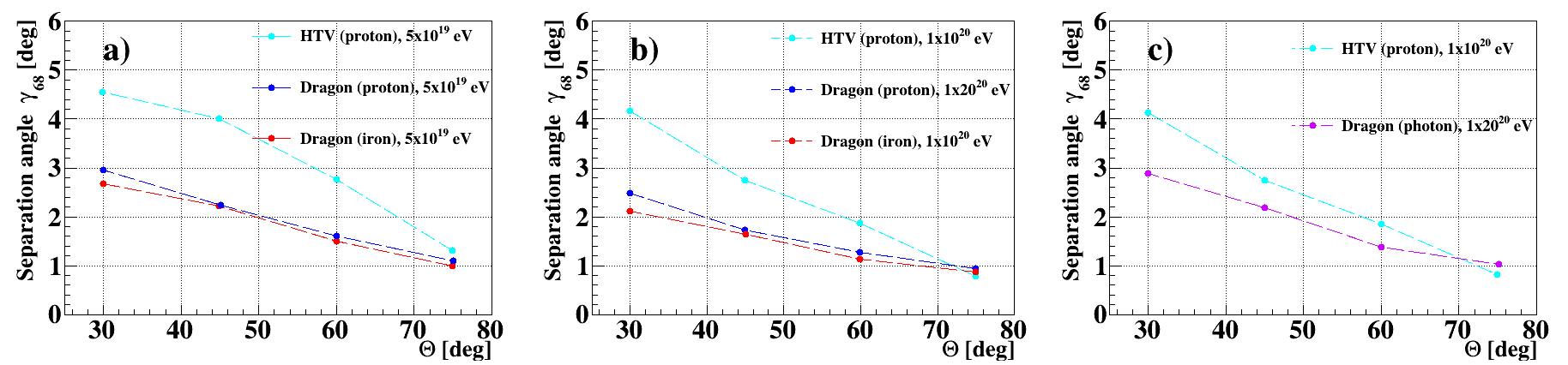}
\caption{ 
The angular resolution as a function of the zenith angle for the Dragon-configuration of the JEM--EUSO mission pointing to nadir for proton
(dark blue) and iron (red) primaries with an energy of 5$\times$10$^{19}$ eV (a), and 10$^{20}$ eV (b). 
Panel c) shows in violet the angular resolution for photons of 10$^{20}$ eV.
Previous results with the 
HTV-configuration are also shown in light blue. Image adapted from~\cite{guzman-icrc2015}.}
\label{fig:jemeuso-arrival}
\end{figure*}

In Fig.~\ref{fig:jemeuso-arrival} we present the angular resolution in nadir mode for proton, iron and 
photon-generated EAS obtained with the Dragon
configuration and adopting the CONEX generator. For proton and iron we have simulated events with energies of 5$\times$10$^{19}$ eV and 10$^{20}$ eV with zenith angles of 
$\theta$ = 30$^\circ$, 45$^\circ$, 60$^\circ$, and 75$^\circ$. For each angle and energy combination, 1000 events are
simulated. Additionally, similar simulations have been performed for photons at the energy 10$^{20}$ eV. Due to possible interactions with the geomagnetic field, special care has to be taken when performing the photon 
simulations. The pair production process depends strongly on the magnetic component transverse to the photon's direction of 
motion, and, therefore, the event simulation is sensitive to the value and direction of the local geomagnetic 
field~\cite{homola}.
We characterize the error in the reconstructed arrival direction as the angle between the simulated
arrival direction vector $\hat \Omega_{\rm Simu}$ and its reconstructed counterpart $\hat \Omega_{\rm Reco}$. We define
$\gamma = \arccos{(\hat \Omega_{\rm Simu} \cdot \hat \Omega_{\rm Reco})}$
as the error in the reconstruction, and the angular resolution as the value where the cumulative distribution of the
reconstruction's error reaches 68\%. We shall refer to this value as $\gamma_{68}$.

The results using the Dragon option indicate that the angular resolution of the detector is not affected too much from the
larger fluctuations of the proton-induced showers as the iron induced EASs have similar reconstruction
performance. 
Fig.~\ref{fig:jemeuso-arrival} displays also results obtained with the HTV configuration which have been studied 
in~\cite{mernik-ea}.
The better performance of the Dragon option compared to the HTV option is mainly due to stricter cuts applied
in the selection performed with the PWISE algorithm particularly visible in the significant improvement obtained for protons at 30$^\circ$. 
However, this comparison shows the possibility to improve the quality of the reconstructed events by applying more stringent
cuts at the price of a reduced statistics 
($\sim$50$\%$ at 30$^\circ$). 
In a real experiment all the triggering events will be reconstructed with coarser selection cuts. 
A series of refinements will allow high quality events to be reconstructed with higher performance still keeping the high statistics. 
The event selection and reconstruction strategy is still in definition. 
In the case of photons as primary particle we experience a low triggering ratio for 
the showers that exhibit a
strong Landau-Pomeranchuk-Migdal (LPM) effect~\cite{lpm1,lpm2}. Showers with the LPM effect appear less bright,
as a consequence of their extended longitudinal profile. 
It works as a selection filter allowing only the brightest photon showers to trigger the detector.
Again this is mostly relevant at the lower zenith angles, whereas for higher zenith angles
the impact is less dramatic. A more detailed discussion of this analysis can be found in~\cite{guzman-icrc2015}.

The angular resolution is studied also for configurations in tilt mode with the Dragon option in case of proton 
showers~\cite{mernik-icrc2015}. The standard 
procedure described in ~\cite{mernik-ea} is applied here. EASs are generated using the SLAST-GIL simulator with fixed
energies between 5$\times$10$^{19}$eV
and 5$\times$10$^{20}$eV 
and fixed zenith angles between 30$^\circ$
and 75$^\circ$. All azimuth angles are picked randomly
between 0$^\circ$ and 360$^\circ$. The shower cores have been placed within a rectangular area of x[-550km;+100km] $\times$ y[-250km;+250km] for $\xi$ = 20$^\circ$ tilt angle and x[-1300km;0km] $\times$ y[-400km;+400km] for $\xi$ = 40$^\circ$.
These areas are considerably larger than the actual FoV of the tilted instrument. For each energy/zenith angle
combination the number of triggering events is of the order of 2000 or higher.

\begin{figure*}[h]
\centering
\includegraphics[width=1.\textwidth]{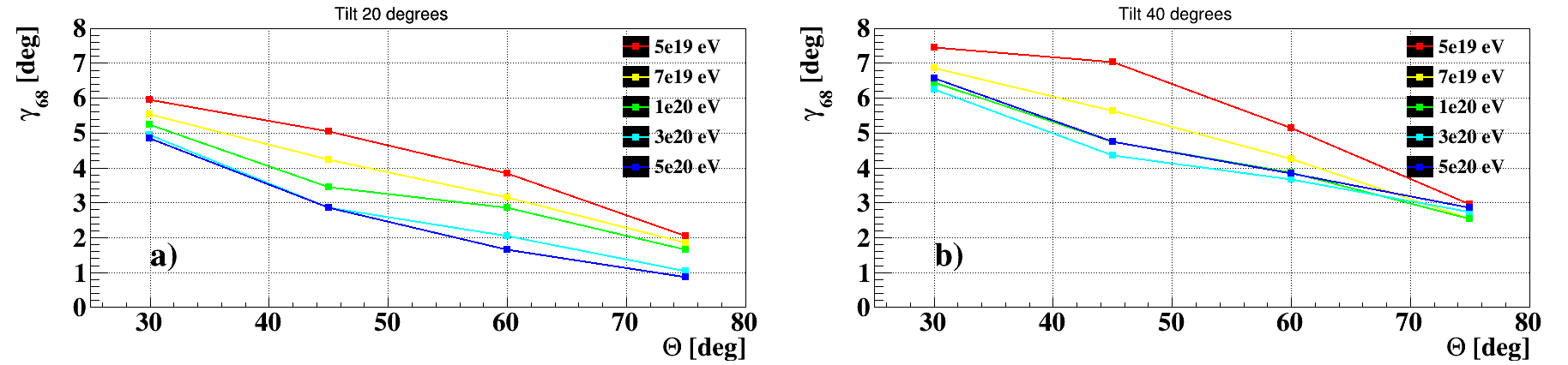}
\caption{ 
The angular resolution of the 20$^\circ$ (a) and 40$^\circ$ (b) tilted JEM--EUSO telescope. 
Image adapted from~\cite{mernik-icrc2015}.}
\label{fig:jemeuso-angreco-tilt}
\end{figure*}
Results are presented in Fig.~\ref{fig:jemeuso-angreco-tilt}. As expected, the
angular resolution of the telescope tilted by $\xi$ = 20$^\circ$ gets worse by approximately 2.5$^\circ$ when compared to
the nadir mode operation (see Fig.~\ref{fig:jemeuso-arrival}). The effect mainly depends on the zenith angle of the showers
and to a smaller extent on the energy. Especially the low zenith angles are affected.
When we tilt the telescope by $\xi$ = 40$^\circ$, the resolution gets worse compared to the previous 20$^\circ$ case.
Again, the effect mainly depends on the zenith angle of the showers and to a smaller extent on the
energy. The loss of the angular resolution is about 1.5$^\circ$ compared to the $\xi$ = 20$^\circ$ tilted case.
As expected from the analysis of the signal
behaviour, we can observe a worsening of the angular resolution due to the fact that less light per
EAS reaches the telescope.
The instrument resolution capability is determined by four limiting factors, three of which 
are related to the distance of the shower from the detector. The first one is the proximity effect. 
Events injected nearby the telescope appear brighter than those farther
away, since the number of photons reaching the telescope is scaled by a factor of $1/d^2$, where $d$ represents the distance between the telescope and the location of the emitted photon.
The second comes from the projected pixel size on ground as it determines the minimum
theoretically reachable air shower resolution of the telescope. Pixels in the outer parts
of the FoV observe a larger volume of air than the ones at the centre.
The third one is due to optics throughput. Events occurring in the outer parts of the FoV face stronger optical
losses, due to a lower transmittance of the optical system. The probability of being attenuated
or defocused by the telescope optics is higher.
The last one is related to the skimming effect. Shower tracks can skim the field of view and appear only partially on the
FS. This effect increases when the nadir FoV is deformed by tilting the detector.

The question whether or not to tilt a JEM--EUSO-like instrument in space strongly depends on the
primary objective of the mission. When the emphasis is put on high exposure for the highest energy events, a tilting of the
instrument might be useful. When the focus lies on accuracy for direction determination,
the nadir mode is the preferred operation mode.

\begin{figure*}[!ht]
\centering
\includegraphics[width=1.\textwidth]{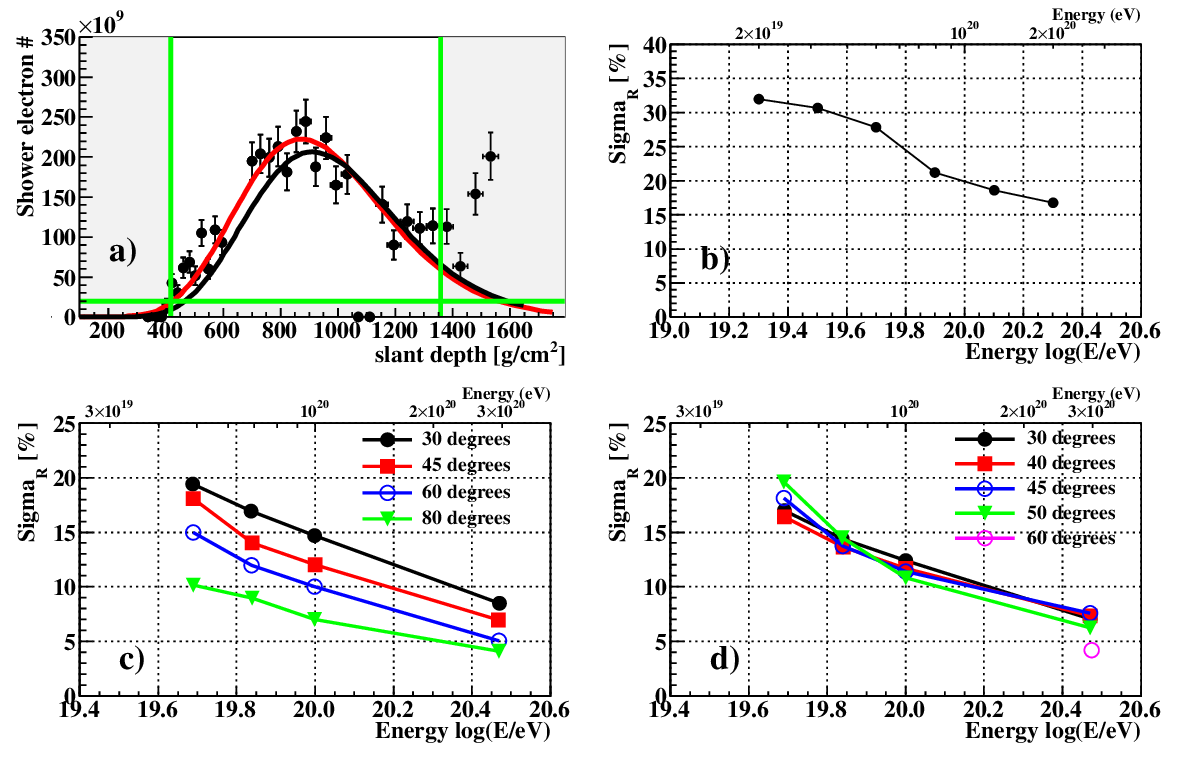}
\caption{Panel a): the simulated (black line) and reconstructed (points) shower electron curve. The GIL fit is shown as a red line. The simulated proton event has an energy of 3$\times$10$^{20}$ 
eV, a zenith angle of 50$^\circ$and an $X_{max}$ of 915 g cm$^{-2}$. 
The reconstructed parameters for this fit are 3.2$\times$10$^{20}$ eV and 873 g cm$^{-2}$. The $\chi^2$/DOF of this event is
0.905. The shaded areas show the points which are excluded from the fit. 
Panel b): the energy reconstruction performance is shown here for the all–event sample extracted on the full FoV ($\pm$270,$\pm$ 200) km and energy range 
2$\times$10$^{19}$ – 2$\times$10$^{20}$ eV. The sample with cuts DOF$>$ 4, $\chi^2$/Ndf $<$ 3 is shown here.
Panel c): The energy resolution in terms
of the $\sigma_R$ factor (see text for details), in percent, is shown. Here, we
plot the results for various zenith angles and energies. All the events are impacting in the central part of the field of 
view (namely in the inner ($\pm$20,$\pm$20) km). The geometry has been reconstructed with the slant depth method. 
Panel d): Same as in c) but adopting the Cherenkov method. 
Image adapted from~\cite{fenu-ea},\cite{fenu-icrc2015}.}
\label{fig:jemeuso-eres}
\end{figure*}

Similar studies have been conducted also in regards to the energy and $X_{\max}$ resolutions. We report here 
those performed for the HTV configuration in nadir mode adopting the SLAST-GIL shower generator~\cite{fenu-ea}, as they were conducted in a more detailed way.
This is the standard procedure adopted also for similar studies conducted with POEMMA and K--EUSO.
An analysis using different primaries (p, Fe and photons) and using CONEX shower simulator can be found 
in~\cite{guzman-icrc2015}.
After the retrieval of the signal identified by the
pattern recognition we start the correction of the inefficiencies of the detector and of the absorption in
atmosphere and compute an estimate of the fluorescence yield. The reconstruction of the geometry is
done following either of two methods: the slant depth or the Cherenkov method. As final result we obtain a shower profile which we will fit with some predefined shower function. In this way
we obtain the energy and $X_{\max}$ of the shower.
An example of the reconstructed profile with the related fit is shown in Fig.~\ref{fig:jemeuso-eres}a. %
This event has an energy of $3\times10^{20}$~eV and a reconstructed energy of
$3.2\times10^{20}$~eV. 
We see the Cherenkov reflection peak in the right part of the profile.

Using the reconstruction procedure discussed in the previous section, a study of the energy resolution of the JEM--EUSO mission has been performed.
The impact point is selected in the central part of the field of view (namely in the inner
($\pm$20,$\pm$20) km). Showers are generated according to the GIL parameterization. We simulated 8000
events for each point and we applied quality cuts DOF$>$ 4, $\chi^2$/DOF $<$ 3 on all the data. The cut on the degrees of freedom implies a minimum number of 7 points to be fit. Such a cut is a minimal requirement to ensure that events with too much light loss because of gaps are rejected.  The cut on the $\chi^2$ rejects instead the catastrophically failing fits. Under such conditions more than 80\% of the events above 10$^{20}$ eV  are selected.

To estimate the resolution, we define the parameter $R$ for each event as: 

\begin{equation}
R = \frac{E_{\rm reco} - E_{\rm real}}{E_{\rm real}}.
\label{Rparameter}
\end{equation}

The distribution of parameter \ref{Rparameter} for all the events which survived the cuts has then been fitted
with a Gaussian curve. The $\sigma_R$ parameter of such Gaussian fit is reported in panels b), c) and d) of
Fig.~\ref{fig:jemeuso-eres}. As can be seen in Fig.~\ref{fig:jemeuso-eres}c), the energy resolution tends to improve toward 
the higher zenith angles and with the increasing energy due to the better quality of the
signal. Generally, the slant depth method will always have a resolution
under 20\%. At the most extreme energies, the resolution reaches 10\% or even lower. In Fig.~\ref{fig:jemeuso-eres}d), the
energy resolution obtained with the Cherenkov method is shown. Again, the highest energies allow
the best performance, while a clear improvement depending on the zenith angle cannot be seen
anymore. This is due to the worsening quality of the Cherenkov peak at the highest zenith angles.
In fact, the Cherenkov peak will be much more difficult to recognize at large zenith angles due
to the larger spread of the reflection spot.
In Fig.~\ref{fig:jemeuso-eres}b) the energy resolution, estimated using the slant depth method, is shown for events
distributed in the range ($\pm$270,$\pm$ 200) km and for energies in the range 2$\times$10$^{19}$ – 2$\times$10$^{20}$ eV. 
The
events have zenith angles between 0$^\circ$ and 90$^\circ$ distributed as sin(2$\cdot \theta$). Here, we also apply
DOF $>$ 4, $\chi^2$/DOF $<$ 3 quality cuts on $\sim$ 10$^4$ events. The resolution ranges from $\sim$30\% at 2$\times$10$^{19}$ eV to 15–20\% at $\sim$10$^{20}$ eV. Systematics have not been corrected and
may still be contributing to the distribution width.

\begin{figure*}[!ht]
\centering
\includegraphics[width=1.\textwidth]{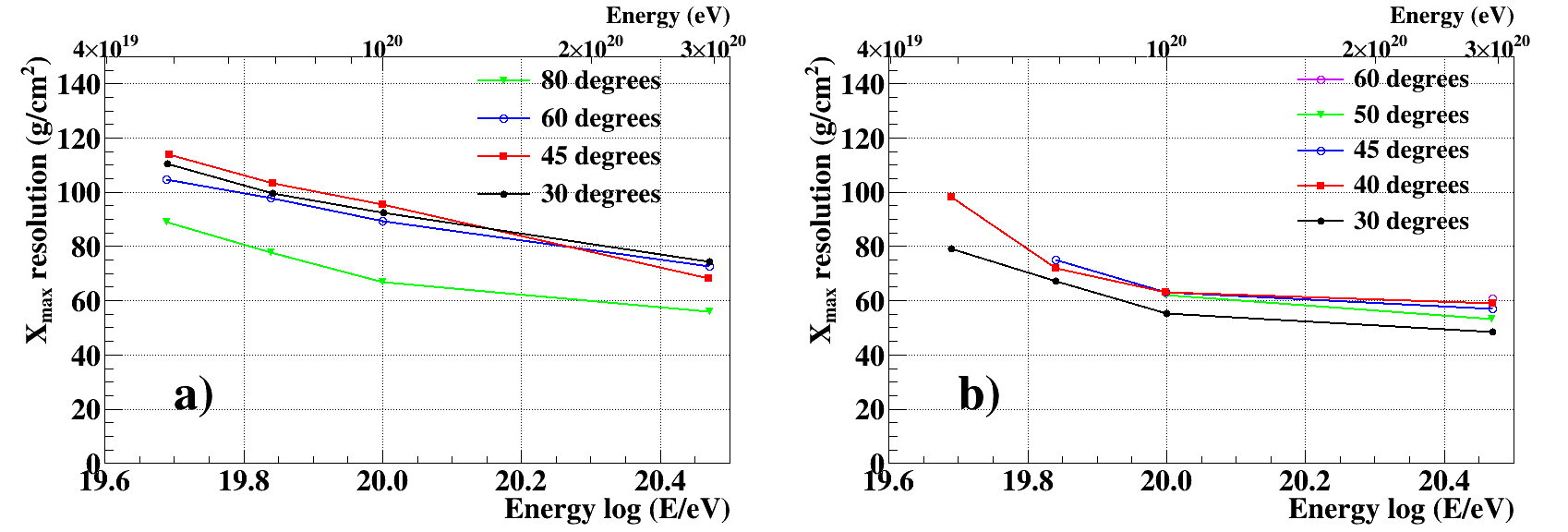}
\caption{The $X_{max}$ resolution is shown for various zenith angles and energies of proton events. All the events are impacting in the central part of the field of view (namely in the inner ($\pm$20,$\pm$20) km). The geometry has been reconstructed with the slant depth method in panel a) and with the Cherenkov method in panel b). For the lowest energy bin the reconstruction procedure adopting the Cherenkov method very often fails due to lack of signal, therefore, it can not be applied. Image adapted from~\cite{fenu-icrc2015}.}
\label{fig:jemeuso-xmax}
\end{figure*}

A similar study has been performed for the $X_{\max}$ parameter. Using the same samples generated for the study of the energy
resolution, we have calculated the distribution of the slant depth of the
maximum. Fig.~\ref{fig:jemeuso-xmax} shows the JEM–-EUSO $X_{\max}$ resolution for fixed conditions of zenith
angle and energy. Similarly to the case for the energy, we evaluate the parameter $X_{\max}^{\rm reco}$ - $X_{\rm \max}^{\rm real}$
for all the events. Then we fit the distribution with a Gaussian and we plot the $\sigma$ parameter. 
Fig.~\ref{fig:jemeuso-xmax}a) displays the reconstruction performance for the slant depth method.
As can be seen, the $X_{\max}$ resolution improves with the energy. At the lowest energies,
it ranges from 90 to 120 g cm$^{-2}$ while at the 
highest energies, from 60 to 80 g cm$^{-2}$. The higher zenith angles also yelds better resolution, due to the better angular resolution. Moreover,
the higher altitude of such events implies higher luminosity at the detector. The complete profile can be totally fitted not being cut by the ground impact.
Fig.~\ref{fig:jemeuso-xmax}b), shows the same plot obtained with the Cherenkov
method. In this case the performance is significantly better ranging from 80–100 g cm$^{-2}$
at the lowest energies to 50–60 g cm$^{-2}$ at the highest ones. At the highest zenith angles the Cherenkov
reflection peak will however not be recognizable. For this reason, the plots will not extend above
60$^\circ$. We reiterate that at these energies the difference between p and Fe $X_{\max}$ is of the order of 100 g cm$^{-2}$ \cite{yushkov-icrc2019,abbasi-composition} 

Comparable results have been obtained on the resolution of the $X_{\rm max}$ distributions using the CONEX simulator for either proton or iron showers, and slightly larger resolution for photon showers in all three cases distributed within the full FoV. 
However, an offset in Xmax was observed in this case indicating a need for a fine tuning of the reconstruction algorithms.
Despite the fact that on a single event the uncertainty on X$_{\max}$ is large, by collecting a large sample the uncertainty
on the average value will reduce significantly (in principle by a factor of 10 with a statistics of 100 events, under the condition of purely statistical errors). This allows obtaining meaningful results in terms of the evolution
of the average mass composition as a function of energy at energies which are not reachable yet by ground-based experiments due to the limited duty cycle and relative exposure. 

\subsection{The K--EUSO configuration}
\label{sec:keuso}

We present in the following the performance results in terms of aperture, exposure, angular, energy and X$_{max}$ resolutions obtained adopting the latest configuration designed for the K-EUSO mission as reported in  Section \ref{sec:missionsoverview}.

For this purpose, N = 4500 showers were simulated in the energy range E = 10$^{19}$ eV - 3$\times$10$^{20}$ eV, from all directions uniformly in azimuth and with $\sin(2\theta)$ zenith dependence in the entire field of view of the detector.
The yearly exposure as a function of energy is shown in Fig.~\ref{fig:comparison-exposure}. As it can be seen, at the plateau,
which is reached at around 10$^{20}$~eV, K--EUSO achieves an exposure of $\sim$18000 km$^2$ sr yr. The 50\%
efficiency is reached at $\sim 4 \times 10^{19}$~eV.
Assuming the spectrum recently published by the Auger collaboration~\cite{auger}, the expected rate
of triggered events has been calculated to be of the order of 4~events/year above 10$^{20}$ eV and 65~events/year above 5$\times 10^{19}$~eV.

As discussed in~\cite{keuso}, and shown in Table~\ref{tab:keuso-ang-en-xmax} by adopting the so called Numerical Exact 1 method~\cite{mernik-ea}, 
K--EUSO achieves an angular resolution between 3$^\circ$ to 7$^\circ$ at small zenith angles and
improves to 1–2$^\circ$ for nearly-horizontal events in the energy range between $5\times10^{19}$ and 3$\times 10^{20}$ eV. There is a clear improvement trend as the energy increases.

\begin{table*}[!ht]
    \begin{center}
    \caption{Angular, energy and $X_{\max}$ resolutions of the K--EUSO detector for proton EASs of different energies and zenith angles. The angular and energy estimations are done at fixed angles and full FoV while in case of $X_{\max}$ they are performed at one single angle (30$^\circ$) and refer to the center of the FoV.}
    \smallskip
    \begin{tabular}{|c|cccc|cccc|c|}
    \hline
        Energy & \multicolumn{4}{|c|}{Energy res. (\%)} & \multicolumn{4}{|c|}{Angular res. ($^\circ)$} & X$_{\max}$ res.\\
        \hline
        [EeV] & 30$^\circ$ & 45$^\circ$ & 60$^\circ$ & 75$^\circ$ & 30$^\circ$ & 45$^\circ$ & 60$^\circ$ & 75$^\circ$ & [g/cm$^2$]  30$^\circ$  \\
    \hline
        50 & 25 & 25 & 22 & 16 & 6.8 & 6.8 & 4.0 & 1.6 & 110 \\
        70 & 27 & 24 & 20 & 14 & 5.6 & 4.6 & 3.0 & 1.6 & 83 \\
        100 & 27 & 24 & 20 & 13 & 4.8 & 3.0 & 2.0 & 1.0 & 69 \\
          300 & 21 & 16 & 13 & 7 & 4.2 & 1.6 & 1.0 & 0.63 & 41 \\
    \hline
    \end{tabular}
    \label{tab:keuso-ang-en-xmax}
    \end{center}
\end{table*}


Estimations of the energy resolution of K--EUSO for UHECRs with different energies arriving
at various zenith angles are shown in Tab.~\ref{tab:keuso-ang-en-xmax}. Similarly to the analysis related to the angular
resolution, 2500 showers were simulated
at fixed energies and zenith angles, both for the center and for the full FoV of the detector.
The resolution was estimated as the standard deviation of the R distribution as shown in the previous section. It
can be seen that the energy resolution is around 25\% at low zenith angles and improves to around
15\% for nearly horizontal events, with a small improvement for higher energies.
No significant improvement of the performance has been observed if events are simulated on the central part of the FoV.


Reconstruction of the $X_{\rm max}$ of an EAS
is also performed according to~\cite{fenu-ea} and is obtained from the
fit of the reconstructed shower profile.
In this work, we only show a few examples of the reconstruction performance obtained in some specific condition.
The method we use here considers only events with a visible Cherenkov reflection peak. In this way, the impact point of the shower can be identified in the profile and therefore a clear constraint can be put onto the shower geometry. An overview of the performance in few conditions is given in
Table~\ref{tab:keuso-ang-en-xmax}. The resolution is always around
50--90~g/cm$^2$ for the center while similar values are obtained
for the whole FoV. 

These results have to be regarded as a first indication of the performance.
Improvements will be possible when the K-EUSO design is frozen and the algorithms
will be optimized for this specific configuration.

\begin{figure*}[h]
\centering
\includegraphics[width=1.\textwidth]{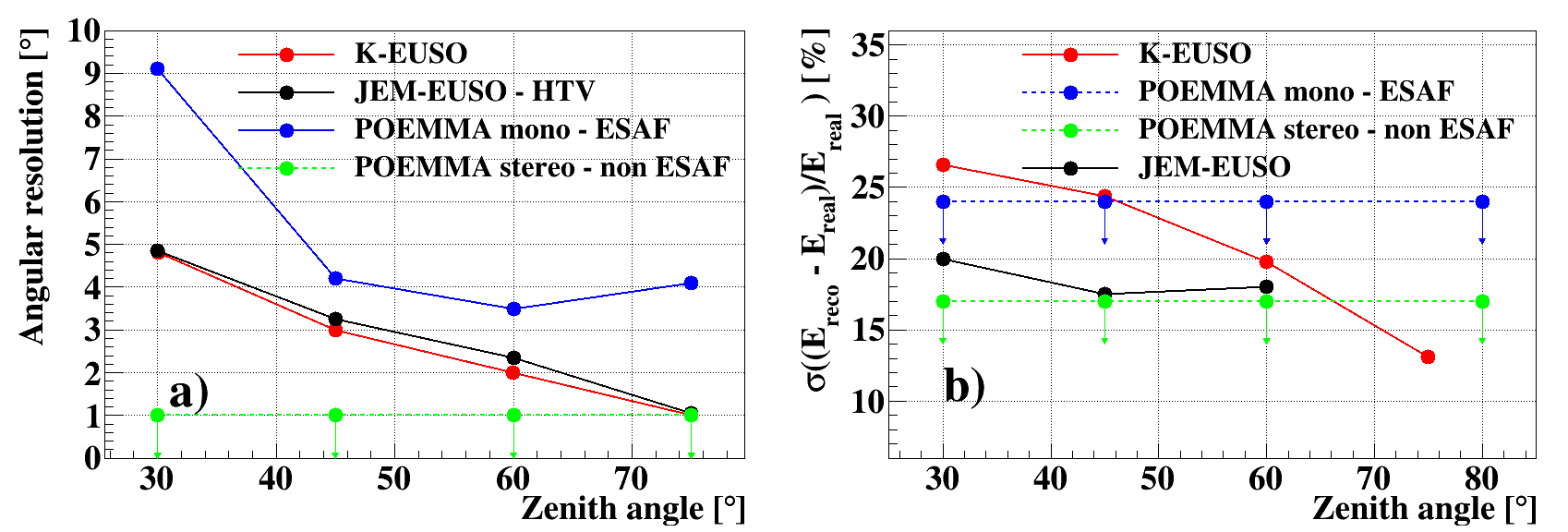}
\caption{
Comparison between JEM--EUSO, K--EUSO and POEMMA performance in terms of angular (panel a) and energy (panel b) resolutions for 10$^{20}$ eV proton EASs with different zenith angles. 
The POEMMA angular resolution in stereo mode is obtained without the use of ESAF, as it implies a stereoscopic vision, while the energy resolution has been extrapolated from the monocular mode derived with ESAF.
The dashed lines represent the result of the cumulative resolution obtained for all theta angles.
The better performance of JEM--EUSO is due to the smaller pixel footprint at ground and by a dedicated optimization
of the reconstruction algorithms. POEMMA and K--EUSO adopted the same algorithms developed for JEM--EUSO with only limited optimizations. Moreover,
in case of POEMMA a parametric simulation of the optics has been considered instead of a ray-tracing code. The comparison between 
JEM--EUSO and POEMMA shows also the margin of improvement which is in principle obtainable with a dedicated fine tuning of the reconstruction algorithms. At the same time the significant improvement of a stereoscopic configuration is clear when the mono and stereo performance of POEMMA are compared.}
\label{fig:comparison-angle-energy}
\end{figure*}

Fig.~\ref{fig:comparison-angle-energy} shows a comparison between JEM--EUSO, K--EUSO and POEMMA performance in terms of angular and energy resolutions at 10$^{20}$ eV for different zenith angles. 

\subsection{The POEMMA configuration}
\label{sec:poemma}

\begin{figure*}[h]
\centering
\includegraphics[width=0.45\textwidth]{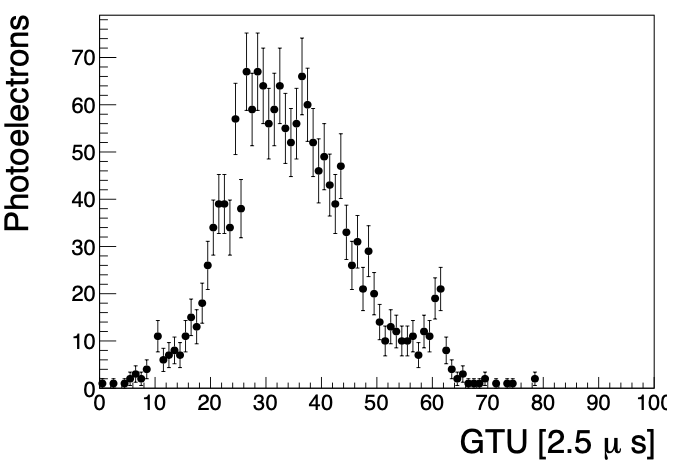}~~
\includegraphics[width=0.50\textwidth]{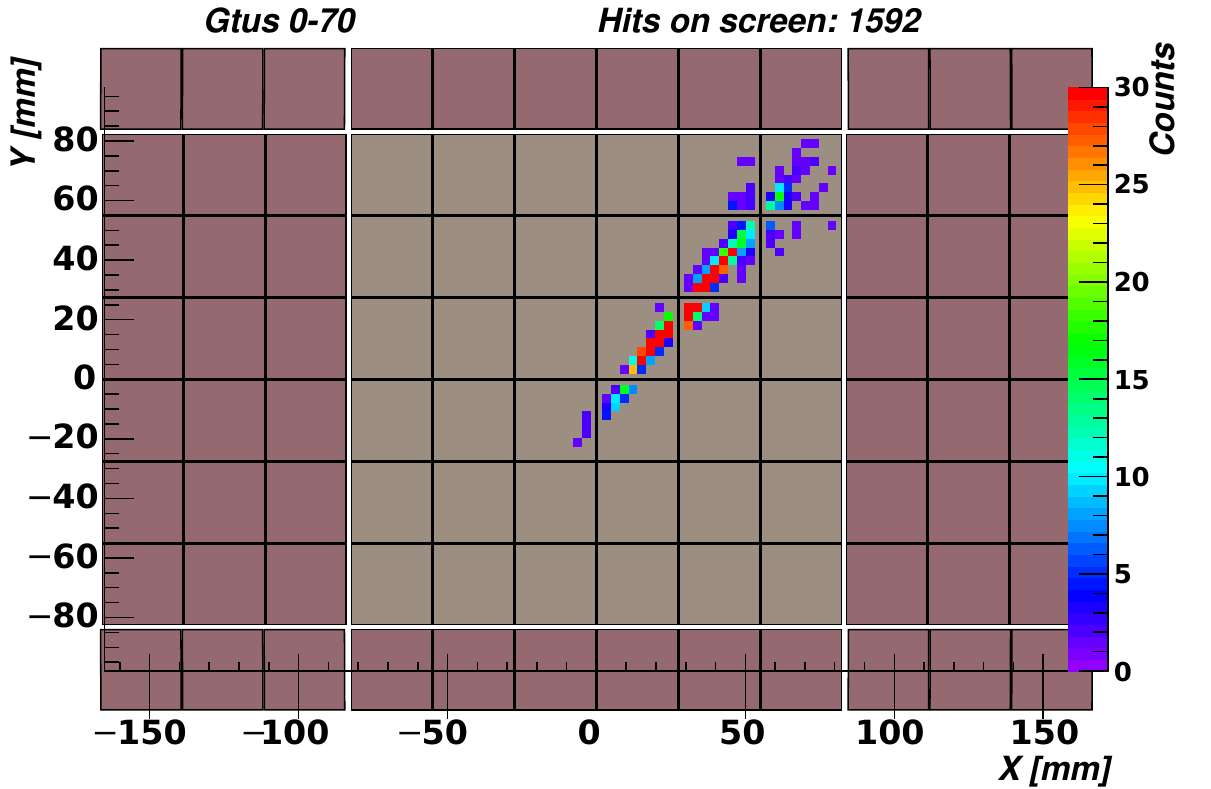}
\caption{A 10$^{20}$ eV, 60$^\circ$ zenith angle proton EAS. On the left: the photoelectron profile for the POEMMA 
detector. On the right: the image on the POEMMA's focal surface.
Image taken from~\cite{fenu-icrc2019}.}
\label{fig:poemma-event}
\end{figure*}

A first estimation of POEMMA performance in terms of trigger exposure and quality of event reconstruction has
been assessed using the ESAF code assuming a clear atmosphere (details can be found in~\cite{poemma-prd}).
Because the POEMMA PFC baseline design employs the PDMs and electronics developed for the JEM--EUSO
mission, the JEM--EUSO trigger algorithms and reconstruction procedures have been considered to evaluate
the POEMMA performance. Proton showers have been simulated using the SLANT-GIL shower generator.
The POEMMA optics response was implemented in ESAF using the parametric model. ESAF doesn't allow yet a stereoscopic
vision, therefore, monocular mode performance has been determined at first and then extrapolated to the stereoscopic view for what concerns the energy resolution. Instead an independent approach (as detailed in \cite{poemma-prd}) has been adopted for the estimation of the angular and Xmax resolutions in stereo mode.
The simulations were performed assuming a standard UV night
glow background level of 500 photons m$^{-2}$ ns$^{-1}$ sr$^{-1}$ in the 300–500~nm band. Taking into account
the POEMMA detector response, this corresponds to an average equivalent count rate of 1.54 counts $\mu$s$^{-1}$ pixel$^{-1}$. 
The study presented here assumes a 2.5 $\mu$s GTU, used as reference for the different projects.
Preliminary tests using the independent approach discussed in~\cite{poemma-prd}, with short time resolution indicate a significant impact on the angular and X$_{max}$ resolutions, while negligible on the energy resolution. 
For this reason the stereo reconstruction adopted the 1 $\mu$s GTU while for the monocular mode studies the 2.5 $\mu$s resolution was considered. The present results have, therefore, to be considered conservative, in terms of POEMMA performance.
In Fig.~\ref{fig:poemma-event} we show the photoelectron (left) and track profile (right) on the PFC focal plane
induced by a proton EAS of 10$^{20}$ eV, inclined 60$^\circ$ from the nadir.
To estimate the exposure curve of POEMMA, an overall set of 20,000 proton EASs was simulated with ESAF
in the energy range 5$\times$10$^{18}$ eV -- 5$\times$10$^{20}$ eV uniformly in azimuth and with the same zenith dependence
as in K--EUSO simulations over a sampling area that is almost twice the size of that in the FoV ($S_{FoV} \sim$145,000 km$^2$).
The exposure $\mathcal{E}(E)$ was then determined using equation~\ref{eqn:exposure} under similar assumptions. In case of
POEMMA only the first level trigger was applied. 
Based on JEM--EUSO simulations, it was estimated that the effect of the
second level trigger is to increase the exposure curve by $\sim$10\% at higher energies. The exposure was
estimated also for different tilt angles as shown in Fig.~\ref{fig:comparison-exposure}. The annual exposure in monocular mode reaches $\sim$70000 km$^{2}$ yr sr, while it reduces by about a factor of two if the stereo mode is applied with the satellites separated by 300 km ~\cite{poemma-prd}.


The triggered EASs were passed through the JEM--EUSO pattern recognition and reconstruction chain using both the slant depth
and Cherenkov methods to evaluate the exposure for reconstructed events. 
To perform a reconstruction, the same cuts as described in the JEM-EUSO section (\ref{sec:jemeuso}) were applied. The slant depth method successfully reconstructed $\sim$84\%
of the triggered events with almost constant efficiency above log(E/eV) $>$ 19.6, while for the Cherenkov method only about half of the events were reconstructed. As
mentioned before, this is due to the fact that this method is usable up to zenith angles $\sim$50$^\circ$ (the value 
depends on the EAS energy and location on the FS). At higher zenith angles the Cherenkov signal is spread and too dim to be
isolated from background fluctuations.

The performance of the angular reconstruction for POEMMA was evaluated at fixed zenith angles ($\theta$ = 30$^\circ$, 45$^\circ$, 60$^\circ$,
and 75$^\circ$) for three different energies (E = 7$\times$10$^{19}$, 10$^{20}$ and 3$\times$10$^{20}$ eV).
The same methodology defined for the JEM--EUSO reconstruction was applied to POEMMA,
with a fine-tuning of the parameters of the PWISE algorithm.
Results are presented in Tab.~\ref{tab:poemma-ang-en} in terms of $\gamma_{68}$ parameter. It is important to underline
that a more detailed study of the bias should be performed. The reduction of the bias would improve the 
overall performance of $\gamma_{68}$ as it includes both statistical and systematic uncertainties. 

\begin{table}[!ht]
    \begin{center}
    \caption{Angular and energy resolutions of the POEMMA detector for proton EASs of different energies and zenith angles obtained with ESAF under different assumptions. The angular resolution is obtained at fixed energy in monocular mode.
    The energy resolution is in monocular mode and is presented for two cases: results in the first column have been obtained with the assumption that the angular reconstruction was provided in stereo mode with an angular resolution of 1$^\circ$ in both zenith and azimuth angles while in the second column we put results without assumption of 1$^\circ$. Assuming a stereo vision the energy resolution improves by $\sqrt2$, assuming no bias or similar bias between the two telescopes. }
    \smallskip
    \large
    \scalebox{0.63}{%
    \begin{tabular}{|c|c|c|cccc|}
    \hline
        Energy & \multicolumn{2}{|c|}{Energy res. (\%)} & \multicolumn{4}{|c|}{Angular res. ($^\circ$)}\\
        \hline
        [EeV] & All angles (1$^\circ$ res.) & All angles & 30$^\circ$ & 45$^\circ$ & 60$^\circ$ & 75$^\circ$\\
    \hline
        50 & 26 & 30 & 12&6.2&5.5&5.2\\
        70 & 25 & - & 9.9&6.4&3.9&5.1\\
        100 & 24 & 27 &9.1&4.2&3.5&4.1\\

        300 & 23 & - &6.2&2.3&2.1&1.4\\
    \hline
    \end{tabular}}
    \label{tab:poemma-ang-en}
    \end{center}
\end{table}


The stereoscopic vision of POEMMA allows a much better angular reconstruction than does monocular vision. 
In stereo mode the reconstruction of the EAS trajectory is robust, yielding a much
better angular resolution, which is 
$\sim1.5^\circ$($\sim1^\circ$) or better above 5$\times$10$^{19}$ (10$^{20}$) eV
(see~\cite{poemma-prd} for more details). 


For the energy resolution study, we present here the results for two situations: assuming that the EAS direction has been pre-determined
using the stereo approach with a 1$^\circ$ resolution in both zenith and azimuth angles and without this assumption. The results are
shown in Tab.~\ref{tab:poemma-ang-en}. With 1$^\circ$ assumption, for 5$\times$10$^{19}$ eV, the energy resolution is 26\%
with a +3.5\% bias, while for 10$^{20}$ eV the energy resolution is 24\% with a -1.5\% bias. Since the two POEMMA
telescopes provide independent measurements of each EAS, the combined resolution is obtained by dividing by
$\sqrt 2$, assuming no bias or similar bias between the two telescopes, yielding 18\% at 5$\times$10$^{19}$~eV and 17\% at 10$^{20}$~eV.


A preliminary study of the $X_{\max}$ resolution was performed as well indicating resolutions of the order of those obtained
for JEM--EUSO. However, similarly to the angular case, an estimate of the $X_{\max}$ resolution in stereo mode was 
conducted. 
The total $X_{\max}$ resolution of POEMMA, including both angular resolution and photoelectron statistics, is about 31 
g cm$^{-2}$ at 3$\times$10$^{19}$~eV for events below 60$^\circ$ (72\% of the data sample)
and 39 g cm$^{-2}$ below 70$^\circ$ (91\% of the data sample). At 10$^{20}$ eV the resolution is 17 and 21 g cm$^{-2}$, 
respectively (see~\cite{poemma-prd} for more details). These results indicate the much better performance of a stereoscopic
detector compared to a monocular vision.

\subsection{The TUS configuration}
\label{sec:tus}



\begin{figure*}[!ht]
\centering
\includegraphics[width=0.42\textwidth]{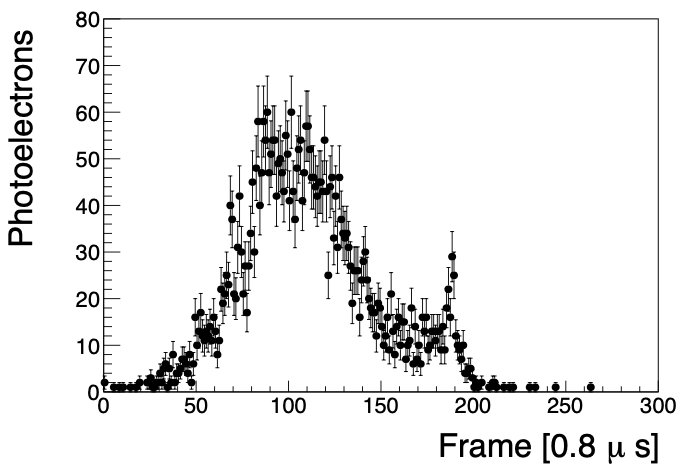}~~
\includegraphics[width=0.53\textwidth]{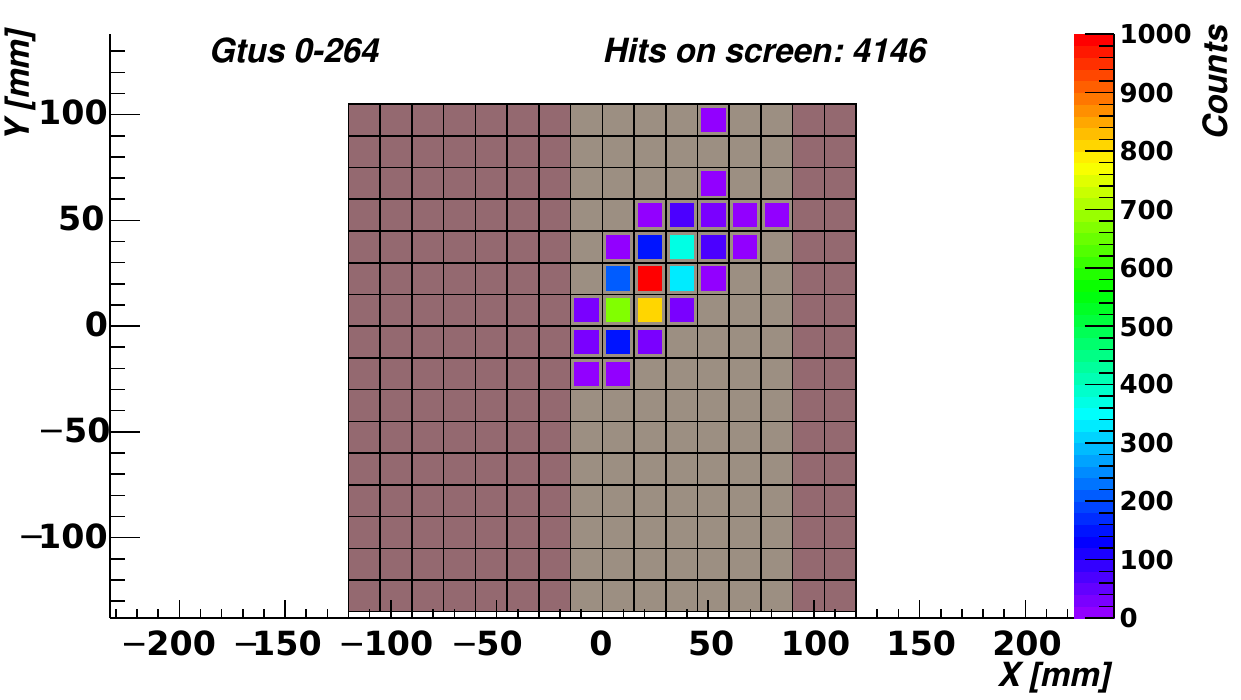}
\caption{ A simulated 10$^{21}$ eV, 60$^\circ$ zenith angle proton event.
Left panel: the photoelectron profile for the TUS
detector. Right panel: the photoelectron image for TUS.
Image taken from~\cite{fenu-icrc2019}.}
\label{fig:tus-event}
\end{figure*}

The ESAF code has been used to study the TUS performance and comparison to data with two different
approaches and simulation codes. The first one is ESAF coupled with the TUSSIM program package developed at
the Joint Institute for Nuclear Research, Dubna (Russia)~\cite{esaf-tussim}. 
ESAF is used to generate the EAS cascade and the fluorescent radiation which is propagated at TUS optics level.
Then, the TUSSIM program simulates the TUS detector performance including the Fresnel mirror optical parameters, the light
concentrator of the photo detector, the front-end and trigger electronics.
In the second approach~\cite{tus-mario}, the TUS detector is implemented directly into the ESAF simulation code.
Regarding the optics simulation, approaches similar to JEM--EUSO have been adopted.
The standard one, which was used in this
work, adopts a parametric simulation module that analytically calculates the position
of the photon on the FS and adds a Gaussian spread around this position.
This is intended to be a fast working
tool to test the features of different optics designs in an approximate way.
Once the photons reach the FS, they are transported through the filter and
the optical adapter before reaching
the photocathode. All the relevant effects including geometrical
losses, inefficiencies of the adapter and of the UV filters are taken into account.
A parameterization of the photo-multiplier response is included in the electronics part.
All the effects like quantum efficiency, dependence on the incident
angle of photons, collection efficiency and cross-talk are also taken into account.
The signal is then amplified
by a parameterized gain and the resulting output current is collected and treated by the
front-end electronics module.
Fig.~\ref{fig:tus-event} shows an example of the light profile and shower track
expected to be detected from a 
10$^{21}$~eV, 60$^\circ$ zenith angle proton EAS.

\begin{figure*}[!ht]
\centering
\includegraphics[width=0.49\textwidth]{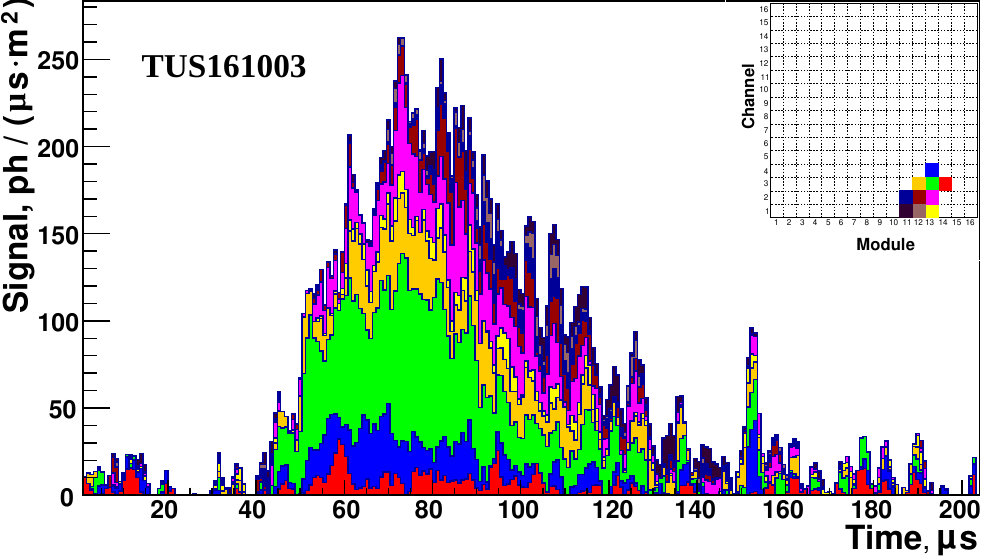}
\includegraphics[width=0.49\textwidth]{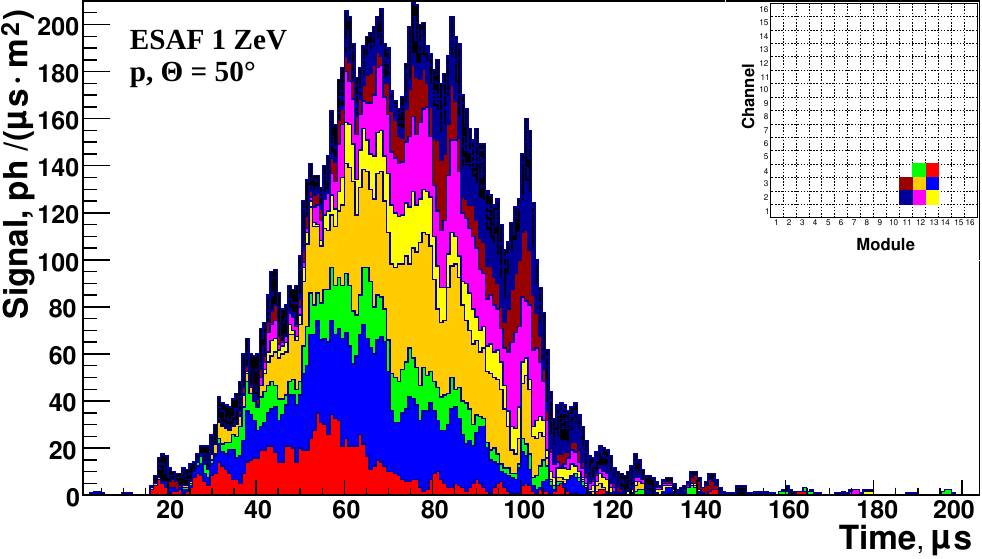}
\caption{Left panel: The light curve of the TUS161003 event as the signal of the ten hit channels stacked together. The insert shows positions of the hit pixels on the focal surface. Right panel:
An expected TUS detector response to an EAS from a 1 ZeV proton arriving at the zenith
angle $\theta \approx 50^\circ$.
Image adapted from~\cite{tus-jcap}.}
\label{fig:tus-minnesota_esaf}
\end{figure*}

During its operation, TUS has detected about $8\times10^4$ events that have been analyzed offline to select those satisfying
basic temporal and spatial criteria of UHECRs. A few events passed this
first screening. Those that exhibited some EAS-like characteristics were
compared to ESAF simulations to understand if they
were consistent with an UHECR origin. One specific event, which was registered in perfect
observational conditions, 
was deeply scrutinized (see Fig.~\ref{fig:tus-minnesota_esaf}). Its phenomenology and the possible interpretations were
reported in detail in~\cite{tus-jcap}. Thanks to the comparison with ESAF simulations, it was possible to demonstrate that the PMT waveforms and the light curve of the event show similarities with expectations from an EAS. However, the amplitude corresponds
to UHECR energies $E$ $\gtrsim$ 10$^{21}$~eV, which makes the cosmic
ray origin of this event highly unlikely as TUS accumulated exposure is two orders of magnitude lower than accumulated exposure collected by ground-based experiments and no event was detected so far above 3$\times$10$^{20}$ eV~\cite{Bird:1994uy}. Another important phenomenological feature of the event 
is that it develops very high in the atmosphere. The duration of the signal and its slow 
attenuation lasting $\sim$60 $\mu$s lead to an estimate of $X_{\max}$ $\sim$550 g cm$^{-2}$, which corresponds to 
the altitude of $\sim$7.5 km. 
On the other hand, if EASs are simulated with an inclination around ~45$^{\circ}$, which corresponds to the reconstructed direction of the TUS event, the X$_{max}$ has values in the range 915 - 985 g cm$^{-2}$ which are much deeper in the atmosphere. 
Moreover, the EAS profile of the TUS event is closer to a ~60$^{\circ}$ inclined event which again doesn’t provide a good matching between shower profile and direction reconstruction. 
An even higher altitude ($\sim$8.5 km) of the signal maximum is obtained if one 
assumes the narrow peak at around 150 $\mu$s to be Cherenkov reflection. These values are in a contradiction with expectations from the ESAF simulation of a ZeV event (see Fig~\ref{fig:tus-minnesota_esaf}).
This inconsistency indicates that the peak around 150 $\mu$s might be due to other reasons, including a random fluctuation of the signal or an artifact of the electronics or due to an anthropogenic source on ground. A similar peak was indeed observed also in other events which showed an apparent movement of the signal. These were later on associated with the presence of airports (\cite{sharakin-tus-airports}). On the ocean 15 EAS-like events in total were found, four of which had at least three active channels and were registered in good observational conditions. However, for none of them it was possible to reconstruct the arrival direction of the light source accurately, either for the limited number of hit pixels or because they occurred at the edge of the FoV (\cite{tus-mario}).
This shows the importance of ESAF not only to understand the performance of
an instrument but also to inspect experimental events through simulations. Mini-EUSO results, as the detector features similar FoV per pixel, even though a higher energy threshold, will help understanding the expected random imitation rate of this kind of EAS-like signals.

ESAF was also used to perform a detailed analysis of the TUS exposure and sensitivity to UHECRs~\cite{tus-francesco}.
Two thousand EASs were injected in an area $A_{\text{simu}}$ larger than the FoV
($\pm150$~km)
to avoid border effects. The TUS trigger logic was implemented in ESAF. Several trigger thresholds used in the mission were 
tested with an airglow rate of $\sim18$ photoelectrons per frame. 
 As a result of our preliminary
estimation,
we obtained a trigger threshold $\gtrsim5\times10^{20}$~eV. 
However, this estimation is affected by large uncertainties due to an accident that occurred during the first days after launch (described in~\cite{tus-jcap}), when 20\% of
the PMTs were destroyed and sensitivities of the remaining PMTs changed compared to the
pre-flight measurements. Despite a number of in-flight calibration attempts,
considerable uncertainties still remain on the PMT gains. 
Moreover, the TUS trigger
algorithm is more efficient for
horizontal showers leading to a higher fraction of high zenith angle events.
The majority of the
events could indeed trigger only above $40^\circ\text{--}50^\circ$. This is a consequence
of the persistence condition
of the trigger that rejects all events lasting for a short time.

Secondly, the efficiency of the trigger in cloudy conditions was evaluated. 
The cloud condition for each trigger has been estimated based on MERRA data (see~\cite{tus-mario} for details).
One thousand EASs at fixed energy have been simulated for each cloud top height condition in similar way as for clear sky. 
An estimate of the overall reduction of the exposure during the whole flight can be given by an average of the trigger efficiency weighted by the fraction of triggers in each condition. This leads to 57\% of what is expected for the clear sky case. By taking into account the above factors the geometrical exposure in clear sky conditions amounts to $\sim$1550~km$^2$~sr yr at plateau energies, reducing to $\sim$884~km$^2$ sr yr at $2\times10^{21}$~eV taking into account the cloud impact. It is important to recall that the estimation of the exposure might have a cloud dependence due to the interplay of the brightness of the shower and the location of its maximum. For the same cloud location, at lower energy a lower value for the exposure is expected.

\subsection{The Mini--EUSO configuration}
\label{sec:minieuso}

\begin{figure*}[!ht]
\centering
\includegraphics[width=0.45\textwidth]{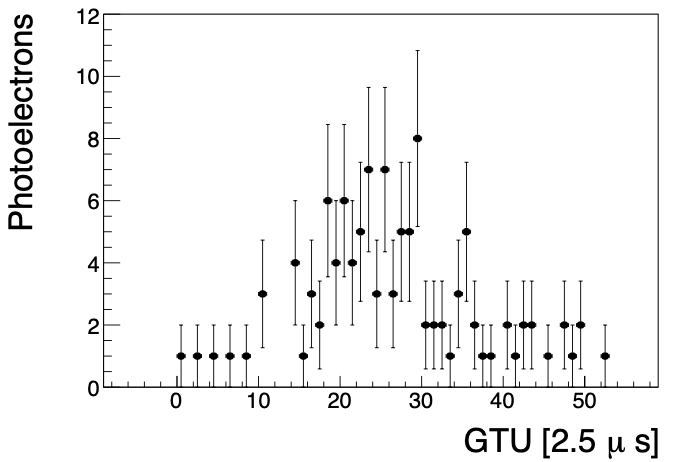}~~
\includegraphics[width=0.53\textwidth]{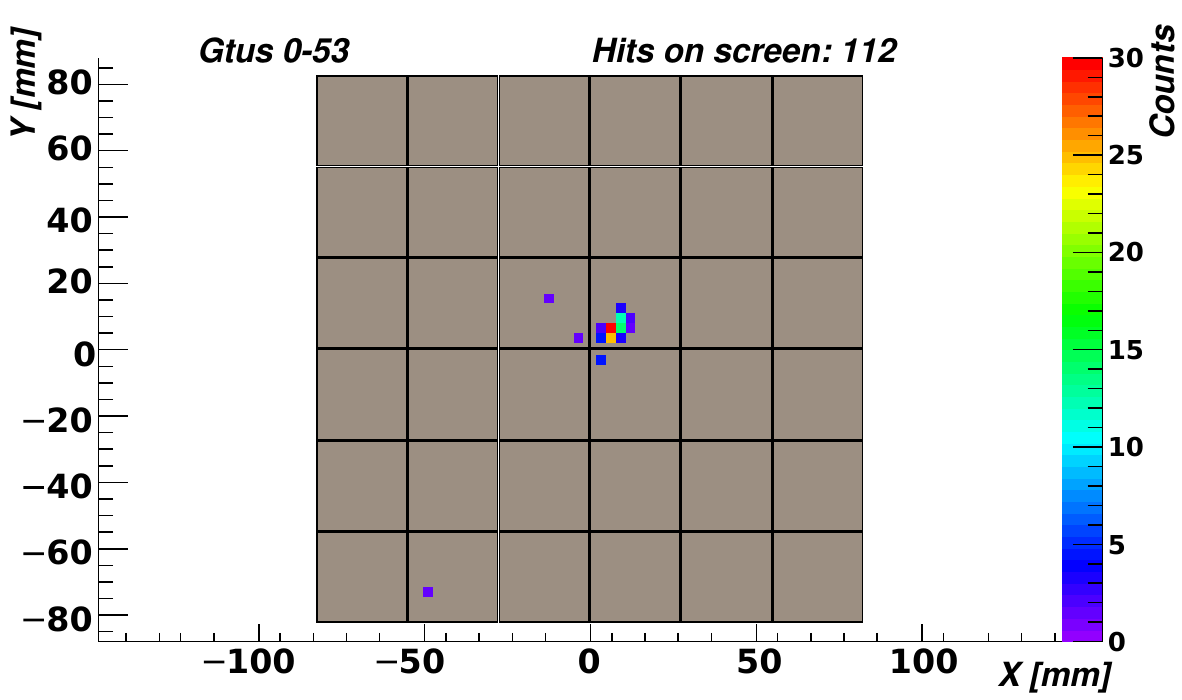}
\caption{A 10$^{21}$ eV, 60$^\circ$ zenith angle proton event. Left: the photoelectron profile for the 
Mini–-EUSO detector. Right: the photoelectron image. Image taken from~\cite{fenu-icrc2019}.}
\label{fig:minieuso-event}
\end{figure*}

The Mini--EUSO configuration has been incorporated in ESAF including its trigger logic~\cite{mini-trigger}. 
Proton showers have been simulated using the SLAST-GIL shower generator. A ray trace code of the Mini-EUSO optics response
is included in ESAF. Fig.~\ref{fig:minieuso-event} shows an example of a 10$^{21}$ eV, 60$^\circ$ zenith angle simulated
proton event. Mini-–EUSO is at the detection threshold at such energies. The signal looks dim. Background
has not been simulated but it is expected to be at a level of $\sim$1 count/pixel/GTU(2.5~$\mu$s) in standard observational conditions.

\begin{figure}[!ht]
\centering
\includegraphics[width=\columnwidth]{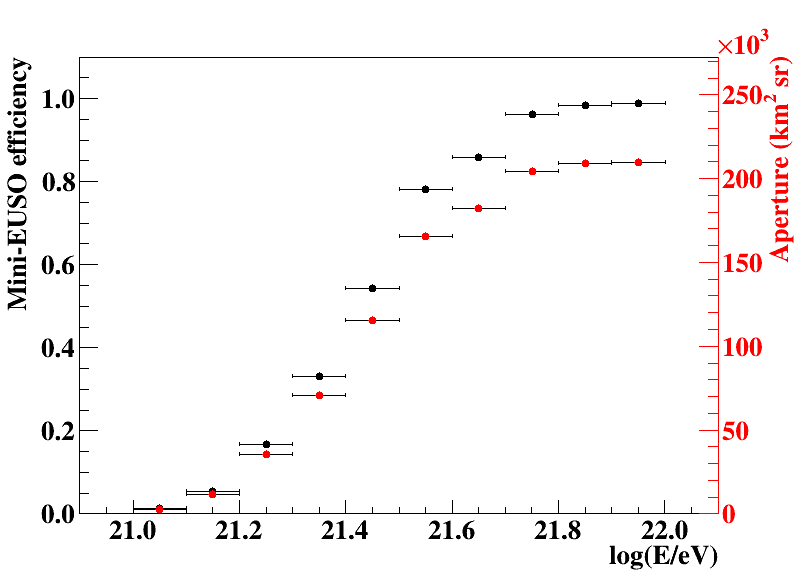}
\caption{The trigger efficiency (on the left axis, in black) and geometrical aperture (on the
right axis, in red) are shown as a function of the EAS energy, E, in eV. A UV background level of 1~count/pixel/GTU(2.5~$\mu$s) was considered in both cases.}
\label{fig:minieuso-exposure}
\end{figure}

The trigger efficiency curve and aperture have been studied in a way
similar to that used for the other
instruments. Fig.~\ref{fig:minieuso-exposure} shows the derived trigger efficiency
and the aperture curves.

The experimental data taken by Mini--EUSO allow a first comparison with the assumed background levels in JEM--EUSO, K--EUSO and POEMMA to verify that the estimated performance is based on justified assumptions. Table~\ref{tab:table_moon_cloud} shows Mini--EUSO results on the average UV emissions in different conditions: clear and cloudy conditions, sea and land, various lunar phases as reported in~\cite{minieuso-uv}. As explained in~\cite{minieuso-uv} the presence of clouds is derived from the US National Weather Service Global Forecast System (GFS)~\cite{GFS}. Assuming no-moon conditions and typical land/ocean and clear/cloudy atmosphere ratios equal to 30/70, the average background level is $\sim$1.3 counts/pixel/GTU.




In order to re-scale this value to the JEM--EUSO (JE) case a full simulation of JEM--EUSO and Mini--EUSO detectors was performed with ESAF. 
In case of Mini--EUSO (ME) the overall efficiency of the detector was fine-tuned in ESAF, mainly acting at the level of MAPMT response, to match the measured one $\epsilon_{ME}$ = 0.080 $\pm$ 0.015 (see~\cite{minieuso-uv}) for a point-like source on ground. 
A flat diffused UV emission in the range $\lambda$ = 300 - 400 nm was simulated at the detector’s aperture either with a range of zenith directions much larger than the FoV of the instrument ($\pm$ 60$^\circ$ for both detectors) or just within the FoV of the detectors ($\pm$30$^\circ$ for JEM--EUSO and $\pm$22$^\circ$ for Mini--EUSO). 
The measured background ratio (R(ME/JE)) between Mini--EUSO and JEM--EUSO at FS level was R(ME/JE) = 0.98 - 1.04 slightly depending on the range of zenith angles. 
This indicates that the expected photon counts for JEM--EUSO should be similar to Mini--EUSO one $\sim$(1.3 $\pm$ 0.2) counts/pixel/GTU taking into account the uncertainty in the estimation of Mini--EUSO efficiency.
This result indicates that the average value of $\sim$1.1 counts/pixel/GTU assumed in JEM--EUSO simulations is within the current estimation and confirms the robustness of the hypotheses undertaken with ESAF.

 \begin{table*}[!ht]
     \centering
     \caption{Average emission values in counts/(pixel$\cdot$GTU) for Mini-EUSO for sea and ground for various lunar phases and cloudiness. In the table ‘cloudy all’ indicates the weighted average of the counts in ‘cloudy land’ and ‘cloudy sea’. Half-moon includes Moon fractions between 0.4 and 0.5, and full-moon includes fractions between 0.9 and 1. The brightest pixels (above the 99th percentile) were excluded when calculating the mean and standard deviation to mitigate the effects from bright anthropogenic sources. For conditions with multi-modal distributions, the mode closest to the average is displayed. Table adapted from~\cite{minieuso-uv}.}
    \begin{tabular}{|c|c|c|c|c|c|}
         \hline
         & clear sea  & clear land  & cloudy sea  & cloudy land  & cloudy all \\\hline
         No-moon   &    0.9 &    1.4 &     1.3 &     1.7 &      1.4 \\
         Half-moon &      1.8 &  2.8 &     13.0 &      8.1 &      9.7 \\
         Full-moon &    37.6 &   35.1 &   50.7 &    51.1 &    51.0 \\\hline
     \end{tabular} 
     \label{tab:table_moon_cloud}
 \end{table*}

\begin{figure*}[!ht]
\centering
\includegraphics[width=\textwidth]{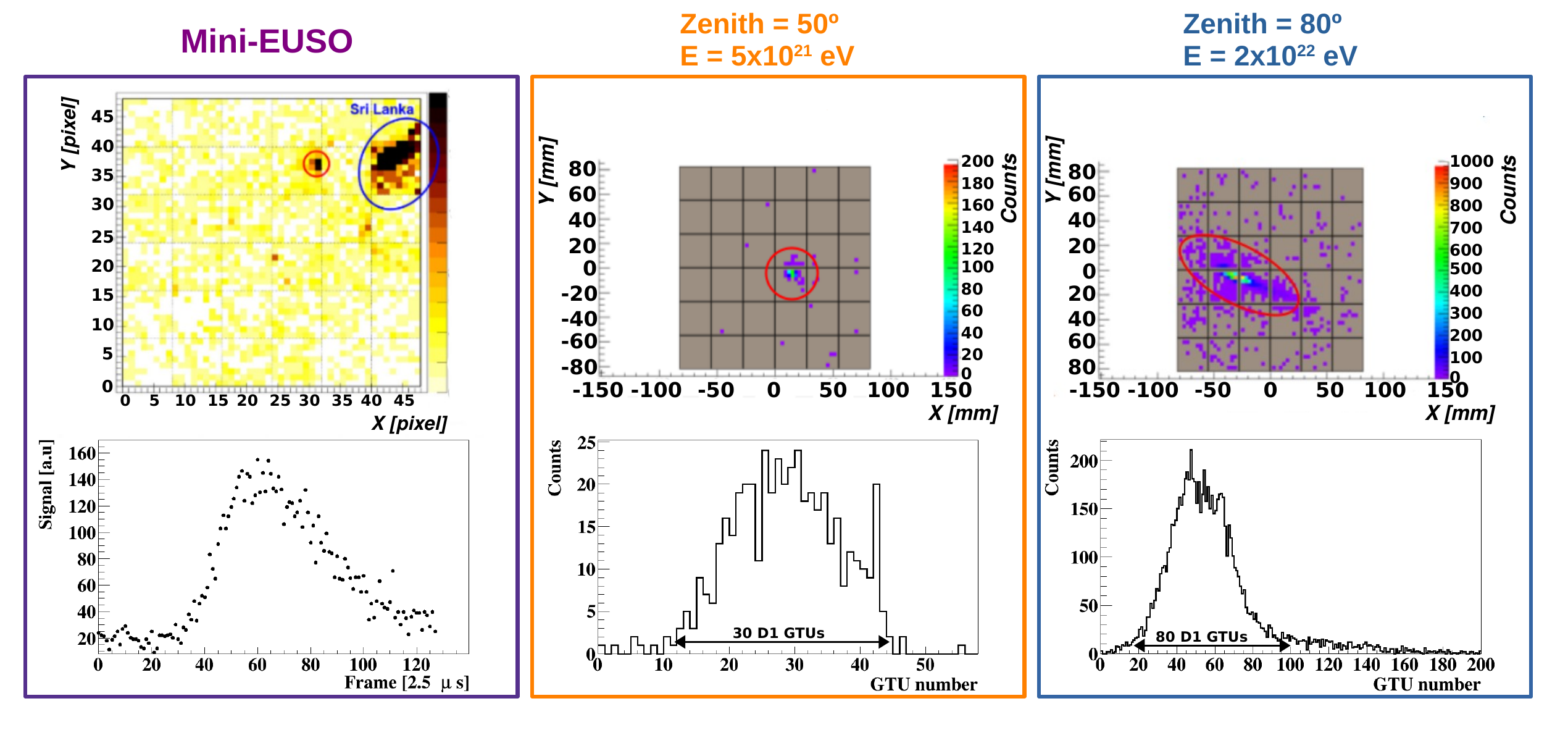}
\caption{Left panel: one frame of an event triggered off the coast of Sri Lanka. The blue circle (upper right) encloses a static, bright light source to be discarded. The trigger was issued by
the event in the red circle. Bottom left pane: the lightcurve of a 3$\times$3 pixel box that 
contains the event. The lightcurve presents
the characteristic bi-gaussian shape, with a faster rise and a slower decay time.
Center and right panels: Two proton EAS simulated through ESAF at two different
energies and zenith angles (top part presents the image of the events while the bottom part the corresponding light profiles). Center panel shows a simulation at zenith $\theta$ = 50$^\circ$ and energy
E = 5$\times$10$^{21}$ eV. The signal persists on few pixels for $\sim$30 GTUs, much shorter than the measured one, the sharp
cutoff is given by the shower reaching the ground. Right side is a simulation at zenith
$\theta$ = 80$^\circ$ and energy E = 2$\times$10$^{22}$ eV. The signal is much longer in time but the
footprint on the focal plane is much more elongated. Image adapted from~\cite{matteo-asr}.}
\label{fig:minieuso-srilanka}
\end{figure*}

ESAF simulations are performed to test the UHECR origin
of short bright signals detected by Mini--EUSO with time duration in the 
range of EASs. So far all the investigated candidates are not compatible in
pulse shape and track image with those expected from EAS. Those showing a periodic behaviour can be immediately discarded as anthropogenic sources, while non-repetitive ones require a more detailed analysis. One of the most
interesting examples in the Mini--EUSO data sample is shown in Fig.~\ref{fig:minieuso-srilanka}.
It has been detected off the coast of Sri Lanka. The trigger was issued by
the event in the red circle. The lightcurve presents
the characteristic bi-gaussian shape of an EAS, with a faster rise and a slower decay time.
The event was compared to different simulated EASs with variable energy and zenith angle. No simulated
EAS is compatible with both the image size and the time duration of the light profile. In fact, the light spot is compatible with a nearly vertical event, but the duration is much longer than the time needed by
a vertical shower to develop in atmosphere and reach the ground. This event has,
therefore, a different nature which is currently under investigation.
A detailed description of the onboard performance of the Mini--EUSO first level trigger and search for EAS-like events can be found in~\cite{matteo-asr}.

%

Among Mini--EUSO scientific objectives there is also the study of slower events such as meteors and fireballs, 
and the proof-of-principle of space debris 
observation with absolute magnitude of M $\lesssim$ +5 using the
albedo from the Sun~\cite{ebisu-debris}. 
In optimal dark conditions, at the detection threshold M=+5, the signal
(integrated at steps of 40.96 ms) will exceed the UV-nightglow level by 3--4$\sigma$.
These events will
be detected using offline trigger algorithms on ground~\cite{hiroko-debris}.

\begin{figure*}[!ht]
    \begin{center}
        \includegraphics[width=1.\textwidth]{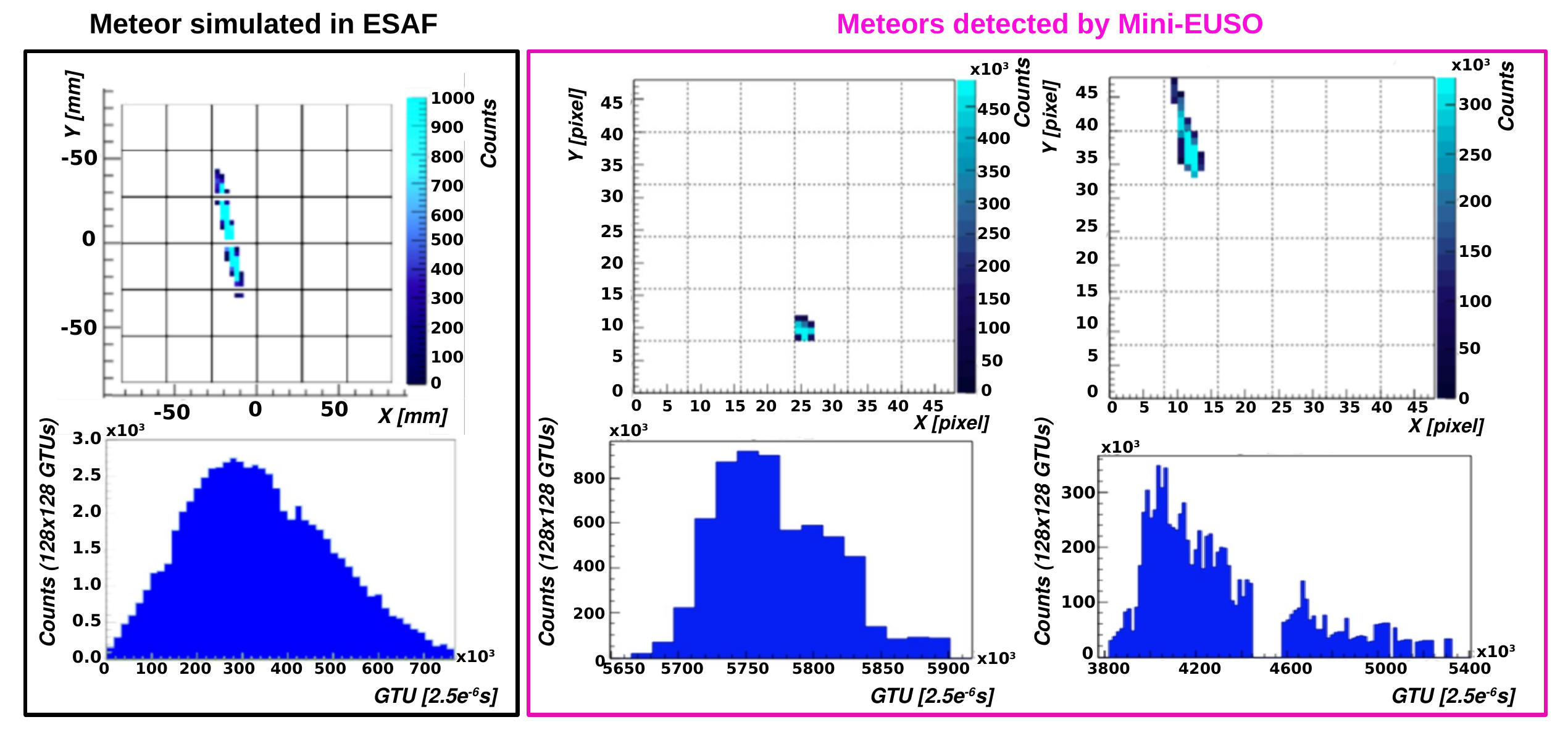}
        \caption{Left panel: ESAF simulated light track of a meteor of absolute magnitude M = +5 and 45$^\circ$ inclination in Mini--EUSO (the
effects of UV-nightglow are not included and the threshold is 30 counts). Bottom: Expected
light profile. Each time bin on the x-axis corresponds to an integration time of 40.96 ms. Center \& right panels: Example of meteors detected by Mini--EUSO. In the
Mini--EUSO data, there are meteors with different brightness and time duration. Image taken from~\cite{hiroko-icrc2021}.}
        \label{fig:minieuso-meteor}
    \end{center}
\end{figure*}

Fig.~\ref{fig:minieuso-meteor} shows an example of a simulated
meteor track having absolute magnitude M = +5 crossing the field of view of
Mini--EUSO with a 45$^\circ$ inclination with respect to the nadir axis. The meteor
speed is 70 km s$^{-1}$ and its duration is 2 s. The center and right panels show two different
meteor candidates detected by Mini--EUSO. In Mini--EUSO data, there are tens of thousands of meteor candidates with different brightness and time duration which can be simulated with ESAF.

As pointed out in~\cite{ebisu-debris}, a detector like Mini--EUSO is also potentially capable of detecting space debris, if
not directly exposed to sunlight.
Under this assumption, Mini--EUSO
would be effectively a high-speed camera with a large FoV and could be used as
a prototype for the detection of space debris
during the twilight periods of observation. It will detect debris when
they are illuminated by the Sun. In the current simulation with
ESAF, the photon flux of the Sun in the 300--400~nm has been considered to be $10^{20}$ photons m$^{-2}$ s$^{-1}$. The debris are assumed
to have a spherical shape of diameter~$d$ and a variable reflectance. Therefore, the event 
appears as a spot of light moving through the field of view with slowly variable light profile.

\begin{figure*}[!ht]
    \begin{center}
        \includegraphics[width=0.36\textwidth]{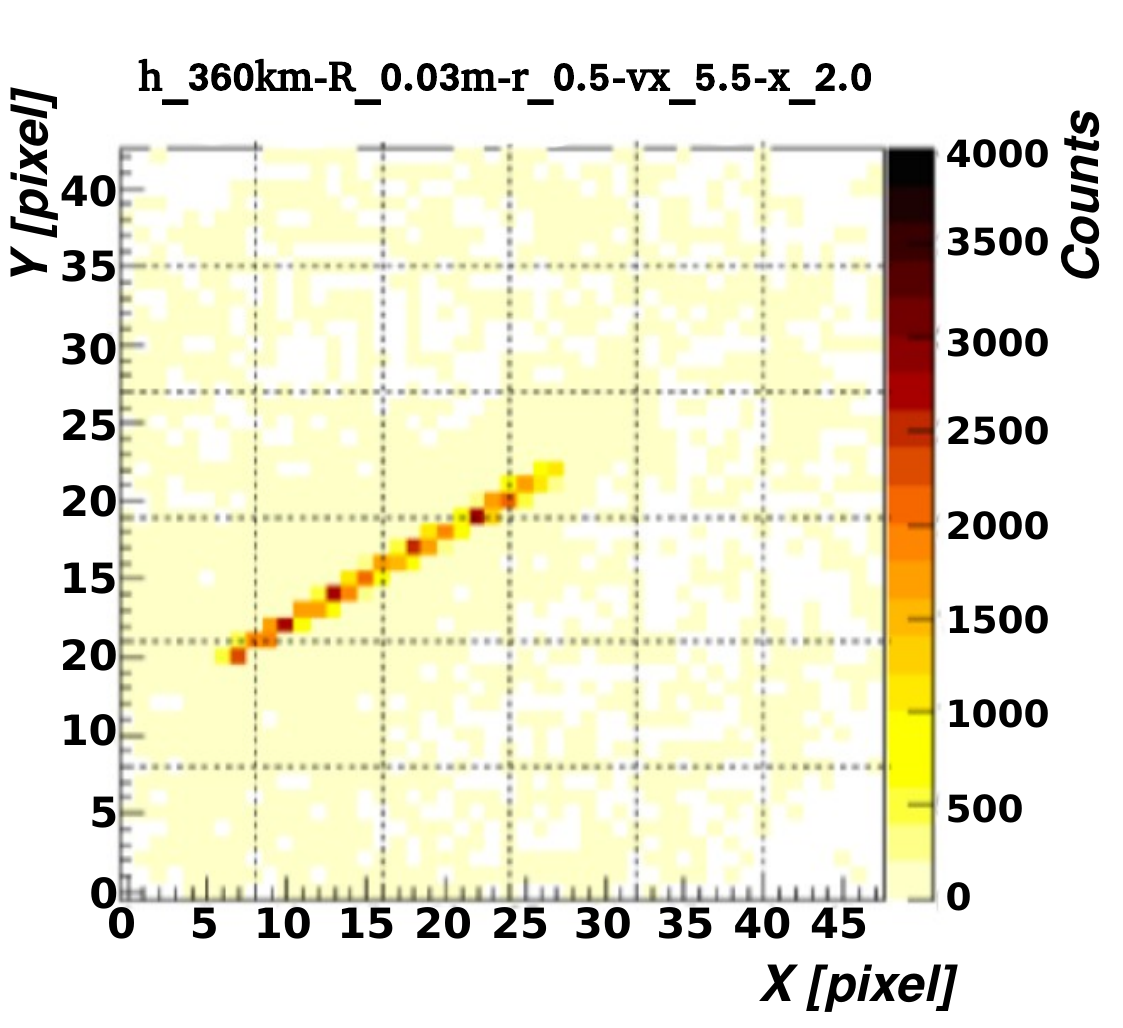}
        \includegraphics[width=0.62\textwidth]{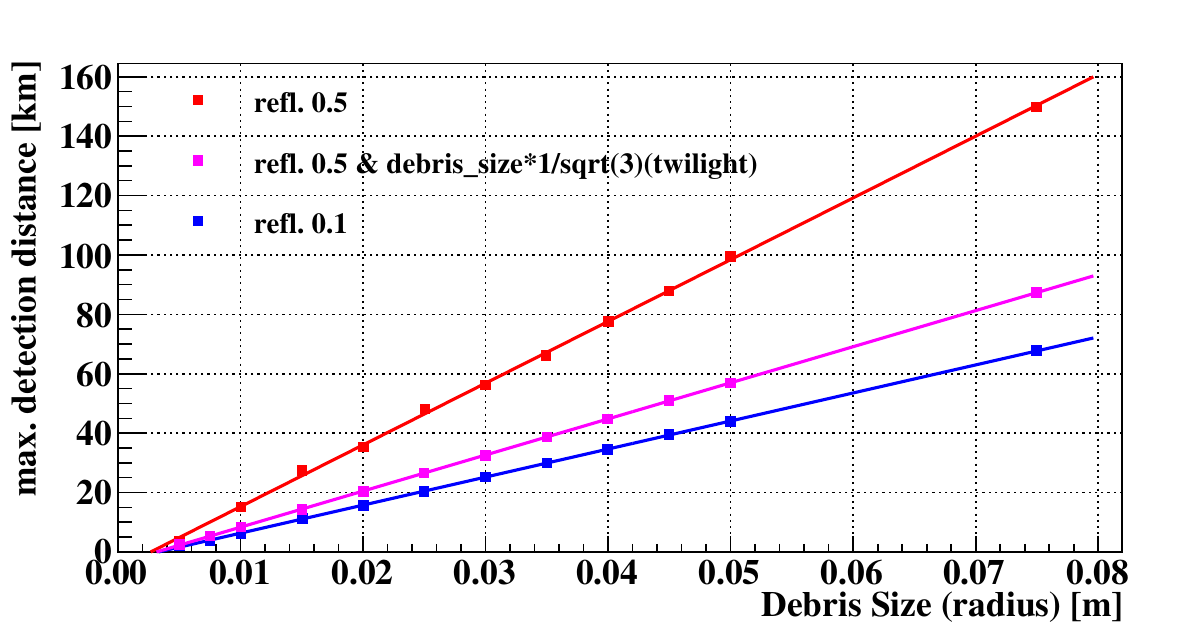}
        \caption{Left panel: An example of space debris detection simulated with ESAF. Right panel: maximum distance of space debris 
        observable by Mini--EUSO from the ISS as a function of reflectance and size of the debris, derived from  
        ESAF simulations using the standard background level adopted for UHECR detections.
        Image adopted from~\cite{hiroko-icrc2021}.}
        \label{fig:minieuso-debris}
    \end{center}
\end{figure*}

Fig.~\ref{fig:minieuso-debris} shows the potential of Mini--EUSO to detect space debris. 
On the left side a simulated object with 3~cm radius flying at 360~km height (40 km from Mini--EUSO) with a speed 7.7~km/s is displayed.
A study of the sensitivity of Mini--EUSO to space debris in terms of size and relative distance has been carried out.
The trigger threshold requires a signal at least
3$\sigma$ above background for at least 5 consecutive frames of 40.96~ms each. The UV background has
been assumed to be the same as in the other simulations at 1 count/pixel/GTU.
However, it is possible that for this specific measurement, the background could
be higher due to the presence of some sunlight. Results are displayed on the right side of Fig~\ref{fig:minieuso-debris}. 

\begin{figure*}[!ht]
    \begin{center}
\includegraphics[width=1.0\textwidth]{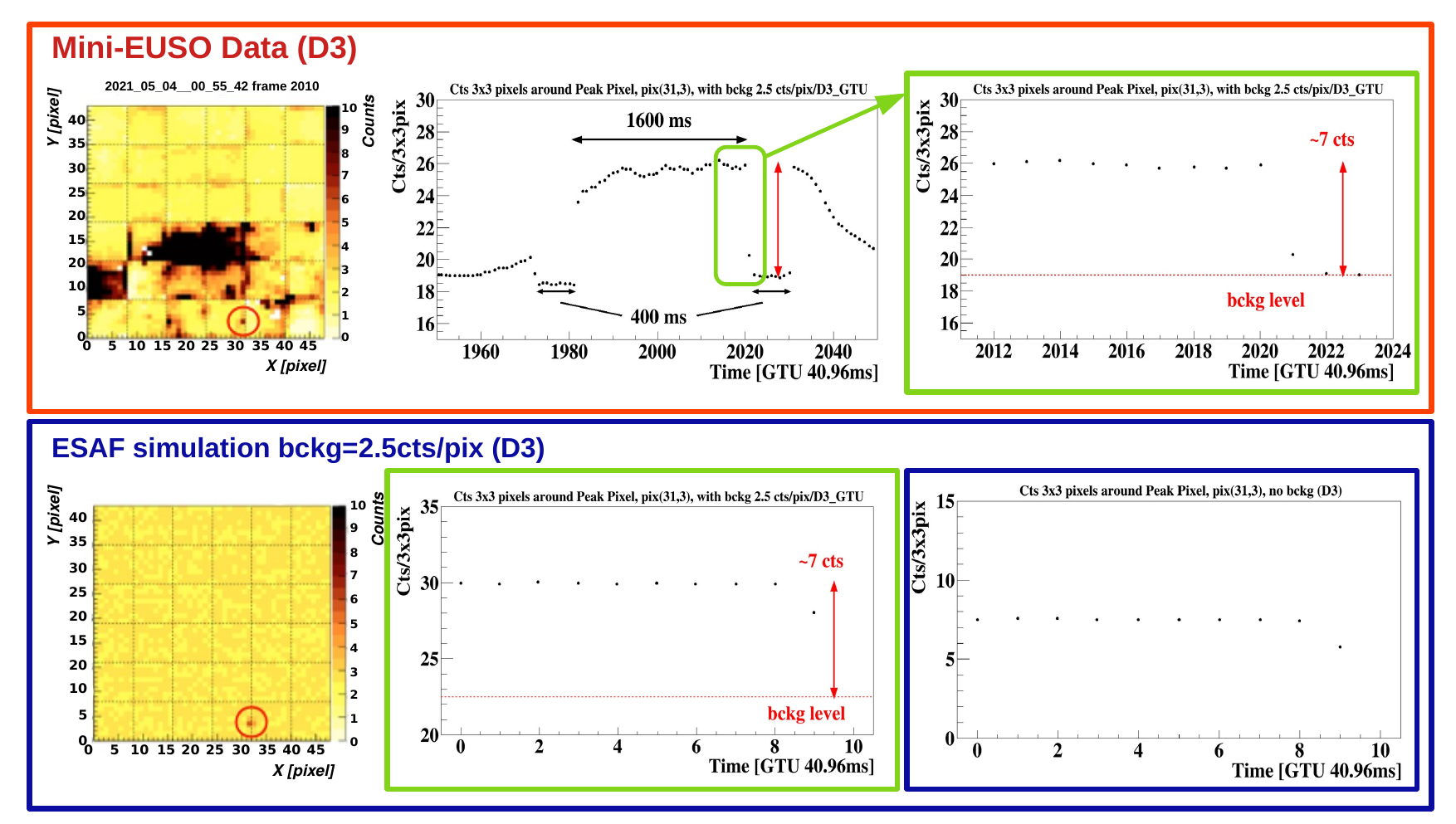}
        \caption{Top: Mini--EUSO flasher campaign event. The central panel shows the light evolution of summed counts of 3$\times$3 pixels around the peak count pixel, indicated by the red-circle on the 
        left plot, for a cycle of 1600 ms pulse (GTU=40.96ms). The transit of the 1600 ms pulse preceded and followed by 400 ms off is clearly seen. Right panel: the zoomed in plot to the duration and timing where ESAF simulated the same event, as shown in the bottom plots.
Bottom: Simulated flasher campaign event by ESAF. The central panel shows the light evolution of summed counts
of 3$\times$3 pixels around the peak count pixel, indicated by the red-circle on the left plot with a background level of 2.5 counts/pixel/GTU, which is resulting in $\sim$15$\%$ difference between simulated and real background. The right panel is the same as the central one without background to see the signal contribution.}
        \label{fig:minieuso-Flasher}
    \end{center}
\end{figure*}

Usually, Mini--EUSO detects on continents ground flashers.
They are often associated
with airport lights (but not always). These types of lights are simulated in ESAF
as point-like sources with a rising and
decaying curve following the specifications of commercial flashers.
In parallel we are also testing the response of ESAF to 
the light emitted by flasher systems that have been built in the framework of the
JEM--EUSO collaboration to provide an in-flight instrument calibration.
The flasher consists of a 100 W COB-UV LEDs, DC power supply and an Arduino circuit.
The flasher campaign has been done at Piana
di Castelluccio, Italy, at 1500 m above sea level (lat. 42.766, lon. 13.190), in the clear night sky conditions of May 3$^{rd}$ - 4$^{th}$, 2021 when
Mini--EUSO was operational while ISS was flying over central Italy.
After including in ESAF the estimation of the detector sensitivity obtained with ground tests, simulations were performed.
The top panel of Fig. \ref{fig:minieuso-Flasher} shows the Mini--EUSO data. The left
plot is the raw data image of one D3 frame (GTU=40.96 ms) while the central one shows the light curve of
the integrated counts of 3$\times$3 pixels around the peak count pixel inside the red circle. The right
plot shows the zoomed in image of the green circled part of the center plot, corresponding to the timing
and duration of the ESAF simulation. 
The bottom panel shows the simulated flasher event with a background level of 2.5 counts/pixel/GTU which is resulting in $\sim$15$\%$ difference between simulated and real background.
The center shows
the integrated counts of 3$\times$3 pixels around the peak pixel inside the red circle in the left plot,
with background photons, while the right shows the integrated counts of the same pixels but without
background photon counts to see the total signal contribution. The tendency of signal  fading as the flasher moves in the FoV is clearly visible in both Mini--EUSO and ESAF data. As
a preliminary result, the number of expected detections obtained by ESAF matches well to
the detected number by Mini--EUSO. 
More details can be found in~\cite{hiroko-icrc2021}.



\subsection{EUSO Balloons configuration}
\label{sec:balloon}

\begin{figure*}[!ht]
\centering
\includegraphics[width=0.45\textwidth]{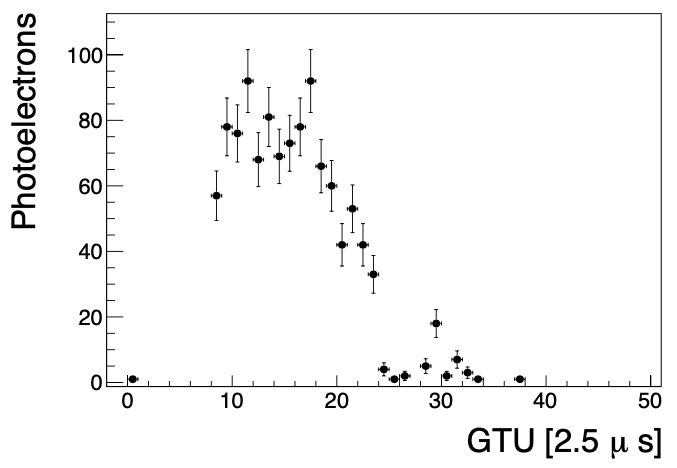}~~
\includegraphics[width=0.50\textwidth]{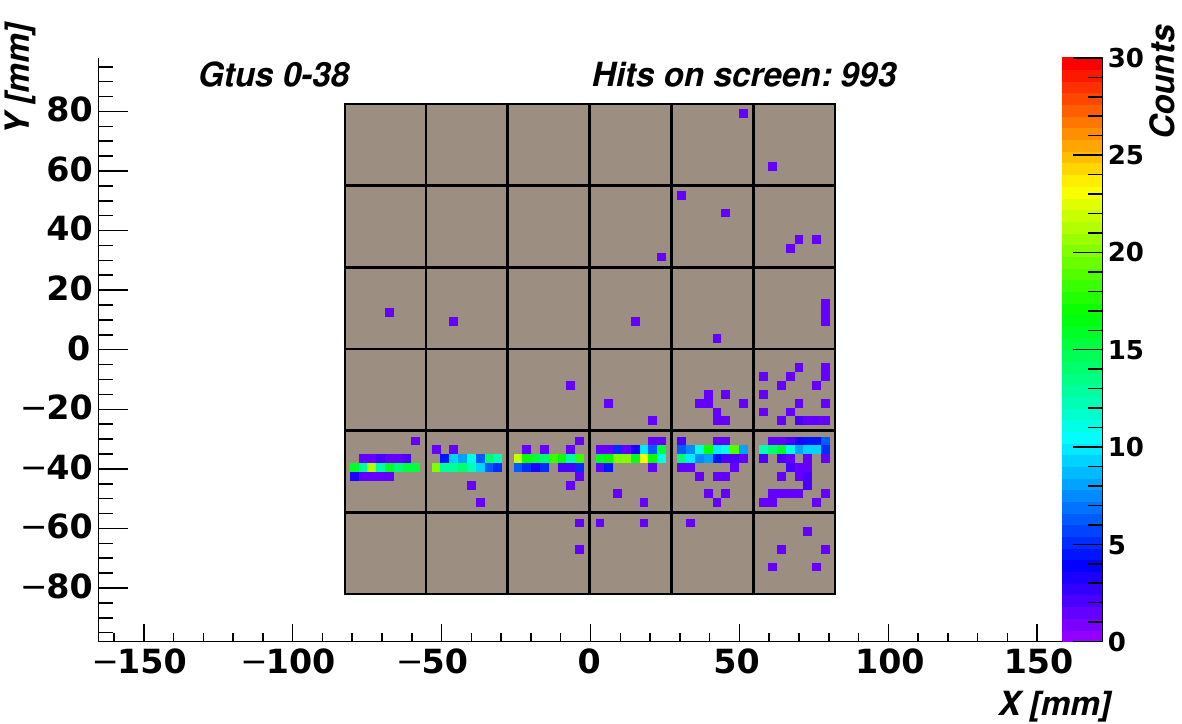}
\caption{A 10$^{19}$ eV, 45$^\circ$ zenith angle simulated proton event. Left: the photoelectron profile for the 
EUSO--SPB1 detector. Right: the photoelectron image for EUSO--SPB1. Image adapted from~\cite{fenu-icrc2019}.}
\label{fig:eusospb1-event}       
\end{figure*}

EUSO--Balloon and EUSO--SPB1 have been implemented in ESAF
while EUSO--SPB2 in its final configuration is currently implemented only in $\overline{\mbox{Off}}\underline{\mbox{Line}}$.
EUSO--Balloon and EUSO--SPB1 share the same configuration in ESAF. Only parameters such as quantum efficiency of MAPMTs, 
balloon height, etc.
change depending on the performed simulation. A ray trace code configuration for the optics response has been implemented
as well together with the parametric option. The adopted trigger logic can be activated too.
Fig.~\ref{fig:eusospb1-event} shows an example of a light profile and shower track for a proton simulated EAS of 10$^{19}$
eV energy and 45$^\circ$ of zenith angle.

\begin{figure*}[h]
\centering
\includegraphics[width=1.00\textwidth]{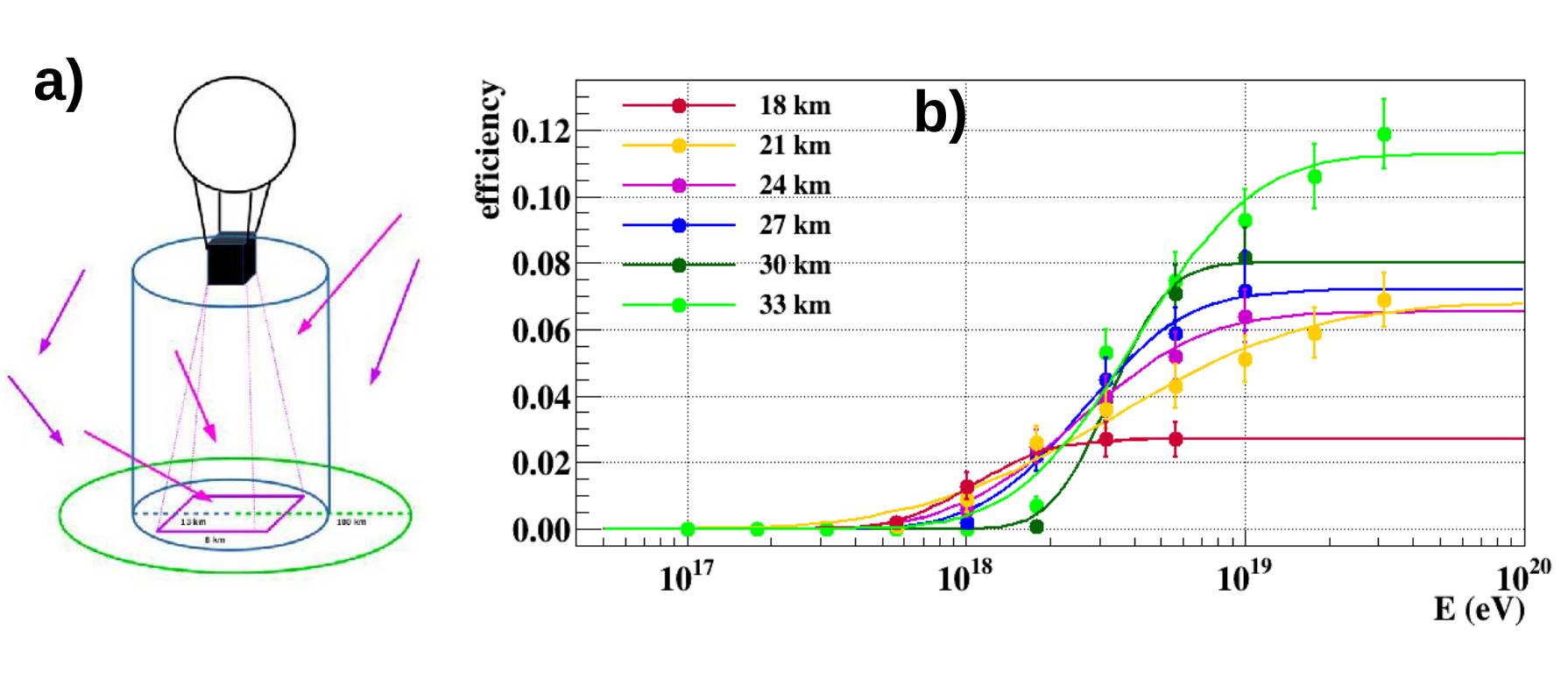}
\caption{Panel a): To speed up the simulation process in balloon configurations, EASs are fully simulated if crossing a cylindrical volume
$\mathcal{V}$ which includes the balloon FoV. EAS not crossing this volume are not simulated.
Panel b): trigger efficiency as a function of proton EASs energy for different heights of the balloon.}
\label{fig:eusospb1-trigger}       
\end{figure*}

A significant difference compared to space-based missions is the much shorter distance between the detector and cascade in the atmosphere. 
As a result, most of the EAS tracks are not fully contained in the FoV. 
To perform an estimation of the energy range of sensitivity to UHECRs and of the expected number of events under specific assumptions of the balloon configuration, a new methodology has been adopted to compute the aperture of the instrument which speeds up the simulation by a factor of 10 at least. We defined a cylindrical volume $\mathcal{V}$ (8 km radius on ground) around the FoV where EAS must pass before being fully simulated by ESAF (see Fig.~\ref{fig:eusospb1-trigger}). 
Showers not passing through this volume are guaranteed to never cross the FoV (which is much smaller, $\pm$4 km side on ground) and therefore are not simulated.


We define the exposure $\mathcal{E}$ as:

\begin{equation}
\mathcal{E}(E) = <\epsilon>(E) \int_{A_{\rm simu}}dA \int_\Omega d \Omega \int_{t_{\rm acq}}dt
\label{eqn:exp-bal}
\end{equation}

The average efficiency $<\epsilon>(E)$ is therefore computed as:

\begin{equation}
<\epsilon>(E) = \frac{\int_{A_{\rm simu}}dA \int_\Omega d \Omega \frac{N_{\rm trigg}(E,x,y,\theta,\phi)}
{N_{\rm simu}(E,x,y,\theta,\phi)}}{\int_{A_{\rm simu}}dA \int_\Omega d \Omega} k,
\label{eqn:efficiency}
\end{equation}
where $k$ represents the average correction factor for the efficiency taking into account the fraction
of events passing through the volume $\mathcal{V}$ with respect to the events on the entire solid angle $N_{\rm test}$:

\begin{equation}
k = \frac{\int_{A_{\rm simu}}dA \int_\Omega d \Omega \frac{N_{\rm simu}}
{N_{\rm test}}}{\int_{A_{\rm simu}}dA \int_\Omega d \Omega},
\label{eqn:kappa}
\end{equation}

\begin{figure}[h]
\centering
\includegraphics[width=1.0\columnwidth]{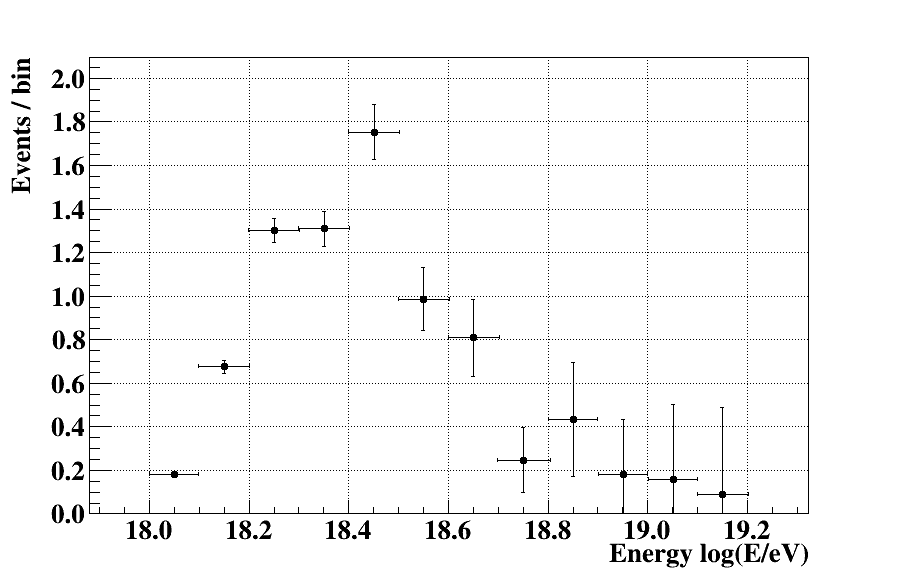}
\caption{The expected number of triggered EAS for EUSO--SPB1 floating at an altitude of 30 km assuming clear sky
condition and a flight duration corresponding to 118 hours of dark time. Image adapted from~\cite{fenu-icrc2017}.}
\label{fig:eusospb1-expected-events}       
\end{figure}

The right panel of Fig.~\ref{fig:eusospb1-trigger} presents an example of the trigger efficiency as a function of 
proton EAS energy for different heights of the balloon.
Finally, the number of events is calculated as

\begin{equation}
N(E) = \mathcal{E}(E)\Psi(E),
\label{eqn:events}
\end{equation}
where $\Psi(E)$ is a fit to the Pierre Auger spectrum~\cite{auger}.
We see in Fig.~\ref{fig:eusospb1-expected-events} the triggered
spectrum as expected by EUSO--SPB1 floating at an altitude of 30 km in clear sky and for a mission duration $t_{\rm acq}$ of 118 hours. This was the total dark time expected prior to launch on the moon phase of March - April 2017. The
total number of events has been calculated by integrating the number of events on the entire energy
range. The spectrum is peaked at energies around 3$\times$10$^{18}$ eV. It is important to point out that
this is the energy at which EUSO--SPB1 was 50\% efficient in EAS
triggering according to the field tests performed at EUSO--TA site in October 2016, prior to flight~\cite{mario-nim2019} (see sec.~\ref{sec:eusota}).

By implementing the cloud distribution in ESAF according to satellite databases along the trajectories of previous NASA-SPB
flights and by assuming 30 hours of acquired data, due to the short EUSO--SPB1 flight, ESAF provided a preliminary expected number of $\sim$1 UHECR along the flight right after the conclusion of the mission. Successive studies which employed both 
$\overline{\mbox{Off}}\underline{\mbox{Line}}$ and ESAF, and took better into account the effective detector performance, the measured background
level and the effective presence of clouds showed that this preliminary estimation
~\cite{eser_icrc19,kenji_icrc19,spb1} has to be decreased by a factor of $\sim$2 (0.4 events in 25 hours of effective good quality data taking). This shows the usefulness of ESAF either as a tool for quick performance studies or for more accurate data analyses.
Additionally, EAS tracks generated with ESAF were extensively used in neural network algorithms to train the networks to 
search for EAS tracks. No track was found in the data~\cite{vrabel_icrc19}.  

\begin{figure*}[h]
\centering
\includegraphics[width=0.40\textwidth]{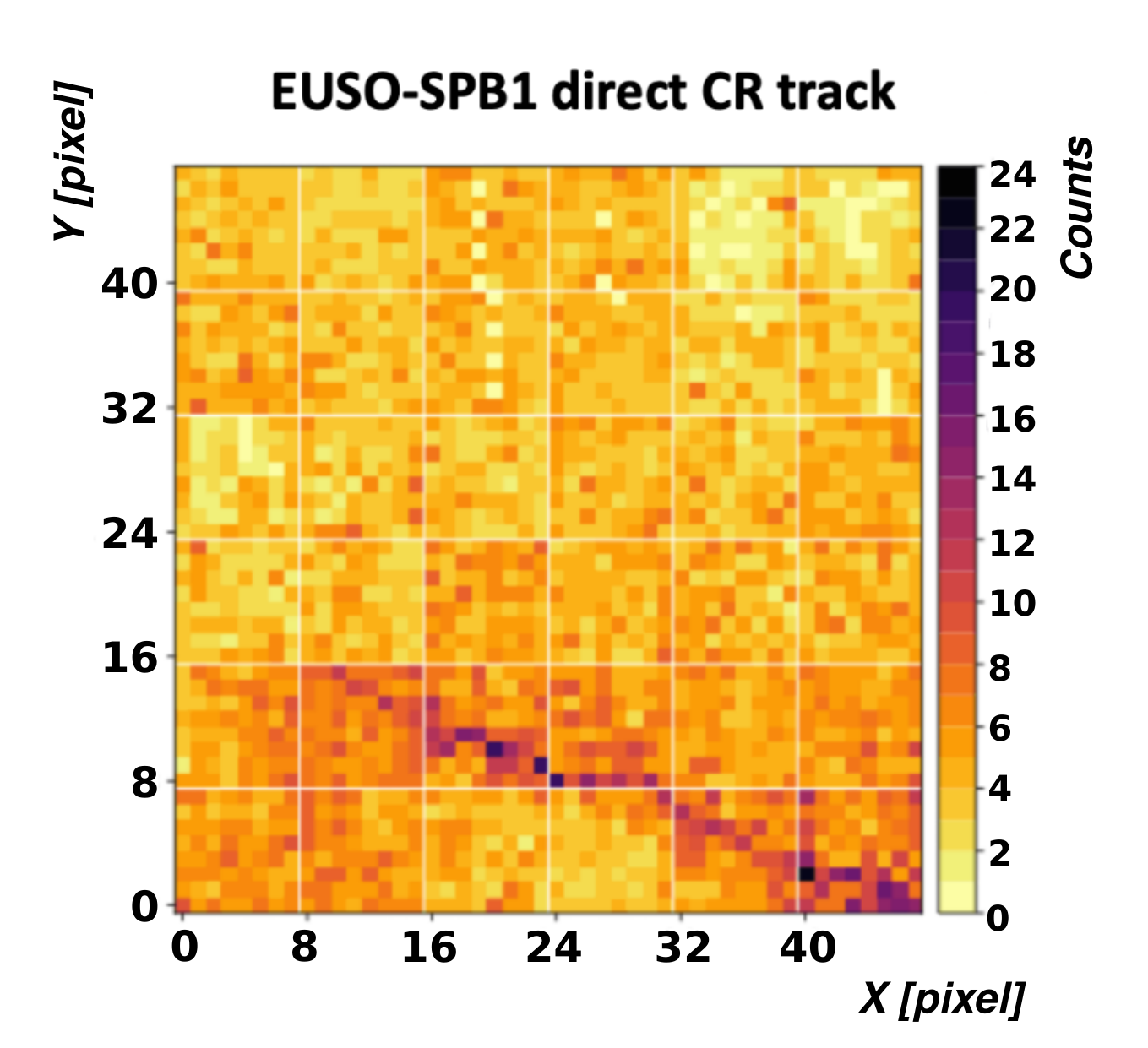}
\includegraphics[width=0.54\textwidth]{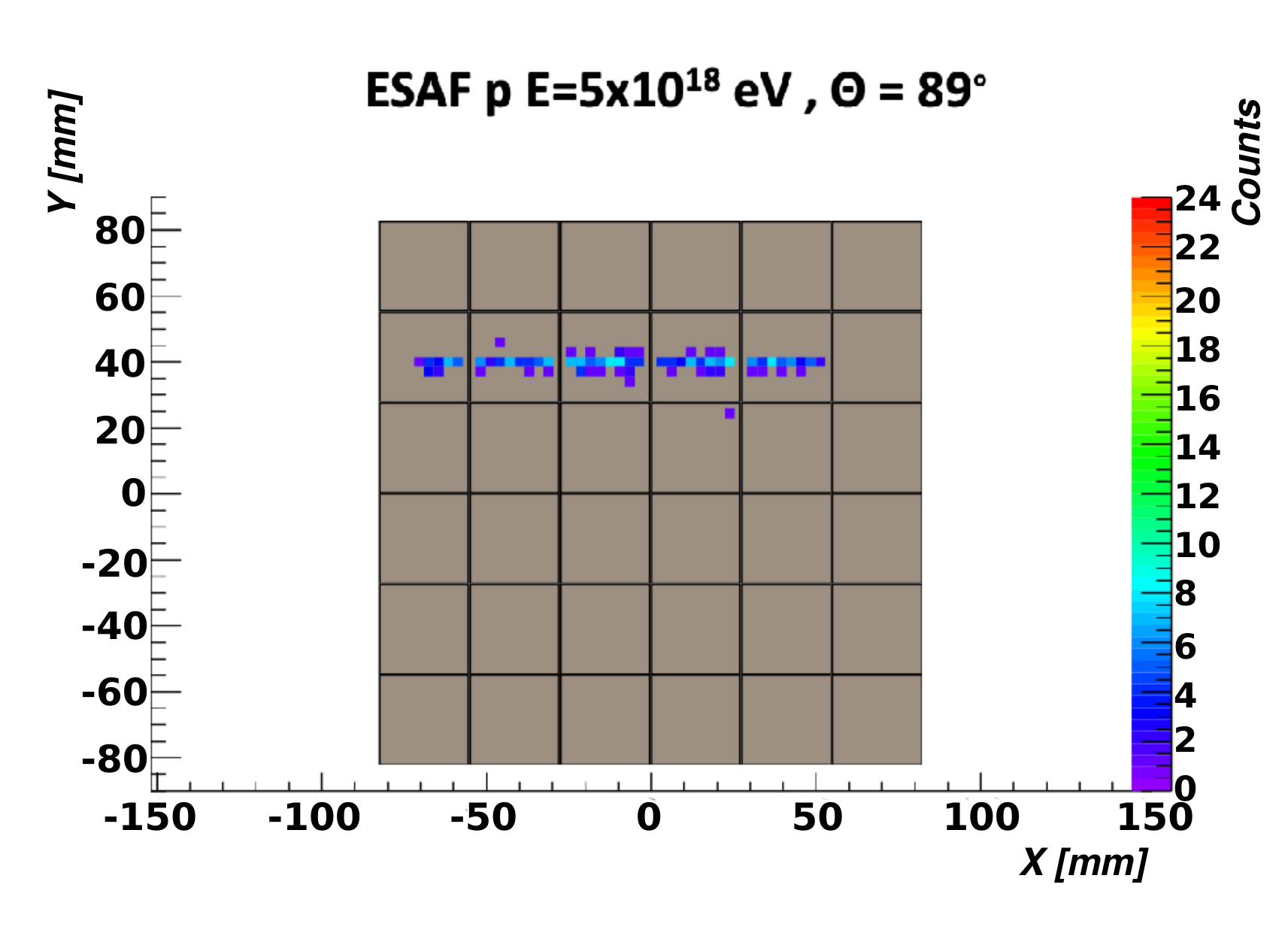}
\caption{Left side: direct cosmic ray triggering the EUSO--SPB1 detector. Right side: simulated proton
EAS of energy E = 5$\times$10$^{18}$ eV crossing the balloon FoV at $\sim$25 km altitude with a zenith
angle of $\theta$ = 89$^\circ$. They both correspond to 1 GTU data.}
\label{fig:spb-crtrack}       
\end{figure*}

At the very beginning of the flight, low energy cosmic rays directly interacting in the detector drew the
attention as they looked like very fast linear tracks, faster than normal EAS events. During one 
single GTU the entire PDM is crossed by such a light pulse. An example of this class of events is
shown in Fig.~\ref{fig:spb-crtrack}. The left side shows the typical light track of a direct cosmic ray quasi-planar to the PDM. The signal lasts one GTU. The right side of the figure shows the attempted production of a similar track with an EAS generated by an UHECR proton. 
The track shown in the figure is generated by a highly inclined EAS with $\theta$ = 89$^\circ$
and energy E = 5$\times$10$^{18}$ eV, with the first interaction point at 280 km from the nadir
position at 30 km altitude, and crossing the detector FoV at a distance of $\sim$4.5 km
from the detector.
The EAS event appears much dimmer than the experimental one. This is due to two different reasons.
The first is that at high altitudes the light emission is smaller due to the much lower atmospheric density. The second is related to the fact that the signal crosses the pixel FoV in $\sim$80 ns as the pixel FoV is $\sim$25 m wide at 4.5 km distance from the balloon. Taking into account a double-pulse resolution of 6 ns, the signal can not exceed a few counts/pixel/GTU as seen in the ESAF event. It is, therefore, extremely difficult to reproduce such a bright and fast signal in the camera with an UHECR event. 
On the other hand, a direct cosmic ray can produce de-excitation in the glass filter or photo-cathode with a decay time comparable to 1 GTU and the signal can be much brighter. 
This study was carried out during the days just after this event was detected and shows the utility of ESAF also as a quasi-online tool to interpret the data.

\begin{figure*}[h]
\centering
\includegraphics[width=1.0\textwidth]{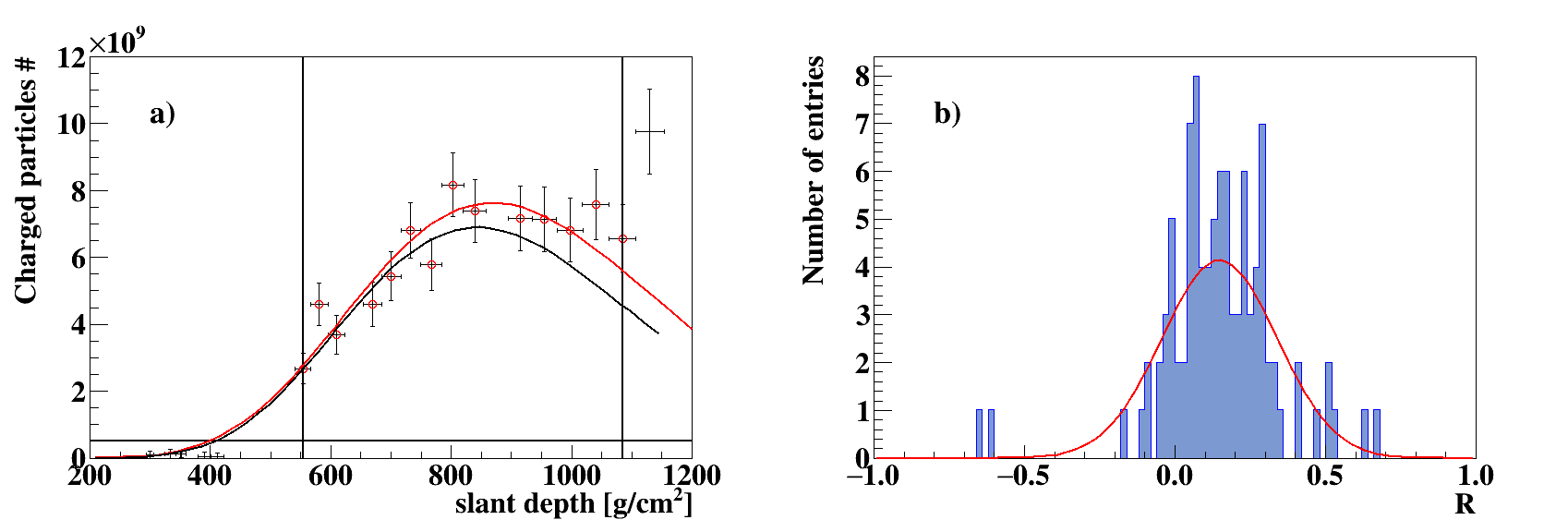}
\caption{Panel a): we show here an example of an event reconstructed by EUSO--Balloon. 
The continuous black line indicates the real 
simulated shower profile (the number of charged particles in the shower at each slant depth). 
The black points show the reconstructed profile as obtained by the algorithms. 
The points which are accepted as 
fit data are indicated by red circles and the GIL fit applied to these points is shown by the continuous red line. 
We clearly see how the automatic fit correctly 
excluded the Cherenkov mark visible at the end of the curve. 
The fitting range and the minimal threshold automatically chosen by the algorithms are also shown as black straight lines. 
Panel b): an example of the $R$ reconstruction parameter fitted over a sample of events with a gaussian function. 
The sigma of the distribution is 0.16, with a mean value of $\sim$0.15 showing a clear overestimation of the energy due to the lack of the backscattered Cherenkov correction. 
Image adapted from~\cite{fenu-balloon-icrc2015}.}
\label{fig:eusoballoon-reco}       
\end{figure*}

A first attempt to adapt the JEM--EUSO reconstruction algorithms to the balloon configuration was performed.
Fig.~\ref{fig:eusoballoon-reco}a shows an example of reconstructed shower profile for EUSO--Balloon. We can see here the real
(black line) and reconstructed profile (points). The GIL fit to the reconstructed points marked in
red is represented as a continuous red line. As the figure shows, the reconstructed profile is slightly overestimated
(as in the case of JEM--EUSO) due to the lack of backscattered Cherenkov correction.
The profile is cut above $\sim$1090~$g/cm^{2}$ due to the impact with the ground. The resulting Cherenkov
reflection feature is also visible in the reconstructed profile. The shower of energy 10$^{19}$ eV and 25$^\circ$ zenith
angle has been reconstructed as 1.1$\times$10$^{19}$ eV. 
The $\chi^2$/DoF is 0.97 and the number of degrees of freedom of the fit is 12.

Fig.~\ref{fig:eusoballoon-reco}b shows a sample of reconstructed events with  
energy 10$^{19}$ eV and 25$^\circ$ zenith angle. As in JEM--EUSO,
we calculated the parameter $R$ (\ref{Rparameter}) for each event. A gaussian 
fit has been performed for the resulting distribution and represented as a red continuous line. The sigma of this distribution is 0.16 providing an estimation of the resolution for this energy and angle. At energies around 3--5$\times$10$^{18}$ the resolution worsens to 
30\% while for 45$^\circ$ inclined EAS a further 10\% decrease in resolution is observed. This is opposite to the JEM--EUSO case, where the reconstruction of inclined EAS gives better
performance. This behaviour is due to the relatively small FoV of the balloon leading to only partial containment within the FoV for inclined showers and hence much larger uncertainties in the reconstruction.
Concerning the angular resolution, the $\gamma_{68}$ analysis provides resolutions which are $\sim$4$^\circ$ for 10$^{19}$ eV and 25$^\circ$ zenith angle. At 3--5$\times$10$^{18}$ eV the resolution worsens to around 
5$^\circ$. On average the angular resolution for 45$^\circ$ inclined EAS worsens by $\sim$2$^\circ$.
These results are consistent with the energy reconstruction, giving again an indication that 
the different kinematics of the signal may be affecting the reconstruction.
The fraction of events that can be reconstructed, even if with modest angular and energy resolutions, is $\sim$70\% at energies
E $>$ 3$\times$10$^{18}$ eV and angles between 25$^\circ$ and 45$^\circ$. 
This is promising
 for the future balloon flights as it demonstrates that for a significant fraction of detected
 events a reconstruction procedure could be applied to derive a reasonable energy
 estimation.

\begin{figure*}[h]
\centering
\includegraphics[width=1.0\textwidth]
{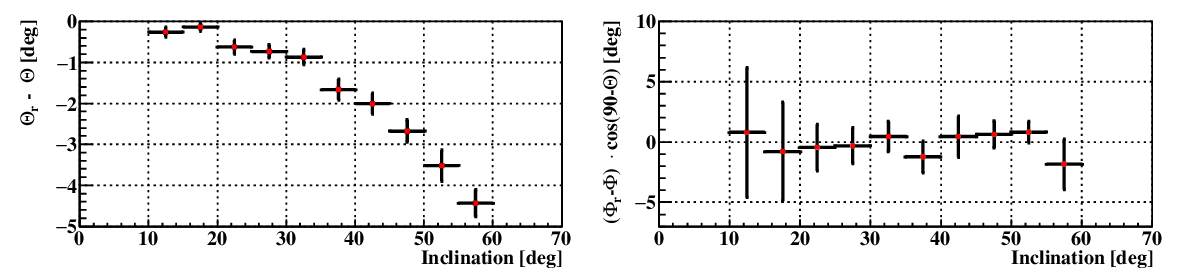}
\caption{Mean value and standard deviation of $\Delta \Theta$  = $\Theta_{rec} - \Theta_{sim}$ (left) and $\Delta \Phi$  = ($\Phi_{rec} - \Phi_{sim}) cos(90-\Theta)$
(right) plotted against the true zenith angle (inclination) for EUSO--Balloon. Image adapted from~\cite{mernik-balloon-icrc2013}.}
\label{fig:balloon-angrec}       
\end{figure*}

We present in Fig.~\ref{fig:balloon-angrec}  results related to the angular reconstruction in EUSO-Balloon. 
Over 12300 EAS were simulated in the energy range between 10$^{18}$ and 10$^{19}$ eV. 
The zenith angles were chosen between 10$^\circ$ and 60$^\circ$. 
All events were distributed randomly having their impact point within an area of 10 $\times$ 10 km$^2$ around nadir position. 
The FoV projected on ground corresponds to an area of 8.4 $\times$ 8.4 km$^2$. 
The lower limit in zenith angle is due to the fact that the track on the FS is too short to be fitted. 
For zenith angles exceeding about 50$^\circ$, the shower track does not fit entirely on the PDM. 
Slightly fewer than 2500 events were successfully triggered and reconstructed. 
The angular reconstruction is evaluated in terms of mean value and standard deviation between reconstructed and simulated angles ($\Delta \Theta$  = $\Theta_{\rm rec} - \Theta_{\rm sim}$ and $\Delta \Phi$  = ($\Phi_{\rm rec} - \Phi_{\rm sim}) \cdot \cos(90-\Theta) $ ). 
The direction of the EAS can be measured sufficiently well when
the zenith angle is between 10$^\circ$ and approximately 50$^\circ$. 
The resolution of $\gamma$, the opening angle between the true shower direction and the reconstructed direction, is found to be $\gamma<5$ for EAS contained within the FoV increasing and saturating at about 8 deg for uncontained showers.
Obviously, the probability that parts of the signal are lost increases at the edge of the FoV, increasing the uncertainty in the arrival direction reconstruction.
More details of this analysis can be found in~\cite{mernik-balloon-icrc2013}.

\subsection{EUSO--TA configuration}
\label{sec:eusota}

\begin{figure*}[h]
\centering
\includegraphics[width=0.44\textwidth]{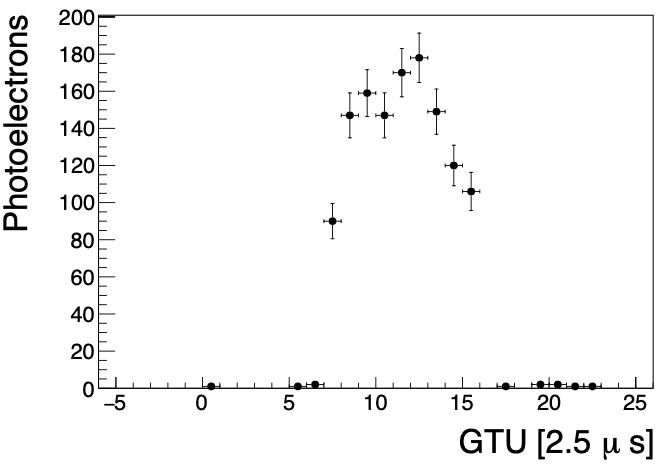}~~
\includegraphics[width=0.53\textwidth]{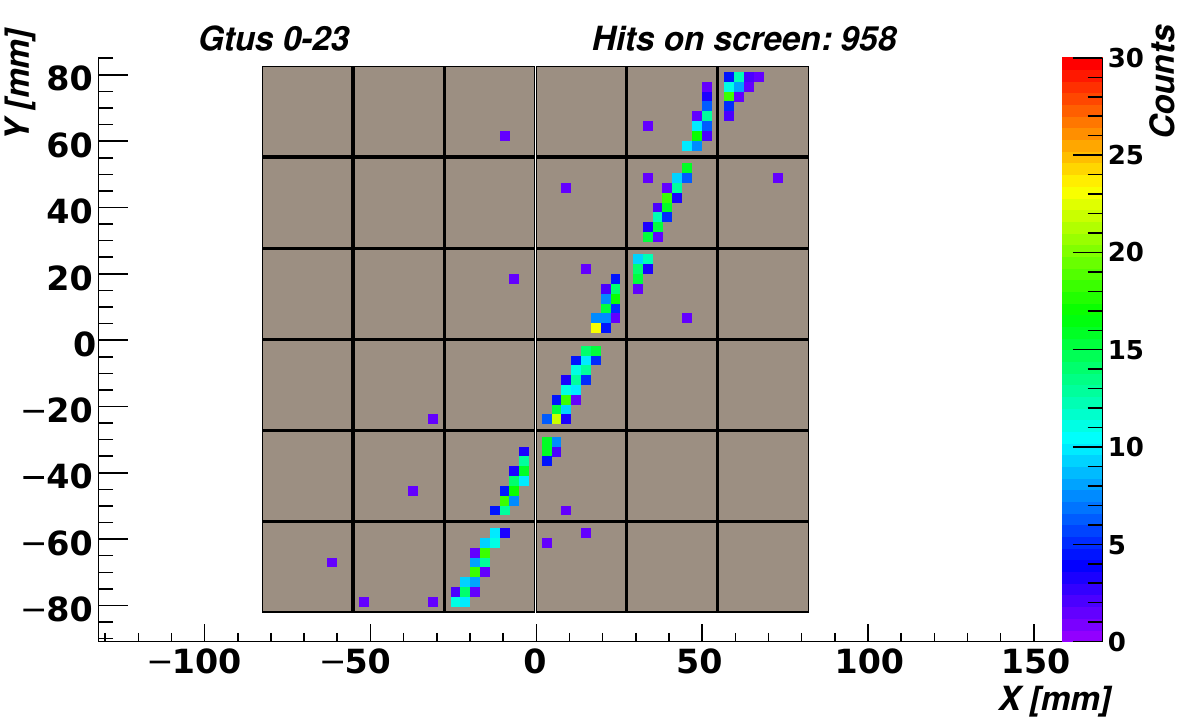}
\caption{A 10$^{19}$ eV, 45$^\circ$ zenith angle simulated proton event impacting at 25 km distance from the detector. Panel a): 
the photoelectron profile for the EUSO-–TA detector. Panel b): the photoelectron image for EUSO–-TA. Image taken 
from~\cite{fenu-icrc2019}.}
\label{fig:eusota-event}       
\end{figure*}

The EUSO--TA configuration has been included in ESAF as well. An example of a simulated 
10$^{19}$ eV, 45$^\circ$ zenith angle event impacting at a distance of 25 km from the detector is shown in 
Fig.~\ref{fig:eusota-event}. 

\begin{figure*}[h]
\centering
\includegraphics[width=1.\textwidth]{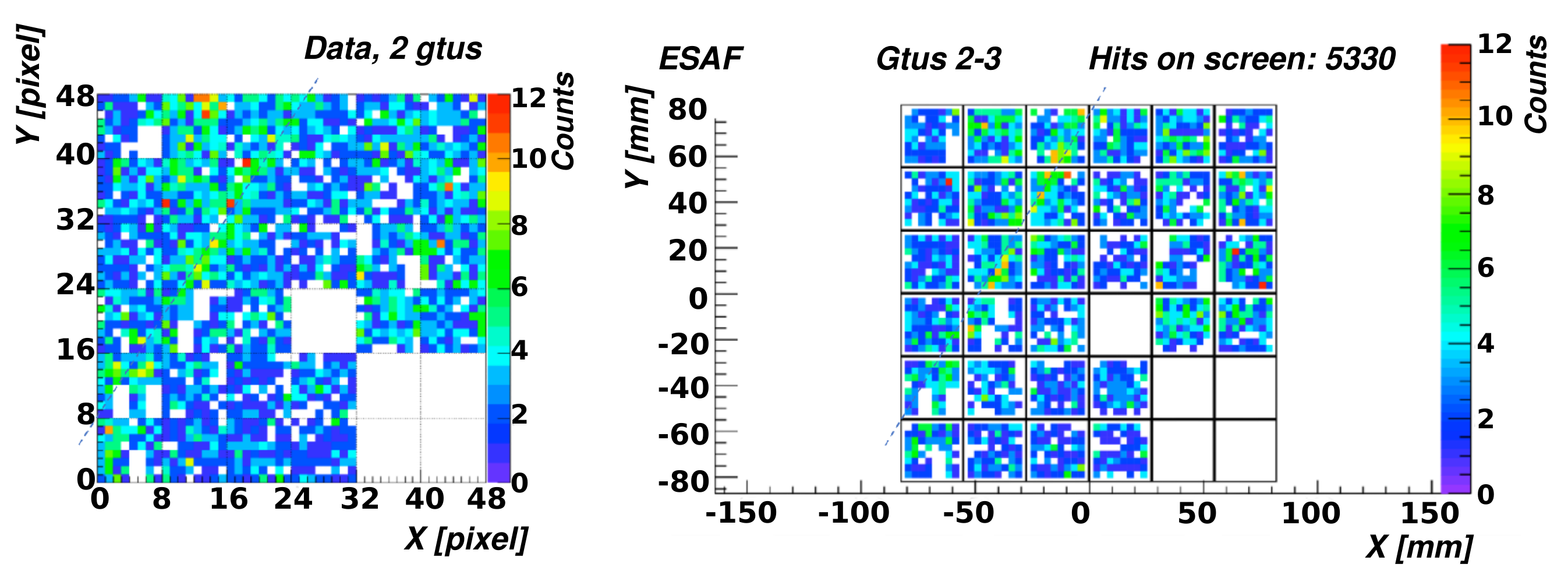}
\caption{Comparison between an UHECR event observed by EUSO--TA (left) and a proton shower simulated with ESAF (right) assuming the EAS parameters provided by the TA reconstruction of the event (E = 2.4$\times$10$^{18}$ eV, zenith $\theta$ = 41$^\circ$). No longitudinal distribution is provided for this event as its duration is confined in 2 GTUs. The background is added to the EAS simulation using real data. The dashed line is just guidance to recognize the event track. Data collected in two following GTUs are added together, reducing signal over noise ratio. The white squares include 1 full EC and 1 MAPMT not operational at that time as well as low efficiency pixels.}
\label{fig:eusota-uhecr_esaf}
\end{figure*}

EUSO--TA was also the first detector implemented in $\overline{\mbox{Off}}\underline{\mbox{Line}}$.
Therefore, at the very
beginning several efforts were made to cross-check the detector implementations and EAS simulations in the two software packages, by comparing the expected number of detected events and their characteristics with those predicted by 
simulations. 
Fig.~\ref{fig:eusota-uhecr_esaf} shows a comparison between an UHECR event observed by EUSO--TA and a proton simulated with ESAF assuming the EAS parameters provided by the TA reconstruction of the event (E = 2.4$\times$10$^{18}$ eV, zenith $\theta$ = 41$^\circ$). In this phase EUSO--TA events were acquired through an external trigger from TA fluorescence detectors. The background is added to the EAS simulation using real data. Taking account of reconstruction uncertainties, the simulated event reasonably reproduces the detected one.

\begin{figure}[h]
\centering
\includegraphics[width=\columnwidth]{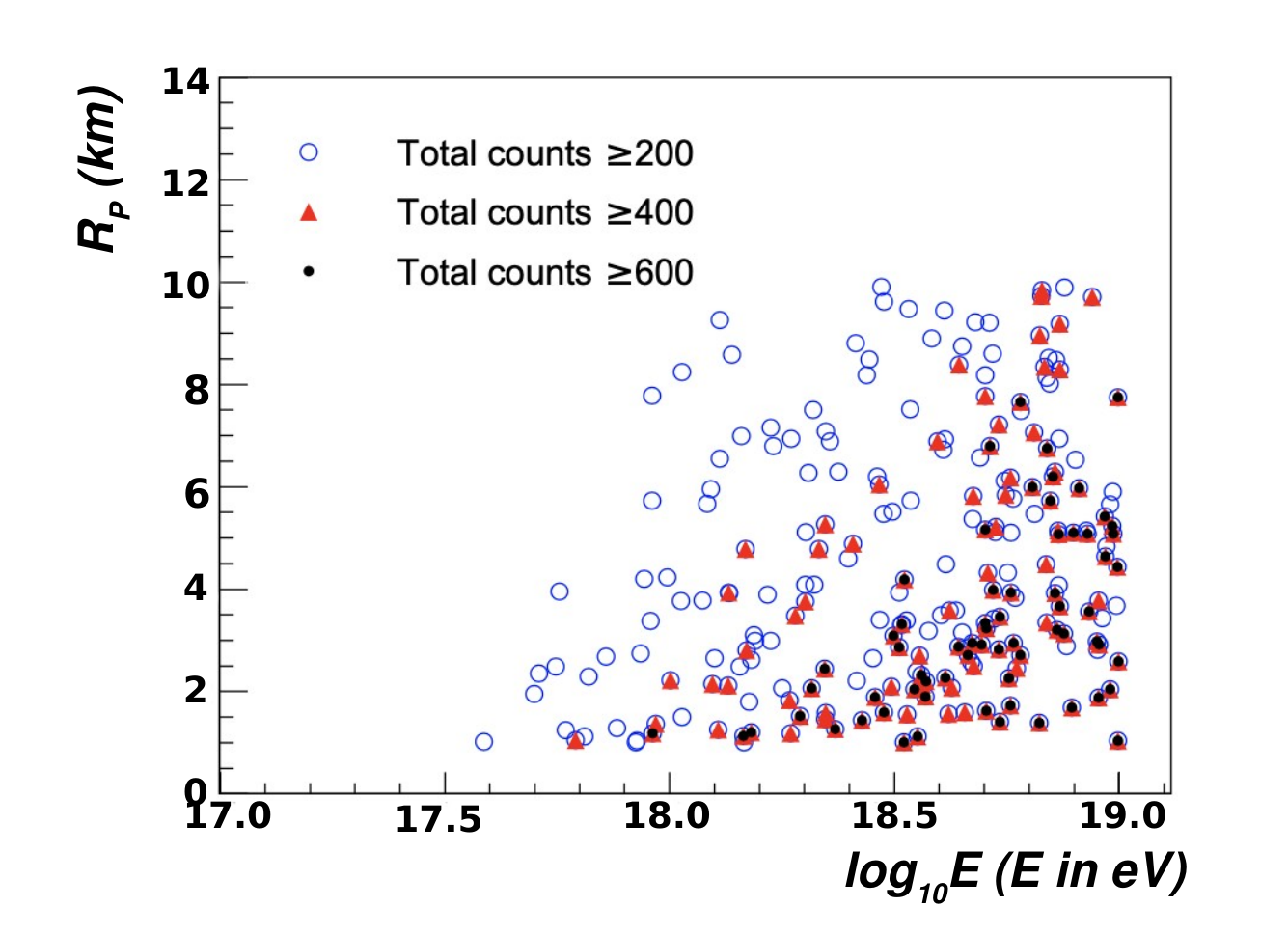}
\caption{Relation between the EAS energy and the impact parameter $R_p$. 1000 proton events were simulated, and among them
229, 118, and 67 events generated more than 200, 400, and 600 counts over 3 consecutive GTUs in the EUSO--TA detector, respectively.
In this plot a cosmic ray power spectrum E$^{-1}$ has been considered: more events at high energies are present in
the plot than expected in reality, for which the average spectrum scales by about E$^{-3}$. Image taken from~\cite{francesca-icrc2017}.}
\label{fig:eusota-detectable}
\end{figure}

The ESAF software was also used at the beginning to cross-check the measured number of UHECRs with expectations from 
simulations~\cite{francesca-icrc2017}.
In this analysis the following assumptions were made: 1) the TA trigger efficiency was assumed to be 100\%; 2) the cosmic ray power spectrum was simulated with an E$^{-1}$ differential spectrum in the range 10$^{17}$ $\leq$ E $\leq$ 
10$^{19}$ eV to have higher statistics of sample for calculating the trigger efficiency for every energy bin. The spectrum was later on re-weighted according to the
IceTop measurements~\cite{icetop} in the 10$^{17}$ - 5$\times$10$^{17}$ eV energy range and with the Auger 
spectrum~\cite{auger} at higher energies; 3) EASs were uniformly generated in azimuth angle with a $\sin(2 \cdot \theta)$ distribution in
zenith angle; 4) the impact point on ground of the shower axis was generated uniformly
within a radius $R$ $\leq$ 50 km around the telescope. \newline The number of EASs within the EUSO--TA FoV
with an impact parameter 1 $\leq$ $R_p$ $\leq$ 10 km was then estimated. A selection was made to require at least
200, 400, or 600 counts in the detector integrated over 3 consecutive GTUs. EASs in EUSO--TA are detected on average
at even closer distances compared to the balloon configuration, so the signal can last no more than 1 GTU.
Fig.~\ref{fig:eusota-detectable} shows the relation between the EAS energy and the impact parameter $R_p$. Among
1000 simulated events with a power spectrum E$^{-1}$, 229, 118, and 67 EASs generated more than 200, 400, and 600 counts in the EUSO--TA detector, respectively.

This number of events was then rescaled by taking into account the IceTop and Auger fluxes.
The results indicate that under the assumption of an acquisition time of 120 hours, the
total acquisition time in 2015, EUSO--TA can detect $\sim$6 events with a signal of at least
600 counts integrated over 3 consecutive GTUs. 
This signal is in the range of the photon-counts
 excess over the background level, typically measured in the events detected by
 EUSO–TA.

These results, extrapolated to 140 hours of effective data taking by EUSO-TA would predict 7 detectable events which is a number compatible with the 9 effectively detected events by EUSO-TA in similar amount of time, thus confirming the capability of ESAF to reproduce the EAS detected by the EUSO-TA.

\section{Conclusions and perspectives}
\label{sec:conclusions}

The JEM--EUSO program is an international effort devoted to the study of \mbox{UHECRs}
from space. The program consists of a 
series of missions some completed and some in preparation, in space, on stratospheric balloons, or on ground. 
All such detectors demand an extensive simulation work to estimate the performance and to support the data analysis.
The original ESAF package developed in the framework of the EUSO project evolved within the JEM--EUSO 
collaboration. A new branch was created which includes the detector configurations of almost all the missions
which have been conceived, performed, or under development in the program: 
JEM--EUSO, K--EUSO, POEMMA, TUS, Mini--EUSO, EUSO--Balloon, EUSO--SPB1, and EUSO--TA. 
Along the years, the ESAF code was an essential tool to assess the expected performance of the
missions, to fine tune their objectives and to drive the technological developments in order to satisfy the scientific
requirements as well as to help in the interpretation of the collected experimental data.
In fact, thanks to the missions of the JEM--EUSO program, for the first time the ESAF simulated data have been compared to experimental measurements.
Moreover, the large panel of scientific objectives of missions 
like Mini--EUSO and TUS,
required the implementation in ESAF of simplified versions of a large variety of
luminous transients
aside from EASs, such as elves, meteors, space debris and flashers which were not present in the original ESAF version.
In conclusion, evaluation of the expected performance of future space-based
missions and interpretation of the experimental data with ESAF simulations 
indicate that a space-based UHECR detector has a satisfactory performance to contribute in unveiling the origin of the extremely energetic particles of the Universe.


ESAF is expected to remain an essential tool for further developments of the planned
missions of the program and for the interpretation of the data they will acquire.

\bmhead{Acknowledgments}
This work was partially supported by Basic Science Interdisciplinary Research Projects of
RIKEN and JSPS KAKENHI Grant (22340063, 23340081, and 24244042), by the ASI-INAF agreement n.2017-14-H.O,
by the Italian Ministry of Foreign Affairs and International Cooperation, 
by the Italian Space Agency through the ASI INFN agreements Mini-EUSO n. 2016-1-U.0, EUSO-SPB1 n. 2017-8-H.0, OBP (n. 2020-26-Hh.0) and EUSO-SPB2 n. 2021-8-HH.0.  
by NASA award 11-APRA-0058, 16-APROBES16-0023, 17-APRA17-0066, NNX17AJ82G, NNX13AH54G, 80NSSC18K0246, 
80NSSC18K0473, 80NSSC19K0626, 80\-NSSC\-18K\-0464 and 80NSSC22K1488 in the USA,
by the French space agency CNES,
by the Deutsches Zentrum f\"ur Luft- und Raumfahrt,
the Helmholtz Alliance for Astroparticle Physics funded by the Initiative and Networking Fund
of the Helmholtz Association (Germany),
by Slovak Academy of Sciences MVTS JEM--EUSO, by National Science Centre in Poland grants 2017/27/B/ST9/02162 and
2020/37/B/ST9/01821,
by Deutsche Forschungsgemeinschaft (DFG, German Research Foundation) under Germany Excellence Strategy - EXC-2094-390783311,
by Mexican funding agencies PAPIIT-UNAM, CONACyT and the Mexican Space Agency (AEM),
as well as VEGA grant agency project 2/0132/17, by grant S2018/NMT-4291 ( TEC2SPACE-CM ) "Desarrollo y explotación de nuevas tecnologías para instrumentación espacial en la Comunidad de Madrid", and by State Space Corporation ROSCOSMOS and the 
Interdisciplinary Scientific and Educational School of Moscow University "Fundamental and Applied Space Research". L. W. Piotrowski acknowledges financing by the Polish National Agency for Academic Exchange within Polish Returns Programme no. PPN/PPO/2020/1/00024/U/00001 and National Science Centre, Poland grant no. 2022/45/B/ST2/02889.

A special thanks goes to the original developers of the ESAF code. Without their effort it would have been a much tougher
job to setup a simulation and analysis framework as complex and as powerful as ESAF. Their help and support during the transition phase between EUSO and JEM--EUSO is deeply acknowledged.



\begin{appendices}

\section{The ESAF design}\label{sec:esaf-design}

The ESAF simulation code is structured in several independent modules with the {\it SimuApplication} at the top of the hierarchy. An instance of this class is created in the
{\it simu\_main.cc} file where the method {\it SimuApplication::DoAll()} is called. This method performs
the iterative call of the {\it SimuApplication::DoEvent ()} method which takes care of the
entire physical process on a single event basis. This method creates an instance of
the {\it LightToEuso} class which executes the entire process from primary particle to photons
on pupil. Several choices are available for which simulator is to be used but the default
option is the {\it StandardLightToEuso} class. By calling the {\it StandardLightToEuso::Get()},
the virtual {\it Get()} methods of the shower generator and of the light production and transport
will be called. Each one of the above-mentioned {\it Get()} methods delivers output objects describing
the shower profile, photons in atmosphere and photons on pupil. The choice of
the object oriented approach shows its power here where the call of several polymorphic
{\it Get()} methods allows great flexibility.
The {\it SimuApplication::DoEvent()} method calls the virtual {\it Detector::Get()} method. This method takes care of the entire detector simulation. At this stage, several choices
are available between various detector configurations. The most important of
them are the {\it EusoDetector} (activating the RIKEN ray trace code), the
{\it G4Detector} (activating the Geant 4 optics) and other testing or debugging detector simulators.
Calling one of the above described methods activates both optics and electronics
simulators. As final output of the entire procedure a {\it Telemetry} object is produced.

\begin{figure*}[h]
\centering
\includegraphics[width=1.\textwidth]{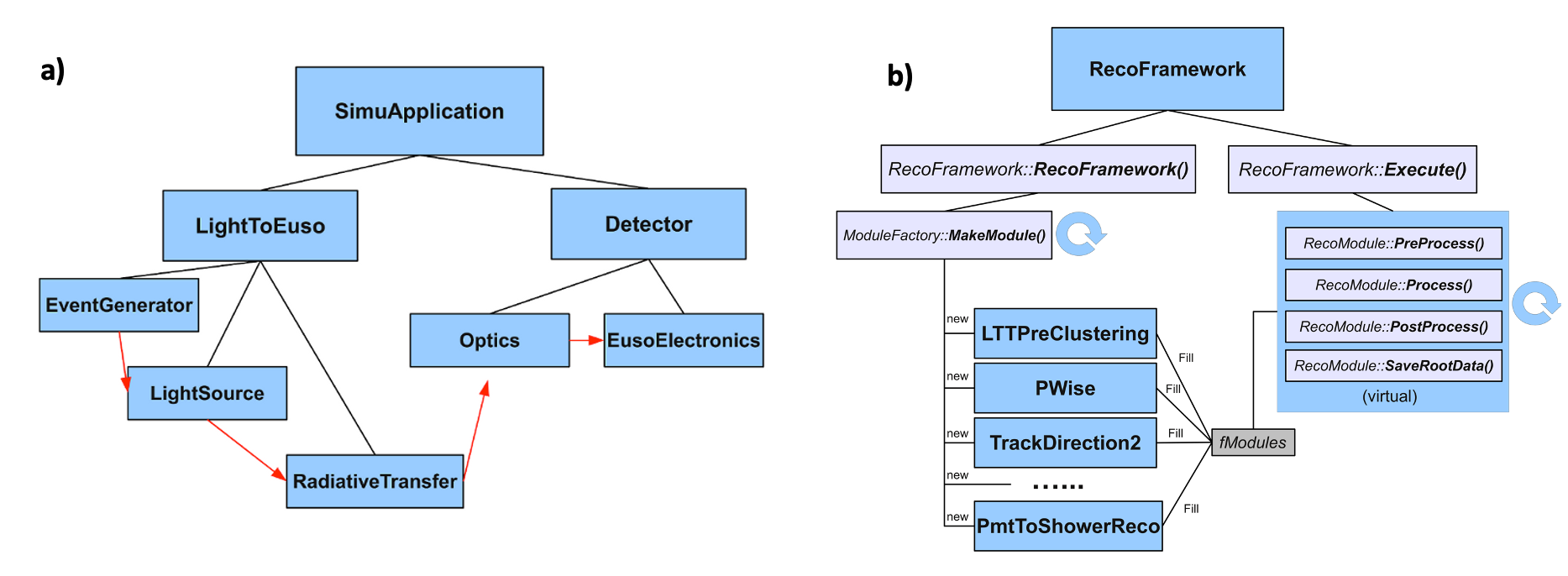}
\caption{Panel a): A schematic view of the ESAF Simu application structure. The main application is the so called {\it SimuApplication}. The {\it LightToEuso} application takes care of all the physical process from shower to detector. The {\it EusoDetector} (Detector in the panel) application performs the simulation of optics and electronics. Panel b): A sketch of the reconstruction framework. The main application {\it RecoFramework} calls iteratively the {\it MakeModule} method which allocates all the required modules. In the {\it Execute} method the operations of all the modules are performed. All the modules are inheriting from the {\it RecoModule} class. The virtual methods {\it PreProcess, Process, PostProcess} and {\it SaveRootData} are called for all the allocated
modules. In this picture, blue boxes represent classes, blue-gray boxes methods, the gray box is a \CC~{\it vector} and the circular arrow indicates iterative repetition of some method or sequence of methods. Image adapted from~\cite{spb2}.}
\label{fig:esaf-design}
\end{figure*}

The reconstruction procedure is activated in the {\it reco main.cc} file. Here an instance
of the {\it RecoFramework} class is created and the method {\it RecoFramework::Execute()} is called.
While in the constructor function {\it RecoFramework::RecoFramework()} the module chain is
built, the {\it RecoFramework::Execute()} method performs the entire sequence of calls to reconstruct
the event. In fact, the module sequence is firstly initialized through an iterative
call of the {\it ModuleFactory::MakeModule()} method which allocates all the {\it RecoModule} objects
requested by parameter files. A vector named {\it fModules} with all the pointers to the {\it RecoModule} objects is created. In the {\it RecoFramework::Execute()} method all the
modules (which inherit from {\it RecoModule}) are initialized, called and cleared. Eventually
all the output data are saved in a ROOT file. For performing all the mentioned
operations, the polymorphic methods {\it RecoModule::PreProcess(), RecoModule::Process(), RecoModule::
PostProcess() and RecoModule::SaveRootData()} are declared in each module. Each
module has a specific function which can be either pattern recognition, direction fitting,
profile reconstruction or $X_{max}$ and energy reconstruction. Several modules have been
implemented in the course of the years but the most current are the
{\it LTTPreClustering} and {\it PWISE} for the pattern recognition, the {\it TrackDirection2} for the direction
reconstruction and the {\it PmtToShowerReco} for the energy reconstruction. A schematic
view of the above mentioned structure is shown in Fig.~\ref{fig:esaf-design}.

\section{The main simulation parameters}\label{sec:esaf-param}
We present in the following the description of the parameters used in the simulations. Such parameters are the standard but a set of alternatives is still present in ESAF. A review of such parameters can be found in Chapter 3 of the bibliographic reference
\cite{fenu}.

The parameterizations for the shower generation, light production and transport are reported in the following:

\begin{itemize}
\item Shower parameterization: GIL \cite{gil}
\item Secondary particle energy distribution: Giller parameterization \cite{Giller}
\item Fluorescence Yield: Nagano et al. \cite{nagano} 
\item Atmospheric model: Standard US atmosphere \cite{US_Atmosphere}
\item Clouds: parametric; uniform layer; optical depth, altitude and thickness by parameter.
\item Rayleigh scattering and Ozone absorption modelled according to the lowtran 7 package \cite{lowtran7}
\end{itemize}

The detector is simulated according to the following assumptions:
\begin{itemize}
    \item A parametric optics in case of POEMMA and TUS. A Gaussian PSF is assumed (ParamOpticalSystem)
    \item The RIKEN ray trace code for: JEM-EUSO, K-EUSO, Mini-EUSO, EUSO-TA, EUSO-Balloon and EUSO-SPB (NOpticalSystem)
    \item A parametric optical adaptor is simulated to collect photons on the photocatode (IdealOpticalAdaptor)
    \item A parametric PMT is simulated. A detection efficiency accounting for quantum and collection efficiency is applied to the photons reaching the photocatode. The gain and gain fluctuations are read in by parameter. A threshold in charge is set on the anode signal to determine the number of counts per pixel per GTU.
\end{itemize}

The event reconstruction is based on a chain of modules that aim at the reconstruction of the primary parameters:
\begin{itemize}
    \item the PWISE method is adopted for trace identification. This is optimal for the angular reconstruction \cite{guzman}.
    \item the LTTPatternRecognition is adopted for trace identification. This is optimal for the profile reconstruction
    \item the TrackDirection2Module is adopted to reconstruct the direction of the shower as indicated in \cite{mernik-ea}.
    \item the PmtToShowerReco is adopted to reconstruct the profile of the shower. After a fit of the profile it is possible to retrieve the energy and the $X_{max}$ (see \cite{fenu-ea}).
\end{itemize}

\section{A comparison between ESAF and \offline~simulation codes}\label{sec:esaf-offline}

The two official software packages adopted by the JEM--EUSO collaboration are the 
Euso Simulation and Analysis Framework (ESAF)~\cite{berat}, originally developed
within the EUSO project, and the 
\offline\ package~\cite{offline} originally designed for the Pierre Auger 
Observatory~\cite{auger-nim}. 
The main motivations to adopt both packages are:
a)~it is straightforward to re-adapt the EUSO code to the JEM--EUSO configuration; b)~\offline\ output is extensively tested  within
the Pierre Auger Observatory and thus with experimental data;
c)~the possibility to adopt both packages gives opportunities for 
cross-checks. 
In the following we present a subset of tests that were done in the past years to cross-check the response of the two simulation codes for a mutual validation. A thorough comparison of the codes is out of the scope of this paper.

\begin{figure}[h]
\centering
\includegraphics[width=\columnwidth]{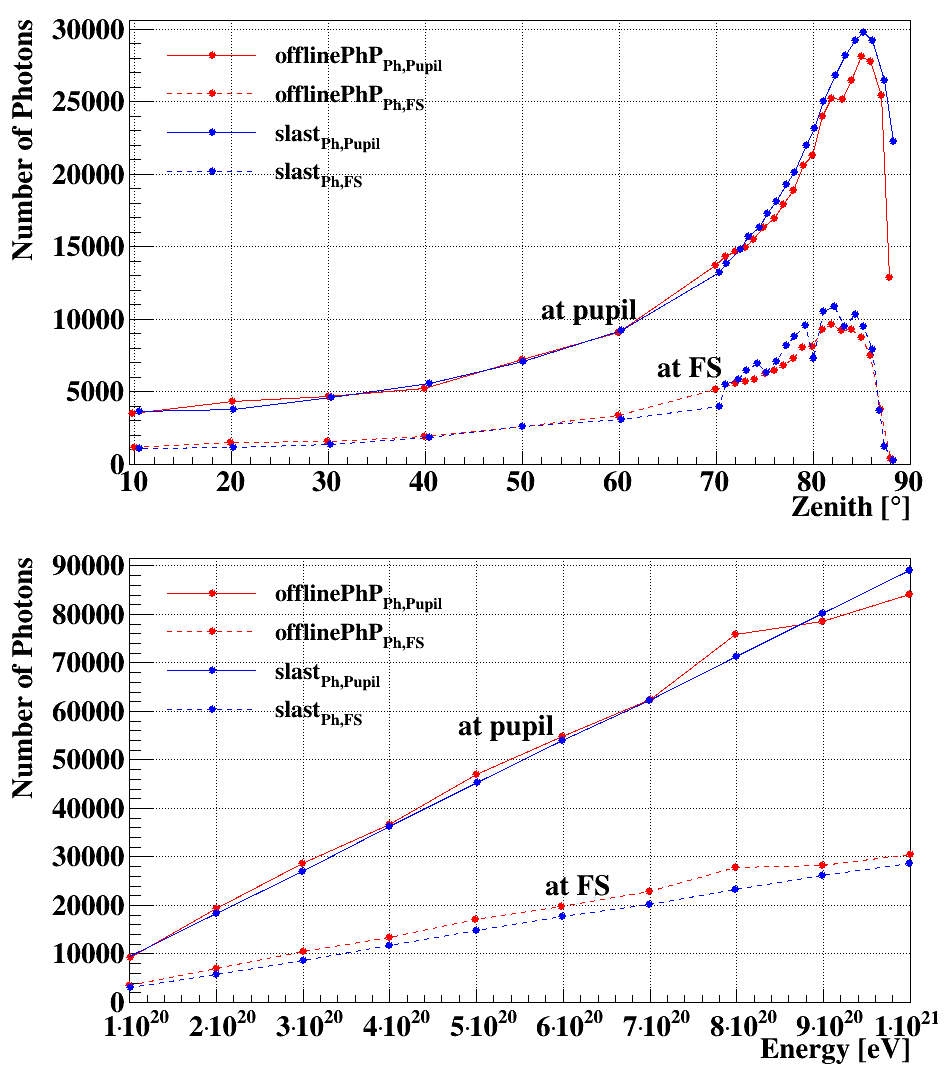}
\caption{Comparison between light signals received by the JEM-EUSO detector from simulated proton showers using CONEX+\offline~or ESAF/SLAST simulation codes. Red color indicates the photon counts obtained using \offline~and blue color the corresponding photon counts obtained with SLAST; the continuous lines refer to the signal at the pupil level, while the dashed lines to the signal collected at the Focal Surface (FS) level.
The top panel shows the dependence of the photon counts on the zenith angle of the EAS for 10$^{20}$ eV events, while the bottom panel shows the energy dependence of the signal for a fixed zenith angle of 60$^\circ$. Figure adapted from~\cite{blaicher}.}
\label{fig:esaf-offline-comparison}
\end{figure}

A first test was performed by comparing the light signal produced in atmosphere by an EAS and its propagation through the atmosphere up to the detector's level. In this case, the light signal generated by showers simulated with SLAST and propagated through the atmosphere using the routines embedded in ESAF was compared to the signal induced by similar showers simulated with CONEX and propagated through the atmosphere according to \offline~routines. Both sets of showers were then passed through the JEM-EUSO detector with ESAF. In this way the possible differences in the results had to be ascribed only to the cascading process and/or light generation in atmosphere.
Fig.~\ref{fig:esaf-offline-comparison} shows the results of this comparison by simulating proton showers with fixed zenith angle (60$^\circ$) at different energies (top panel) and proton showers at fixed energy (10$^{20}$ eV) with variable zenith angle (bottom panel). The comparison is performed at both the pupil (=aperture) and the focal surface levels. Red color indicates the photon counts obtained using the CONEX+\offline~ package and blue color the corresponding photon counts obtained with ESAF/SLAST. There is a general agreement in the light intensity at all energies and angles with a slightly higher signal by ESAF/SLAST at high zenith angles, but the differences remain within 10\% level. More details can be found in~\cite{blaicher}.

\begin{figure*}[h]
\centering
\includegraphics[width=1.\textwidth]{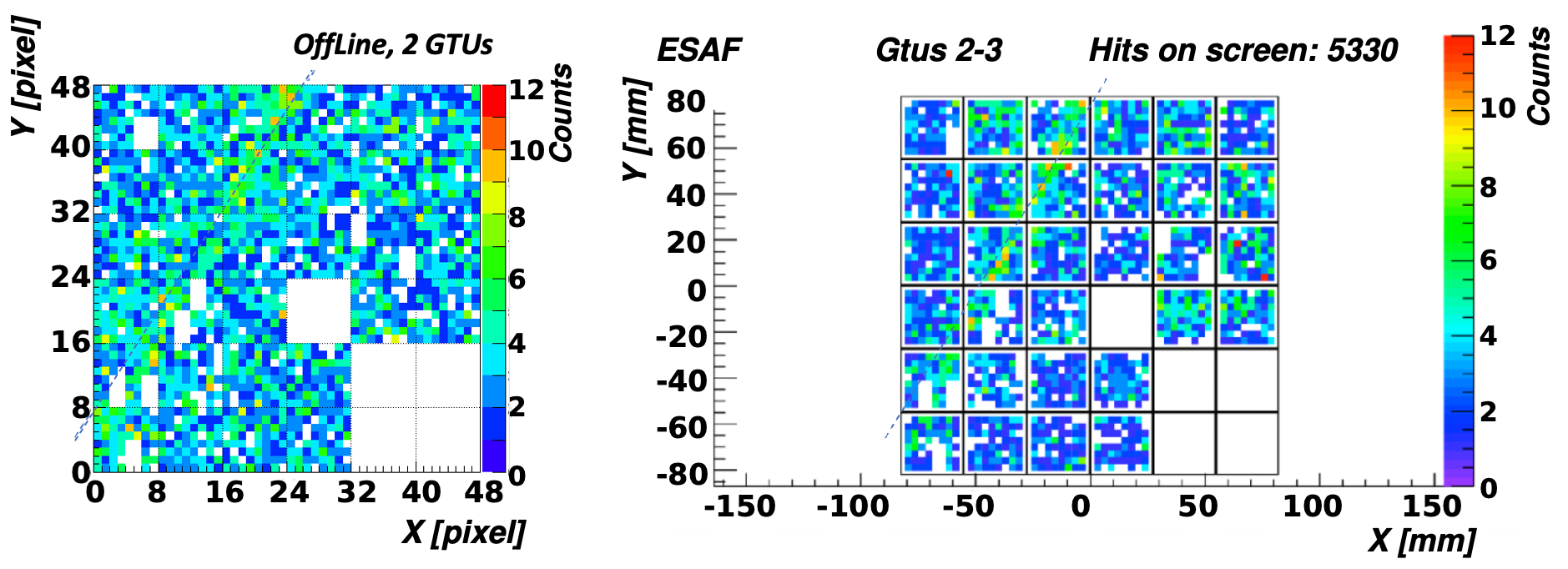}
\caption{Comparison between a proton shower simulated with \offline~(left) and ESAF (right) assuming the EAS parameters provided by the TA reconstruction of the event (E = 2.4$\times$10$^{18}$ eV, zenith $\theta$ = 41$^\circ$) shown on the left side of Fig.~\ref{fig:eusota-uhecr_esaf}.}
\label{fig:eusota-esaf-offline}
\end{figure*}

The second test refers to the capability of reproducing through an end-to-end simulation the experimentally detected EAS events by EUSO-TA, including the detector response. Fig.~\ref{fig:eusota-esaf-offline} shows this time the comparison of the event already displayed in Fig.~\ref{fig:eusota-uhecr_esaf} (left panel) simulated with \offline~and ESAF, each one simulating the detector response. Also in this case, both ESAF and \offline~provide a quite similar simulation of the event. 





\end{appendices}


\bibliography{sn-bibliography}





\end{document}